\documentclass[a4paper,11pt]{article}
\pdfoutput=1 

\usepackage{jheppub} 

\usepackage[T1]{fontenc} 
\usepackage{dcolumn}
\usepackage{bm}
\usepackage{mathrsfs}
\usepackage{multirow}
\usepackage{amsmath}
\usepackage{slashed}
\usepackage{slashbox}
\newcommand{\mathsym}[1]{{}}

\newcommand{\baz}{\begin{array}{cc}}
\newcommand{\bad}{\begin{array}{ccc}}
\newcommand{\ba}{\begin{array}{c}}
\newcommand{\ea}{\end{array}}
\newcommand{\be}{\begin{equation}}
\newcommand{\ee}{\end{equation}}
\newcommand{\bea}{\begin{eqnarray}}
\newcommand{\eea}{\end{eqnarray}}

\newcommand{\bi}{\begin{itemize}}
\newcommand{\ei}{\end{itemize}}
\newcommand{\bmt}{\begin{pmatrix}}
\newcommand{\emt}{\end{pmatrix}}
\newcommand{\bt}{\begin{tabular}}
\newcommand{\et}{\end{tabular}}

\newcommand{\benu}{\begin{enumerate}}
\newcommand{\eenu}{\end{enumerate}}
\newcommand\T{\rule{0pt}{2.8ex}} 
\newcommand{\obb}{0\nu\beta\beta}

\newcommand{\mee}{\langle m_{ee}\rangle}

\newcommand{\bav}{\begin{array}{cccc}}

\usepackage{amstext,amssymb}
\usepackage{amsmath}
\usepackage{graphicx}
\usepackage{xspace}
\usepackage{color}
\usepackage{units}
\usepackage{slashed} 

\title{\boldmath Neutrinoless Double Beta Decay in LRSM with Natural Type-II seesaw Dominance}
\author[a]{Prativa Pritimita}
\author[a]{\hspace*{-0.1cm}, Nitali Dash}
\author[a]{\hspace*{-0.1cm}, Sudhanwa Patra}

\affiliation[a]{Center of Excellence in Theoretical and Mathematical Sciences, \\
Siksha 'O' Anusandhan University, Bhubaneswar-751030, India}

\emailAdd{pratibha.pritimita@gmail.com}
\emailAdd{nitali.dash@gmail.com}
\emailAdd{sudha.astro@gmail.com}

\abstract{
We present a detailed discussion on neutrinoless double beta decay within a class of left-right symmetric 
models where neutrino mass originates by natural type-II seesaw dominance. The spontaneous symmetry breaking 
is implemented with doublets, triplets and bidoublet scalars. The fermion sector is extended with an 
extra sterile neutrino per generation that helps in implementing the seesaw mechanism. The presence of 
extra particles in the model exactly cancels type-I seesaw and allows large value for Dirac neutrino mass 
matrix $M_D$. The key feature of this work is that all the physical masses and mixing are expressed in terms 
of neutrino oscillation parameters and lightest neutrino mass thereby facilitating to constrain light neutrino 
masses from $0\nu\beta\beta$ decay. With this large value of $M_D$ new contributions arise due to; 
i) purely left-handed current via exchange of heavy right-handed neutrinos as well as sterile neutrinos, 
ii) the so called $\lambda$ and $\eta$ diagrams. 
New physics contributions also arise from right-handed currents with right-handed gauge boson  
$W_R$ mass around $3$~TeV. From the numerical study, we find that the new contributions 
to $0\nu\beta\beta$ decay not only saturate the current experimental bound but also give 
lower limit on absolute scale of lightest neutrino mass and favor NH pattern of light neutrino 
mass hierarchy. 
}

\begin{document} 
\maketitle
\flushbottom

\section{Introduction}
The discovery that neutrinos have mass and they mix with each other has put before us another vital question 
to speculate over; whether they are Dirac or Majorana~\cite{Majorana:1937vz} particles. Even more intriguing is the theoretical origin 
of such a tiny mass and the mass hierarchy among them. The different seesaw mechanisms like 
type-I~\cite{Minkowski:1977sc,Yanagida:1979as,GellMann:1980vs,Mohapatra:1979ia}, 
type-II~\cite{Cheng:1980qt, Lazarides:1980nt,Magg:1980ut,Schechter:1980gr,Wetterich:1981bx} and 
others {\bf} 
that appropriately explain this tiny mass further require them to be Majorana particles. 
On the contrary, Majorana nature of neutrinos violates global lepton number by 2 units that is regarded as an accidental 
symmetry within the Standard Model (SM). This leads to the search of a rare process called Neutrinoless Double Beta Decay 
($0\nu\beta\beta$) that only can assuredly endorse the Majorana nature of neutrinos and lepton number violation in 
nature~\cite{Schechter:1981bd}. While new theories are trying to find new physics contributions to $0\nu\beta\beta$ 
decay, the experiments are looking for lower limits on the half-lives being decayed. Of yet, GERDA~\cite{Agostini:2013mzu} using $Ge^{76}$ 
gives lower limit on half life of $0\nu\beta\beta$ decay as 
$T_{1/2}^{0 \nu} > 2.1 \times {10}^{25}$ yrs at 90\% C.L. whereas the limits provided by EXO-200~\cite{Gando:2012zm} 
and KamLAND~\cite{Auger:2012ar} are 
$T_{1/2}^{0 \nu} > 1.6 \times {10}^{25}$ yrs and $T_{1/2}^{0 \nu} > 1.9 \times {10}^{26}$ yrs respectively. 
The combined limit from KamLAND-Zen comes to be $T_{1/2}^{0 \nu} > 3.4 \times {10}^{26}$ yrs at 90\% C.L.
This process can be mediated by the exchange of a light Majorana neutrinos or by new particles appearing 
in various extensions of 
SM~\cite{Mohapatra:1986su,Babu:1995vh,Hirsch:1995vr,Hirsch:1995ek,Hirsch:1996ye,
Deppisch:2012nb,Pas:1999fc,Humbert:2015yva,Allanach:2009xx,Pas:2000vn,Deppisch:2006hb,Pas:2015eia,Helo:2013dla,Deppisch:2012nb,Ge:2015bfa}.

Within preview of BSM physics, left-right symmetric models (LRSM) \cite{Mohapatra:1974gc, Pati:1974yy, Senjanovic:1975rk,Senjanovic:1978ev,
Mohapatra:1979ia,Mohapatra:1980yp} are found to be best suited frameworks for explaining the origin of maximal parity violation 
in weak interactions and the origin of small neutrino mass. 
This class of models, based on the gauge group $SU(2)_L \times SU(2)_R \times U(1)_{B-L} \times SU(3)_C$, 
when studied at TeV scale interlinks high energy collider physics to low energy phenomena like neutrinoless 
double beta decay and other LFV processes (see refs.~\cite{Ge:2015yqa,Tello:2010am,Barry:2013xxa,Patra:2015bga,Lindner:2016lpp,Deppisch:2016scs,
Deppisch:2014zta,Deppisch:2014qpa,Awasthi:2013ff,Patra:2014goa,Dev:2014xea,Bambhaniya:2015ipg,Borah:2013lva,Patra:2012ur,
Chakrabortty:2012mh,Dev:2013vxa,Nemevsek:2011hz,Dev:2014iva,Keung:1983uu,Das:2012ii,Bertolini:2014sua,Beall:1981ze,
Deppisch:2015cua,Hirsch:1996qw,Dev:2013oxa,Dev:2015pga,Dhuria:2015cfa}). Moreover, the left-right symmetric models 
can also accommodate stable dark matter candidate contributing $25\%$ energy budget of the Universe
~\cite{Heeck:2015qra,Garcia-Cely:2015quu,Borah:2016ees,Patra:2015vmp,Patra:2015qny}. 
In conventional left-right symmetric models where symmetry breaking is implemented with scalar triplets and bidoublet, the 
light neutrino mass is governed by type-I plus type-II seesaw mechanisms 
$$m_\nu= - M_D M^{-1}_R M^T_D + M_L = m^{I}_\nu + m^{II}_\nu\,.$$
Here $M_L (M_R)$ is the Majorana mass term for light left-handed (heavy right-handed) Majorana neutrinos 
arising from respective VEVs of left-handed (right-handed) scalar triplets and $M_D$ is the Dirac neutrino 
mass matrix connecting light-heavy neutrinos. The scale of $M_R$ is decided by the vacuum expectation value 
of right-handed scalar triplet which spontaneously breaks LRSM to SM. Thus, the smallness of light neutrino 
mass is connected to high scale of parity restoration i.e, $10^{15}~$GeV clearly making it inaccessible 
to current and planned accelerator experiments. Moreover when LRSM breaks around TeV scale, 
the gauge bosons $W_R$, $Z_R$, right-handed neutrinos $N_R$ and scalar triplets $\Delta_{L,R}$ get mass 
around that scale allowing several lepton number violating signatures at high energy as well as low energy 
experiments. A wide range of literature provides discussions on neutrinoless double beta decay within 
TeV scale LRSM assuming type-I seesaw dominance~\cite{Chakrabortty:2012mh} or type-I plus type-II~
\cite{Chakrabortty:2012mh,Dev:2013vxa,Barry:2013xxa,Das:2012ii,Bertolini:2014sua,Borah:2016iqd,Borah:2015ufa} seesaw mechanisms. 
Some more scenarios have been studied in \cite{Deppisch:2014zta,Tello:2010am,Patra:2014goa,
Chakrabortty:2012mh,Ge:2015yqa,Barry:2013xxa,Awasthi:2016kbk,Awasthi:2015ota} where type-II seesaw 
dominance relates the light and heavy neutrinos with each other. Other works that discuss 
complementarity study of lepton number, lepton flavour violation and collider signatures in LRSM 
with spontaneous D-parity breaking mechanism also embed the framework in a non-SUSY $SO(10)$ GUT
~\cite{Awasthi:2013ff,Deppisch:2014zta,Deppisch:2014qpa,Nayak:2013dza}. One should bear in mind that 
the new physics contributions to neutrinoless double beta decay mainly involves left-right mixing 
(or light-heavy neutrino mixing) which crucially depends on Dirac neutrino mass $M_D$. Necessarily 
$M_D$ should be large in order to expect LNV signatures at colliders. Contrary to this, the type-II 
seesaw dominance can be realized with suppressed value of $M_D$ or with very high scale of parity 
restoration. Studies that assume $M_D \to 0$ therefore miss to comment on LNV, LFV and Collider aspects 
involving left-right mixing. We thus feel motivated to explore alternative class of left-right symmetric 
models which allows large value of $M_D$ and carries light and heavy neutrinos proportional to each 
other. 

This work considers a TeV scale LRSM where symmetry breaking is implemented 
with scalar bidoublet $\Phi$, doublets $H_{L,R}$ and triplets $\Delta_{L,R}$. The scalar bidoublet 
carrying $B-L$ charge $0$ provides Dirac masses to charged fermions as well as to neutrinos. 
The scalar triplets with $B-L$ charge 2 units provide Majorana masses to light and heavy neutrinos. 
One extra sterile fermion $S_L$ per generation also finds place in the model that help in implementing 
extended type-II seesaw mechanism. The scalar doublets $H_{L,R}$ play the same role as $S_L$. 
An interesting feature of this new class of LRSM is that it provides possibility of achieving type-II 
seesaw dominance when parity and $SU(2)_R$ break at same scale. Moreover this framework allows large 
value for Dirac neutrino mass matrix $M_D$ thereby leading to new physics contributions to 
neutrinoless double bea decay i.e, i) from purely left-handed currents via exchange of heavy right-handed 
and extra sterile neutrinos, ii) from purely right handed currents via exchange of heavy right-handed neutrinos, 
iii) from so called $\lambda$ and $\eta$ diagrams. This work aims to carefully analyze the new contributions 
to $0\nu\beta\beta$ in order to derive the absolute scale of light neutrino masses and mass hierarchy. 

The complete work is structured as follows. In Sec.\ref{sec:lrsm}, we briefly discuss the generic and TeV scale LRSMs 
in context of neutrino mass and associated lepton number violation. Sec.\ref{sec:seesaw-type-II} highlights the natural 
realization of type-II seesaw dominance. Sec.\ref{sec:0nu2beta_LRSM}
lays out the basic ingredients for neutrinoless double beta decay and the calculation of Feynman amplitudes. 
Sec.\ref{sec:halflife-effLNV} and Sec.\ref{sec:numerics} are devoted towards the numerical study of LNV $0\nu\beta\beta$ 
contributions within the present framework. 
In Sec.\ref{sec:conclusion} we summarize our results. 
      
\section{The Left-Right Symmetric Model and Lepton Number Violation}
\label{sec:lrsm}
The left-right symmetric model~\cite{Mohapatra:1974gc, Pati:1974yy,Senjanovic:1975rk,Senjanovic:1978ev, Mohapatra:1979ia,
Mohapatra:1980yp} is based on the gauge group 
\begin{equation}
\mathcal{G}_{LR} \equiv SU(2)_L \times SU(2)_R \times U(1)_{B-L} \times SU(3)_C \,.
\end{equation}
In this class of models, the difference between baryon $B$ and lepton $L$ number is defined 
as a local gauge symmetry. The electric charge $Q$ is defined as 
\begin{equation}
	Q = T_{3L} + T_{3R} + \frac{B-L}{2} = T_{3L} + Y \,.
\end{equation}
Here, $T_{3L}$ and $T_{3R}$ are, respectively, the third component of isospin of the gauge 
groups $SU(2)_L$ and $SU(2)_R$, and $Y$ is the hypercharge. The usual leptons and quarks are 
given by
\begin{eqnarray}
 \ell_{L} &= \begin{pmatrix}\nu_L \\ e_L\end{pmatrix} \, \sim
(\mathbf{2},\mathbf{1},\mathbf{-1}, \mathbf{1})\, , & \ell_{R} =
\begin{pmatrix}\nu_R \\ e_R\end{pmatrix} \sim
(\mathbf{1},\mathbf{2},\mathbf{-1}, \mathbf{1}) \, , \\[1mm]
 q_{L} &= \begin{pmatrix}u_R \\ d_R\end{pmatrix} \, \sim
(\mathbf{2},\mathbf{1},\mathbf{\frac{1}{3}},\mathbf{3})\, , & q_{R} =
\begin{pmatrix}u_R \\ d_R\end{pmatrix} \sim
(\mathbf{1},\mathbf{2},\mathbf{\tfrac{1}{3}},\mathbf{3})\,.
\end{eqnarray}
The left-right symmetry calls for the presence of right-handed neutrinos and this makes the model suitable for explaining light neutrino masses. 
For generating fermion masses one needs a scalar bidoublet $\Phi$ with the following matrix representation
\begin{equation}
 \Phi \equiv \begin{pmatrix} \phi_1^0 & \phi_2^+ \\ \phi_1^- & \phi_2^0 \end{pmatrix} \sim (\mathbf{2},\mathbf{2},\mathbf{0},\mathbf{1})\, ,
\end{equation}
The relevant Yukawa interactions are expressed as, 
\begin{eqnarray}
-\mathcal{L}_{Yuk} &\supset& \overline{q_{L}} \left[Y_1 \Phi + Y_2 \widetilde{\Phi} \right] q_R 
+\,\overline{\ell_{L}} \left[Y_3 \Phi + Y_4 \widetilde{\Phi} \right] \ell_R+ \mbox{h.c.}\,,
\label{eqn:LR-Yuk}
\end{eqnarray}
where $\widetilde{\Phi} = \sigma_2 \Phi^* \sigma_2$ and $\sigma_2$ is the second Pauli matrix. 
The scalar bidoublet takes a non-zero VEV as,
\begin{equation}
\langle \Phi \rangle
=
\begin{pmatrix}
 v_1 & 0 \\
0 & v_2 
\end{pmatrix}\,,
\end{equation}
it yields masses for quarks and charged leptons as
\begin{eqnarray}
&&  M_u =  Y_1 v_1 + Y_2 v^*_2\,, \quad \quad 
    M_d =  Y_1 v_2 + Y_2 v^*_1\,, \quad \quad \nonumber \\
&&  M_e =  Y_3 v_2 + Y_4 v^*_1 \,.\quad \quad
\end{eqnarray}
One can generate Dirac masses for light neutrinos using scalar bidoublet as
\begin{equation}
M^\nu_D\equiv M_D = Y_3 v_1 + Y_4 v^*_2 \,.
\end{equation}
However, the Majorana masses for neutrinos depend crucially on how spontaneous symmetry breaking 
of LRSM down to the SM i.e, $\mathcal{G}_{LR} \to \mathcal{G}_{SM}$ 
is implemented. 
\subsection{Lepton number violation and the origin of neutrino mass}
\label{subsec:numass}
The spontaneous symmetry breaking of LRSM to SM goes in favor of neutrino mass generation and 
associated lepton number violation. This happens in the following three ways
\begin{itemize}
\item with Higgs doublets $H_L(\mathbf{2},\mathbf{1},\mathbf{-1},\mathbf{1}) \oplus H_R(\mathbf{1},\mathbf{2},\mathbf{-1},\mathbf{1})$, 
\item with scalar triplets $\Delta_L(\mathbf{3},\mathbf{1},\mathbf{2},\mathbf{1}) \oplus \Delta_R(\mathbf{1},\mathbf{3},\mathbf{2},\mathbf{1})$,
\item with the combination of doublets and triplets $H_L \oplus H_R$ and $\Delta_L \oplus \Delta_R$. 
\end{itemize}
In the first case, $H_R$ breaks the LR symmetry while the left-handed counterpart is required for left-right 
invariance. Though this framework holds a minimal scalar spectrum it lacks Majorana mass for neutrinos and thus 
forbids any signature of lepton number violation or neutrinoless double beta decay. Since the light neutrinos 
here owe their identity to Dirac fermions, their masses can only be explained by adjusting Yukawa couplings through 
the non-zero VEVs of scalar bidoublet. Other important roles that this scalar bidoublet plays are to break the SM 
gauge symmetry to low energy theory and provide the masses to charged fermions. Using the Yukawa interactions given 
in Eq.(\ref{eqn:LR-Yuk}) and with $Y_3 \ll Y_4$, $v_2 \ll v_1$ and $\theta_1,\theta_2=0$, the masses for charged leptons 
and the light neutrinos are given by
\begin{eqnarray}
&&  M_e \simeq Y_4 v^*_1\,, \quad \quad M_D \simeq v_1\left(Y_3+M_e \frac{v_2}{v^2_1} \right)\,.
\end{eqnarray}
However, a pleasant situation arises in the second case where $\Delta_R$ carrying $B-L$ charge $2$ breaks the 
LR symmetry to SM. The inclusion of $\Delta_L$ and $\Delta_R$ in the framework generate Majorana masses 
for light as well as heavy neutrinos and thus violate lepton number by two units. This calls for a 
possibility of smoking-gun same-sign dilepton signatures at collider as well as neutrinoless double beta decay 
in low energy experiments. The interaction terms involving scalar triplets and leptons are given by
\begin{eqnarray}
-\mathcal{L}_{yuk} &\supset& f_{ij}\left[\overline{(\ell_{Li})^c}\ell_{Lj}\Delta_L+
\overline{(\ell_{Ri})^c}\ell_{Rj}\Delta_R\right]+ \mbox{h.c.}\,.
\label{eqn:LR-Yuk-triplets}
\end{eqnarray}
Using Eq.(\ref{eqn:LR-Yuk}) and Eq.(\ref{eqn:LR-Yuk-triplets}), the resulting mass matrix for neutral leptons 
in the basis $\left(\nu_L, N^c_R\right)$ reads as
\begin{equation}
\mathcal{M}_\nu= \left( \begin{array}{cc}
              M_{L} & M_{D}   \\
              M^T_{D} & M_{R}
                      \end{array} \right) \, ,
\label{eqn:numatrix}       
\end{equation}
where, $M_D$ is the Dirac neutrino mass matrix, $M_L (M_R)$ is the Majorana mass matrix arising 
from the non-zero VEV of LH (RH) scalar triplet.
After diagonalization, the resulting light neutrino mass can be written as a combination 
of canonical type-I and type-II seesaw formula
\begin{eqnarray}
m_{\nu} = -M_{D}\, M_{R}^{-1}\, M_{D}^{T} + M_L = m_{\nu}^I + m_{\nu}^{II}\, ,
\label{type2a}
\end{eqnarray}
where, $m_{\nu}^I$ ($ m_{\nu}^{II}$) is denoted as the type-I (type-II) contribution to light neutrino masses,
$$m_{\nu}^I=-M_{D}\, M_{R}^{-1}\, M_{D}^{T}, \quad \quad m_{\nu}^{II}= f\, v_L=f\, \langle \Delta^0_L \rangle\,. $$

In conventional left-right symmetric models, where parity and $SU(2)_R$ break at same scale, the analytic 
formula for induced VEV of left-handed scalar triplet $\Delta_L$ is given by, 
$$v_{L} \simeq\gamma \frac{v^2 }{v_{R}}\,.$$ 
In the above expression $v= \sqrt{v^2_1+ v^2_2}$ lies around electroweak scale, $v_R$ is the VEV of right-handed scalar 
triplet $\langle \Delta_R \rangle$ and $\gamma$ is dimensionless Higgs parameter. In order to be consistent with 
oscillation data $m^{II}_\nu = f v_L$ should be order of $ 0.1~$eV and assuming natural values of $f$ and $\gamma$, 
this sub-eV scale of $v_L$ can be attained only if $v_R$ lies around $10^{14}~$GeV. However such a high scale is inaccessible 
to LHC and thus urges to look for TeV scale LRSM. These frameworks offer numerous opportunities like low scale seesaw mechanism, 
LNV like neutrinoless double beta decay and its collider complementarity and have been already explored by the works mentioned in 
refs~\cite{Ge:2015bfa,Deppisch:2014zta,Tello:2010am,Deppisch:2014qpa,Patra:2014goa,
Borah:2013lva,Awasthi:2013ff,Patra:2012ur,Chakrabortty:2012mh,Dev:2013vxa,Barry:2013xxa,Nemevsek:2011hz,
Dev:2014iva,Keung:1983uu,Das:2012ii,Bertolini:2014sua,Beall:1981ze,Ge:2015yqa,Deppisch:2015cua,Hirsch:1996qw,Dev:2013oxa,
Dev:2015pga}. 
Many of the works considered either type-I seesaw dominance or type-II seesaw dominance for en extensive study of $0\nu\beta\beta$ decay. 
In manifest left-right symmetric model, where right-handed scale lies at TeV range, the neutrino mass mechanism 
 via type-I plus type-II seesaw gives negligible value to the left-right mixing. As a result of this the production cross-section of heavy neutrinos 
and the lepton number violating processes at LHC get suppressed. However, the extension of type-I plus type-II seesaw scheme 
by the inclusion of another sterile neutrino per generation changes the scenario which results large left-right mixing. Now the neutrino mass 
arises only from type-II seesaw dominance since type-I seesaw contribution gets exactly canceled out. We propose a new framework 
where type-II seesaw dominance is achieved naturally and allows large value of Dirac neutrino mass which additionally contributes to $0\nu\beta\beta$ decay 
from purely left-handed current via exchange of heavy neutrinos as well as from the so called $\lambda~$type and $\eta~$type diagrams.

\section{Extended Seesaw Mechanism and Natural type-II seesaw dominance}
\label{sec:seesaw-type-II}
\subsection{Extended Seesaw Mass Matrix}
In order to implement the extended seesaw mechanism\footnote{The discussion of extended seesaw 
mechanism can be found in refs.\cite{Barry:2011wb,Zhang:2011vh}.} within left-right symmetric models, 
one has to add a complete left-right gauge symmetry singlet neutral fermion $S_L$ per generation 
to the usual quarks and leptons. Along with this the Higgs sector includes scalar bidoublet $\Phi$ 
with $B-L=0$, scalar triplets $\Delta_L \oplus \Delta_R$ with $B-L=2$ and scalar doublets $H_L \oplus H_R$ with $B-L=-1$. 
The complete particle spectrum is given in Table.\ref{tab:LRSM-typeII}\,. 
\begin{table}[h]
\begin{center}
\begin{tabular}{c|c|c|c||c|c}
          & Fields      & $ SU(2)_L$ & $SU(2)_R$ & $B-L$ & $SU(3)_C$ \\
\hline
 Fermions &$q_L$     &  2         & 1         & 1/3   & 3   \\
 & $q_R$     &  1         & 2         & 1/3   & 3   \\
 & $\ell_L$  &  2         & 1         & -1    & 1   \\
 & $\ell_R$  &  1         & 2         & -1    & 1   \\
 & $S_{L}$   &  1         & 1         & 0     & 1   \\
\hline \hline
Scalars & $\Phi$   &  2         & 2         &  0     & 1   \\
 & $H_L$    &  2         & 1         & -1     & 1   \\
 & $H_R$    &  1         & 2         & -1     & 1   \\
 & $\Delta_L$    &  3         & 1         & 2     & 1   \\
 & $\Delta_R$    &  1         & 3         & 2     & 1   \\
\end{tabular}
\end{center}
\caption{LRSM representations of extended field content.}
\label{tab:LRSM-typeII}
\end{table}

The relevant leptonic Yukawa interaction terms for extended seesaw mechanism are given by 
\begin{eqnarray}
-\mathcal{L}_{Yuk} &=& \,\overline{\ell_{L}} \left[Y_3 \Phi + Y_4 \widetilde{\Phi} \right] \ell_R
+ f\, \left[\overline{(\ell_{L})^c} \ell_{L} \Delta_L+\overline{(\ell_{R})^c}\ell_{R}\Delta_R\right] \, \nonumber \\
&&+F\, \overline{(\ell_{R})} H_R S^c_L + F^\prime\, \overline{(\ell_{L})} H_L S_L + \mu_S \overline{S^c_L} S_L\ + \mbox{h.c.}\,. \\
&\supset& M_D \overline{\nu_L} N_R + M_L \overline{\nu^c_L} \nu_L + M_R \overline{N^c_R} N_R \nonumber \\
&&+M \overline{N_R} S_L + \mu_L \overline{\nu^c_L} S_L + \mu_S \overline{S^c_L} S_L
\label{yukaw}
\end{eqnarray}
After spontaneous symmetry breaking, the resulting neutral lepton mass matrix for extended seesaw mechanism in 
the basis $\left(\nu_L, N^c_R, S_L\right)$ is given by
\begin{equation}
\mathbb{M}_\nu= \left( \begin{array}{ccc}
              M_L                   & M_D   & \mu_L  \\
              M^T_D                 & M_R   & M^T \\
              \mu^T_L          & M     & \mu_S
                      \end{array} \right) \, ,
\label{eqn:numatrix}       
\end{equation}
where $M_D=Y\langle \Phi\rangle$ is the Dirac neutrino mass matrix connecting left-handed light neutrinos 
with right-handed heavy neutrinos, $M_N=f\, v_R=f\, \langle \Delta_R \rangle$ ($M_L=f\, v_L=f\, \langle \Delta_L \rangle$) 
is the Majorana mass term for heavy (light) neutrinos, $M=F\,\langle H_R\rangle$ is the $N-S$ mixing matrix, 
$\mu_L= F^\prime \langle H_L \rangle$ is the small mass term connecting $\nu-S$ and $\mu_S$ is the bare Majorana mass 
term for extra singlet fermion. 

\noindent
{\bf Inverse Seesaw:-} In Eq.(\ref{eqn:numatrix}), if we assume $M_L, M_R, \mu_L \to 0$ and the mass hierarchy $M \gg M_D \gg \mu_S$, 
we will arrive at the inverse seesaw mass formula for light neutrinos~\cite{Dev:2009aw}
$$m^{}_\nu =\left(\frac{M_D}{M}\right) \mu \left(\frac{M_D}{M}\right)^T\,.$$
The light neutrino mass can be parametrized in terms of model parameters of inverse seesaw framework as, 
$$\left( \frac{m_\nu}{\mbox{0.1\, eV}}\right) = \left(\frac{M_D}{\mbox{100\, GeV}} \right)^2 
 \left(\frac{\mu}{\mbox{keV}}\right) \left(\frac{M}{10^4\, \mbox{GeV}} \right)^{-2}\,.$$
This expression bears $M$ of few TeV which allows large left-right mixing 
and thus leads to interesting testable collider phenomenology. Extension 
of such a scenario has been discussed in the context of allowing large LNV and LFV 
in the work~\cite{Awasthi:2013ff}.

\noindent
{\bf Linear Seesaw:-} Similarly in Eq.(\ref{eqn:numatrix}), if we assume $M_L, M_R, \mu_S \to 0$, the linear seesaw mass formula 
for light neutrinos is given by~\cite{Deppisch:2015cua}
\begin{align}
\label{eq:mnu}
	m_\nu =M^T_D\, M^{-1} \mu_L \mbox{+\,transpose}\, ,
\end{align}
whereas the heavy neutrinos form pair of pseudo-Dirac states with masses 
\begin{equation}
	M_\pm \approx \pm M + m_{\nu}\,.
\end{equation} 
The following discussion considers the same Eq.(\ref{eqn:numatrix}) with the assumption that 
$\mu_L, \mu_S \to 0$ which leads to natural realization of type-II seesaw dominance allowing 
large left-right mixing.
\subsection{Natural realization of type-II seesaw}
The natural realization of type-II seesaw dominance is considered here within a class of left-right 
symmetric models where both discrete left-right parity symmetry and $SU(2)_R$ gauge symmetry break 
at same scale. The scalar sector is comprising of $SU(2)$ doublets $H_{L,R}$, triplets $\Delta_{L,R}$ and bidoublet 
$\Phi$ whereas the fermion sector is extended with one neutral fermion $S_L$ per generation which is complete singlet 
under both LRSM as well as SM gauge group. We denote this class of LR model as {\em Extended LR models} and thus, the corresponding 
seesaw formula which is type-II dominance in this case is termed as {\em Extended type-II seesaw mechanism}. 
In principle, there could be a gauge singlet mass term in the Lagrangian for extra fermion singlet, i.e, 
$\mu_S \overline{S^c} S$ which can take any value. But we have taken this mass parameter to be either zero 
or very small so that the generic inverse seesaw contribution involving $\mu_S$ is very much suppressed. 
In addition, we have assumed the induced VEV for $H_L$ is taken to be zero, i.e, $\langle H_L \rangle \to 0$. 

The relevant interaction terms necessary for realizing natural type-II seesaw dominance is given by
\begin{eqnarray}
\hspace*{-0.5cm}-\mathcal{L}_{Yuk} &=& \,\overline{\ell_{L}} \left[Y_3 \Phi + Y_4 \widetilde{\Phi} \right] \ell_R
+ f\, \left[\overline{(\ell_{L})^c} \ell_{L} \Delta_L+\overline{(\ell_{R})^c}\ell_{R}\Delta_R\right]
+F\, \overline{(\ell_{R})} H_R S^c_L+ \mbox{h.c.}\, \\
&& \supset M_D \overline{\nu_L} N_R + M_L \overline{\nu^c_L} \nu_L + M_R \overline{N^c_R} N_R 
+M \overline{N_R} S_L + \mbox{h.c.}\,.
\label{yukaw}
\end{eqnarray}

With $\langle H_L \rangle \to 0$ and $\mu_S \to 0$, the complete $9 \times 9$ neutral fermion mass 
matrix in the flavor basis of $\left(\nu_L, S_L, N^c_R \right)$ is read as
\begin{eqnarray}
\mathbb{M} = \left(\begin{array}{c|ccc}   & \nu_L & S_L  & N^c_R   \\ \hline
\nu_L  & M_L       & 0       & M_D \\
S_L    & 0         & 0       & M \\
N^c_R  & M^T_D     & M^T     & M_R
\end{array}
\right).
\label{eq:numatrix-complete}
\end{eqnarray}

Using standard formalism of seesaw mechanism and using mass hierarchy $M_R > M > M_D \gg M_L$, 
we can integrate out the heaviest right-handed neutrinos as follows
\begin{eqnarray}
\mathbb{M}^\prime &=& \begin{pmatrix}
                     M_L & 0 \\
                     0   & 0 
                    \end{pmatrix} 
    -  \begin{pmatrix}
         M_D \\
         M 
        \end{pmatrix} M^{-1}_R
                  \begin{pmatrix}
                   M^T_D & M^T 
                  \end{pmatrix} \nonumber \\
&=&   \begin{pmatrix}
       M_L-M_D M^{-1}_R M^T_D    & - M_D M^{-1}_R M^T \\
       M M^{-1}_R M^T_D          &  -M M^{-1}_R M^T
       \end{pmatrix} 
       \label{eq:nuS}
\end{eqnarray}
where the intermediate block diagonalised neutrino states modified as 
\begin{eqnarray}
&&\nu^\prime =    \nu_L - M_D M^{-1}_R N^c_R  \, , \nonumber \\
&&S^\prime =      S_L   - M_D M^{-1}_R N^c_R \, ,  \nonumber \\
&&N^{\prime} =  N^c_R + (M^{-1}_R M^T_D)^* \nu_L + (M^{-1}_R M^T)^* S_L\,. 
\end{eqnarray}
Thus, the intermediate block diagonalised neutrino states are related to flavor eigenstates in the 
following transformation,
\begin{eqnarray}
\begin{pmatrix}
 \nu^\prime \\ S^\prime \\ N^{\prime} 
\end{pmatrix}
=                              
      \begin{pmatrix}
      \mathbb{I}       & \mathbb{O}   & -M_D M^{-1}_R \\
      \mathbb{O}       & \mathbb{I}   & -M M^{-1}_R   \\
      (M_D M^{-1}_R)^\dagger & (M M^{-1}_R)^\dagger & \mathbb{I}
      \end{pmatrix} 
          \begin{pmatrix}
           \nu_L \\ S_L \\ N^{c}_R 
           \end{pmatrix}
\end{eqnarray}
It is found that the $(2,2)$ entries of mass matrix $\mathbb{M}^\prime$ is larger than other 
entries in the limit $M_R > M > M_D \gg M_L$. As a result of this, we can repeat the same 
procedure in Eq.(\ref{eq:nuS}) to integrate out $S^\prime$. Thus, the light neutrino mass 
formula becomes
\begin{eqnarray}
m_\nu &=& \left[ M_L-M_D M^{-1}_R M^T_D \right] 
    - \left(- M_D M^{-1}_R M^T \right) \left(- M M^{-1}_R M^T \right)^{-1} \left(- M M^{-1}_R M^T_D \right)  \nonumber \\
&=&\left[ M_L-M_D M^{-1}_R M^T_D \right] + M_D M^{-1}_R M^T_D\, \nonumber \\
&=&M_L = m^{\rm II}_\nu\, ,
\end{eqnarray}
and the physical block diagonalised states are
\begin{eqnarray}
&&\hat{\nu} = \nu_L - M_D M^{-1} S_L \nonumber \\
&&\hat{S} = S_L - M M^{-1}_R N^c_R + (M_D M^{-1})^\dagger S_L\, 
\end{eqnarray}
with the corresponding block diagonalised transformation as
\begin{eqnarray}
\begin{pmatrix}
 \hat{\nu} \\ \hat{S} 
\end{pmatrix}
=                              
      \begin{pmatrix}
      \mathbb{I}           & -M_D M^{-1} \\
      (M M^{-1})^\dagger   & \mathbb{I}
      \end{pmatrix} 
          \begin{pmatrix}
           \nu^\prime \\ S^\prime
           \end{pmatrix}
\end{eqnarray}
With this block diagonalization procedure and after few simple algebra, the flavor eigenstates are 
related to mass eigenstates in the following transformation,
\begin{eqnarray}
\begin{pmatrix}
 \nu_L \\ S_L \\ N^{c}_R 
\end{pmatrix}
=                              
      \begin{pmatrix}
      \mathbb{I}             & M_D M^{-1}             & M_D M^{-1}_R \\
      (M_D M^{-1})^\dagger   & \mathbb{I}             & M M^{-1}_R   \\
      \mathbb{O}             & -(M M^{-1}_R)^\dagger  & \mathbb{I}
      \end{pmatrix} 
          \begin{pmatrix}
           \nu^\prime \\ S^\prime \\ N^{\prime} 
           \end{pmatrix}
\end{eqnarray}
Subsequently, the final block diagonalised mass matrices can be diagonalised 
in order to give physical masses by a $9 \times 9$ unitary matrix $\mbox{V}_{9 \times 9}$. 
The transformation of the block diagonalised neutrino states in terms of 
mass eigenstates are given by
\begin{eqnarray}
\hat{\nu}_\alpha = {U_\nu}_{\alpha i} \nu_i\, , \quad 
\hat{S}_\alpha = {U_S}_{\alpha i} S_i\, , \quad 
\hat{N}_\alpha = {U_N}_{\alpha i} N_i\,. \quad 
\end{eqnarray}
while the block diagonalised mass matrices for light left-handed neutrinos, heavy right-handed 
neutrinos and extra sterile neutrinos are
\begin{eqnarray}
&&m_\nu = M_L \,, \nonumber \\
&&M_N \equiv M_R = \frac{v_R}{v_L} M_L\,, \nonumber \\
&&M_S = -M M^{-1}_R M^T \,.
\end{eqnarray}
These block diagonalised mass matrices can be further diagonalised by respective $3 \times 3$ 
unitarity matrices as follows
\begin{eqnarray}
&&m^{\rm diag}_\nu=U^\dagger_\nu m_\nu U^*_\nu = \mbox{diag.}\{m_1, m_2, m_3 \}\, , \nonumber \\
&&M^{\rm diag}_S=U^\dagger_S M_S U^*_S = \mbox{diag.}\{ M_{S_1}, M_{S_2}, M_{S_3} \}\, , \nonumber \\
&&M^{\rm diag}_N=U^\dagger_N M_N U^*_N = \mbox{diag.}\{ M_{N_1}, M_{N_2}, M_{N_3} \}\,.
\end{eqnarray}
Finally, the complete block diagonalization yields
\begin{eqnarray}
\widehat{\mathbb{M}} &=& \mbox{V}^\dagger_{9 \times 9} \mathbb{M} \mbox{V}^*_{9 \times 9} 
    = \left(\mathbb{W} \cdot \mathbb{U} \right)^\dagger \mathbb{M} \left(\mathbb{W} \cdot \mathbb{U} \right) \nonumber \\
   &=& \mbox{diag.}\{m_1, m_2, m_3;\, M_{S_1}, M_{S_2}, M_{S_3}; M_{N_1}, M_{N_2}, M_{N_3} \}
\end{eqnarray}
Here the block diagonalised mixing matrix $\mathbb{W}$ and the unitarity matrix $\mathbb{U}$ are given by
\begin{eqnarray}
\mathbb{W}=       \begin{pmatrix}
      \mathbb{I}             & M_D M^{-1}             & M_D M^{-1}_R \\
      (M_D M^{-1})^\dagger   & \mathbb{I}             & M M^{-1}_R   \\
      \mathbb{O}             & -(M M^{-1}_R)^\dagger  & \mathbb{I}
      \end{pmatrix} \,, \quad 
\mathbb{U} = \begin{pmatrix}
      U_\nu          & \mathbb{O}    & \mathbb{O} \\
      \mathbb{O}     & U_S         & \mathbb{O}   \\
      \mathbb{O}     & \mathbb{O}    & U_N 
      \end{pmatrix}\,.
\end{eqnarray}
Thus, the complete $9 \times 9$ unitary mixing matrix diagonalizing the neutral leptons 
is as follows
\begin{equation}
\mbox{V} = \mathbb{W} \cdot \mathbb{U}
   =      \begin{pmatrix}
      U_\nu           & M_D M^{-1} U_S           & M_D M^{-1}_R U_N   \\
      (M_D M^{-1})^\dagger U_\nu  & U_S            & M M^{-1}_R U_N  \\
      \mathbb{O}             & -(M M^{-1}_R)^\dagger U_S  & U_N
      \end{pmatrix} 
\end{equation}

\subsection{Expressing Masses and Mixing in terms of $U_{\rm PMNS}$ and light neutrino masses.}
The light neutrinos are generally diagonalised by standard PMNS mixing matrix 
$U_{\rm PMNS}$ in the basis where charged leptons are already diagonal i.e,$m^{\rm diag}_\nu= U^\dagger_{\rm PMNS} m_\nu U^*_{\rm PMNS}$. 
The Dirac neutrino mass matrix $M_D$ in general is a complex matrix. 
The structure of $M_D$ in LRSM can be approximately taken to be up-quark type 
mass matrix whose origin can be motivated from high scale Pati-Salam symmetry 
or SO(10) GUT. If we consider $M$ to be diagonal and degenerate 
i.e, $M=m_S \mbox{diag}\{1,1,1\}$, then the mass formulas for neutral 
leptons are given by
\begin{eqnarray}
& &m_\nu = M_L = f v_L = U_{\rm PMNS} m^{\rm diag}_\nu U^T_{\rm PMNS}\, , \nonumber \\[2mm]
& &M_N \equiv M_R = f v_R = \frac{v_R}{v_L}\, M_L = \frac{v_R}{v_L} U_{\rm PMNS} m^{\rm diag}_\nu U^T_{\rm PMNS} \,, \nonumber  \\[1mm]
& &M_S = -M M^{-1}_R M^T = -m^2_S\, \left[\frac{v_R}{v_L} U_{\rm PMNS} m^{\rm diag}_\nu U^T_{\rm PMNS} \right]^{-1}, 
\end{eqnarray}
After some simple algebra, the active LH neutrinos $\nu_L$, active RH neutrinos $N_R$ and 
heavy sterile neutrinos $S_L$ in the flavor basis are related to their mass basis as
\begin{eqnarray}
\begin{pmatrix}
\nu_{L} \\ S_{L} \\ N^c_{R}
\end{pmatrix}_\alpha 
&=&
\begin{pmatrix}
{\mbox V}^{\nu\nu} & {\mbox V}^{\nu{S}} & {\mbox V}^{\nu {N}} \\
{\mbox V}^{S\nu} & {\mbox V}^{SS} & {\mbox V}^{SN} \\
{\mbox V}^{N\nu} & {\mbox V}^{NS} & {\mbox V}^{NN} 
\end{pmatrix}_{\alpha i} 
\begin{pmatrix}
\nu_i \\ S_i \\ N_i
\end{pmatrix}  \nonumber \\
&=& \begin{pmatrix}
      U_{\rm PMNS}           & \frac{1}{m_S} M_D U^*_{\rm PMNS}           & \frac{v_L}{v_R} M_D U^{-1}_{\rm PMNS} {m^{\rm diag.}_{\nu}}^{-1}   \\
      \frac{1}{m_S} M^\dagger_D U_{\rm PMNS} & U^*_{\rm PMNS}             & \frac{v_L}{v_R} m_S U^{-1}_{\rm PMNS} {m^{\rm diag.}_{\nu}}^{-1} \\
      \mathbb{O}             & \frac{v_L}{v_R} m_S U^{-1}_{\rm PMNS} {m^{\rm diag.}_{\nu}}^{-1}  & U_{\rm PMNS} 
      \end{pmatrix}_{\alpha i} 
\begin{pmatrix}
\nu_i \\ S_i \\ N_i
\end{pmatrix}  \nonumber \\
 \label{eq:numixing}
\end{eqnarray}
\section{Neutrinoless Double Beta Decay in LRSM}
\label{sec:0nu2beta_LRSM}
In this section, we shall present a detailed discussion on Feynman amplitudes 
for neutrinoless double beta decay within TeV scale LRSM where light neutrino 
mass mechanism is governed by natural type-II seesaw dominance. 
The basic charge current interaction Lagrangian for leptons as well quarks are 
given by
\begin{eqnarray}
\mathcal{L}^{\rm \ell}_{\rm CC} &=& \sum_{\alpha=e, \mu, \tau}
\bigg[\frac{g_L}{\sqrt{2}}\, \overline{\ell}_{\alpha \,L}\, \gamma_\mu {\nu}_{\alpha \,L}\, W^{\mu}_L 
      +\frac{g_R}{\sqrt{2}}\, \overline{\ell}_{\alpha \,R}\, \gamma_\mu {N}_{\alpha \,R}\, W^{\mu}_R \bigg] + \text{h.c.} 
      \nonumber \\
&=& \frac{g_L}{\sqrt{2}}\,
\overline{e}_{\,L}\, \gamma_\mu {\nu}_{e \,L}\, W^{\mu}_L  
   + \frac{g_R}{\sqrt{2}}\,\overline{e}_{\,R}\, \gamma_\mu {N}_{e \,R}\, W^{\mu}_R + \text{h.c.} + \cdots 
\label{eqn:ccint-flavor-lepton}
\\
\mathcal{L}^{\rm q}_{\rm CC} &=& \bigg[\frac{g_L}{\sqrt{2}}\, \overline{u}_{\,L}\, \gamma_\mu {d}_{\,L}\, W^{\mu}_L 
      +\frac{g_R}{\sqrt{2}}\, \overline{u}_{R}\, \gamma_\mu d_{R}\, W^{\mu}_R \bigg] + \text{h.c.} 
\label{eqn:ccint-flavor-quark}
\end{eqnarray}
Using Eq.(\ref{eq:numixing}) of Sec.\ref{sec:seesaw-type-II}, the flavor eigenstates $(\nu_L)$ and $N^c_R$ are 
expressed in terms of admixture of mass eigenstates $(\nu_i, S_i, N_i)$ in the following way,
\begin{eqnarray}
&&\nu_{eL}= \mbox{V}^{\nu\nu}_{e\, i}\, \nu_i + \mbox{V}^{\nu\, S}_{e\, i}\, S_i + \mbox{V}^{\nu\, N}_{e\, i}\, N_i, \nonumber \\
&&N_{eR} = \mbox{V}^{N\, S}_{e\, i}\,S_i +
                      \mbox{V}^{NN}_{e\, i}\, N_i\, .
\end{eqnarray}
This modifies the charged current interaction for leptons as
\begin{eqnarray}
\mathcal{L}^{\rm mass}_{\rm CC}&=& \frac{g_L}{\sqrt{2}}\,
\bigg[ \overline{e}_{\,L}\, \gamma_\mu 
        \{\mbox{V}^{\nu\nu}_{e\, i}\, \nu_i + \mbox{V}^{\nu\, S}_{e\, i}\, S_i +
                      \mbox{V}^{\nu\, N}_{e\, i}\, N_i \}\, 
              W^{\mu}_L \bigg] +\mbox{h.c.} \nonumber \\
    & &  + \frac{g_R}{\sqrt{2}}\,
\bigg[ \overline{e}_{\,R}\, \gamma_\mu 
      \{\mbox{V}^{N\, S}_{e\, i}\,S_i +
                      \mbox{V}^{NN}_{e\, i}\, N_i \}\, 
              W^{\mu}_R \bigg] + \mbox{h.c.}
\label{eqn:ccint-mass}
\end{eqnarray}
In the above charged-current interaction, there is a possibility that both left-handed $W_L$ 
and right-handed $W_R$ gauge bosons can mix with each other which can eventually contribute to 
$0\nu\beta\beta$ transition amplitude. In the present framework, the resulting mass matrix for 
LH (RH) charged gauge bosons ($W_L, W_R$) is given by 
\begin{eqnarray}
 \mathbb M_W
&=&\frac{1}{4}
		\left(\begin{array}{c|cc}&W^{+}_L&W^{+}_R \\
		\hline
W^{-}_L&g^2_L \left(v^2_1+v^2_2+2 v_L^2+ u^2_L\right)        &    -2 g_L g_R v^*_1 v_2\\
W_R^{-}&- 2 g_L g_R v_1 v^*_2 & g^2_R \left(2 v_R^2+ u^2_R+ v^2_1+v^2_2\right) 
\end{array}
\right)
\end{eqnarray}
The physical masses of the charged gauge bosons derived with $g_L = g_R$ 
after diagonalization are given by
\begin{eqnarray}
&&M_{W_1}^2 \approx \frac{1}{4} g_L^2 \bigg[ \left( v^2_1 + v^2_2 \right) 
   - \frac{4 v^2_1 v^2_2}{u^2_R+ 2 v^2_R} \bigg]\, ,\nonumber \\
&&M_{W_2}^2 \approx \frac{1}{4} g_R^2 \bigg[ u^2_R + 2 v^2_R + v^2_1 + v^2_2 \bigg] \,.
\end{eqnarray}
The physical gauge boson states $W_1$ and $W_2$ are related to the mixture of weak eigenstates 
$W_L$ and $W_R$ as 
\begin{equation}
  \mathbb R_W
\equiv
\left\lgroup
\begin{matrix}
  \cos \xi & \sin \xi \\
- \sin \xi & \cos \xi
\end{matrix}
\right\rgroup \,,
\end{equation}
where,  
\begin{equation}
|\tan\, 2\xi | \sim \frac{2 v_1 v_2}{u^2_R + 2 v^2_R - u^2_L - 2 v^2_L}\,.
\end{equation}
Thus, one can express physical states in terms of $W_L$ and $W_R$ as follows
\begin{equation}
\label{eqn:LRmix}
\left\{ 
\begin{array}{l} 
W_1 = \phantom{-}\cos\xi ~W_L + \sin \xi~W_R \\ 
W_2 = - \sin\xi ~W_L + \cos \xi ~W_R 
\end{array} \right. 
\end{equation}

We classify all contributions to neutrinoless double beta decay in the present TeV 
scale LRSM as:
\begin{itemize}
\item  due to standard mechanism mediated by purely left-handed currents ($W_L-W_L$ mediation) via exchange 
       of light neutrinos $\nu_i$, 
\item  due to purely left-handed currents via $W_L^--W_L^-$ mediation through the exchange of the heavy RH Majorana 
       neutrino $N_i$ and heavy sterile neutrinos $S_i$, 
\item  due to purely right-handed currents ($W_R-W_R$ mediation) via exchange 
       of heavy right-handed Majorana neutrinos $N_i$, 
\item  due to purely right-handed currents via $W_R^--W_R^-$ mediation through the exchange of the light neutrinos 
       $\nu_i$ and extra sterile neutrinos $S_i$,
\item due to mixed helicity so called $\lambda$ and $\eta$ diagrams through mediation of $\nu_i, S_i, N_i$ neutrinos.
\end{itemize}

Before deducing Feynman amplitudes for various contributions to neutrinoless double beta decay, it is 
desirable to discuss few points regarding the chiral structure of the matrix element with the neutrino 
propagator as
\cite{Pas:1999fc}
\begin{eqnarray}
& &P_{L}\frac{\slashed{p}+m_i}{p^2-m_i^2}P_{L} = \frac{m_i}{p^2-m_i^2}\,  \quad\, , \quad 
   P_{R}\frac{\slashed{p}+m_i}{p^2-m_i^2}P_{R} = \frac{m_i}{p^2-m_i^2}\,, \nonumber \\
& &P_{L}\frac{\slashed{p}+m_i}{q^2-m_i^2}P_{R} = \frac{\slashed{p}}{p^2-m_i^2}\,  \quad\, , \quad 
P_{R}\frac{\slashed{p}+m_i}{p^2-m_i^2}P_{L} = \frac{\slashed{p}}{p^2-m_i^2}\,, \nonumber \\
\end{eqnarray}
\bea \label{eq:mee}
\frac{m_i}{p^2-m^2_i} \simeq \left\{
\baz 
\frac{m_i}{p^2}\, ,
&  m^2_i \ll p^2\\[0.2cm]
-\frac{1}{m_i}
& m^2_i \gg p^2
\ea \right. 
\eea
and
\bea \label{eq:mee}
\frac{\slashed{p}}{p^2-m^2_i} \propto \left\{
\baz 
\frac{1}{|p|}\, ,
&  m^2_i \ll p^2\\[0.2cm]
-\frac{|p|}{m^2_i}
& m^2_i \gg p^2\, .
\ea \right. 
\eea

\subsection{Feynman amplitudes for $0\nu\beta\beta$ decay due to purely left-handed currents}
\label{subsec:ampl-onu2beta_LL}
\begin{figure}[htb!]
\centering
\includegraphics[scale=0.55]{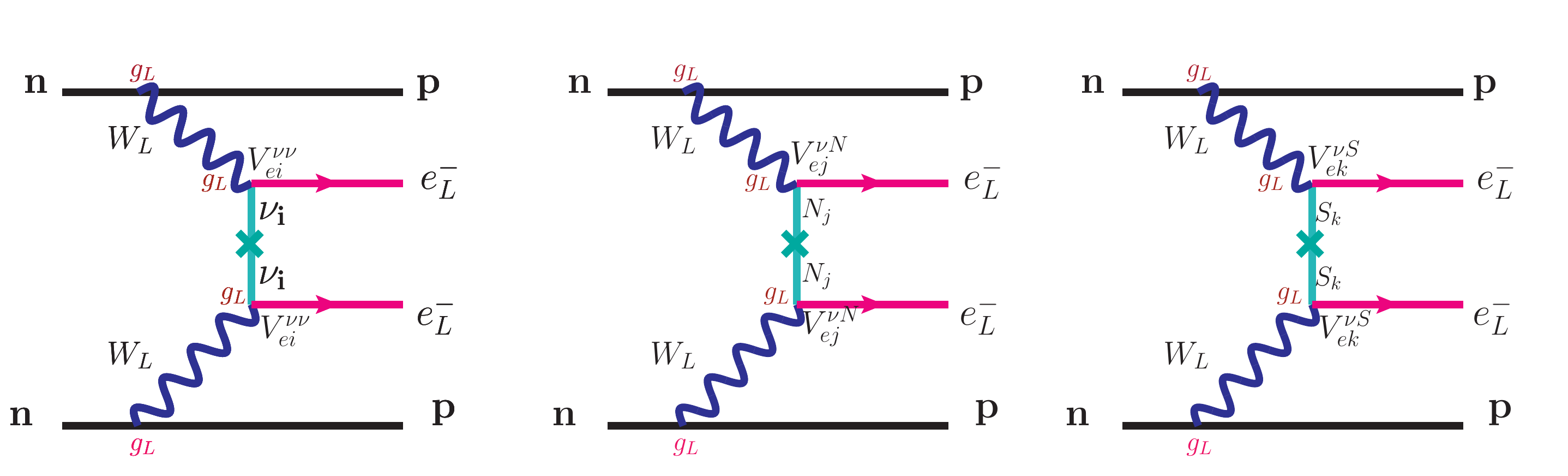}
\caption{Feynman diagrams for neutrinoless double beta decay via $W^-_L - W^-_L$ mediation with the exchange of 
         virtual Majorana neutrinos $\nu_{i}$, $N_j$ and $S_k$.}
\label{feyn:lrsm-WLL}
\end{figure}
%
The Feynman amplitudes for $W^-_L - W^-_L$ mediated diagrams shown in Fig.\ref{feyn:lrsm-WLL} 
with the exchange of Majorana neutrinos $\nu_i$, $N_j$ and $S_k$, respectively, are given by
\begin{eqnarray}
\label{eq:amp_LL} 
& &\mathcal{A}_{LL}^{\nu} \propto G^2_F \sum_{i=1,2,3} \frac{{\mbox{V}^{\nu \nu}_{ei}}^2\, 
                      m_{\nu_i}}{p^2} \,, \nonumber \\
& &\mathcal{A}_{LL}^{N} \propto G^2_F \sum_{j=1,2,3} \left(-\frac{{\mbox{V}^{\nu N}_{ej}}^2}{M_{N_j}} \right)\,,\nonumber \\
& &\mathcal{A}_{LL}^{S} \propto G^2_F \sum_{k=1,2,3} \left(-\frac{{\mbox{V}^{\nu S}_{ek}}^2}{M_{S_k}} \right)\,,                       
\end{eqnarray}
where $p$ is the typical momentum exchange of the $0\nu\beta\beta$ decay process and $G_F =1.2 \times 10^{-5}\, \mbox{GeV}^{-2}$ 
is the Fermi coupling constant. The analytic expressions for suitably normalized dimensionless lepton number violating 
particle physics parameters for these contributions are as follows
\begin{eqnarray}
\label{eq:eta_LL} 
& &\mid \mathcal{\eta}_{LL}^{\nu} \mid = \sum_{i=1,2,3} \frac{{\mbox{V}^{\nu \nu}_{ei}}^2\, m_{\nu_i}}{m_e} \,, 
  \mid \mathcal{\eta}_{LL}^{N} \mid = m_p \sum_{j=1,2,3} \frac{{\mbox{V}^{\nu N}_{ej}}^2}{M_{N_j}} \,,                    
  \mid \mathcal{\eta}_{LL}^{S} \mid = m_p \sum_{k=1,2,3} \frac{{\mbox{V}^{\nu S}_{ek}}^2}{M_{S_k}} \,.
\end{eqnarray}
Though we shall discuss in detail about the lepton number violating effective mass parameters and half-life 
in the following section, it will be better if one can express normalized effective mass parameters representing 
LNV due to these above mentioned Feynman diagrams and are given below
\begin{eqnarray}
\label{eq:eta_LL} 
& &|\mee^{\nu}_{L}| = \sum_{i=1,2,3} {\mbox{V}^{\nu \nu}_{ei}}^2\, m_{\nu_i} \,, 
  |\mee^{N}_{L}| = \langle p^2 \rangle \sum_{j=1,2,3} \frac{{\mbox{V}^{\nu N}_{ej}}^2}{M_{N_j}} \,,                   
  |\mee^{S}_{L}| = \langle p^2 \rangle \sum_{k=1,2,3} \frac{{\mbox{V}^{\nu S}_{ek}}^2}{M_{S_k}} \,. \nonumber
\end{eqnarray}

\begin{figure}[htb!]
\centering
\includegraphics[scale=0.55]{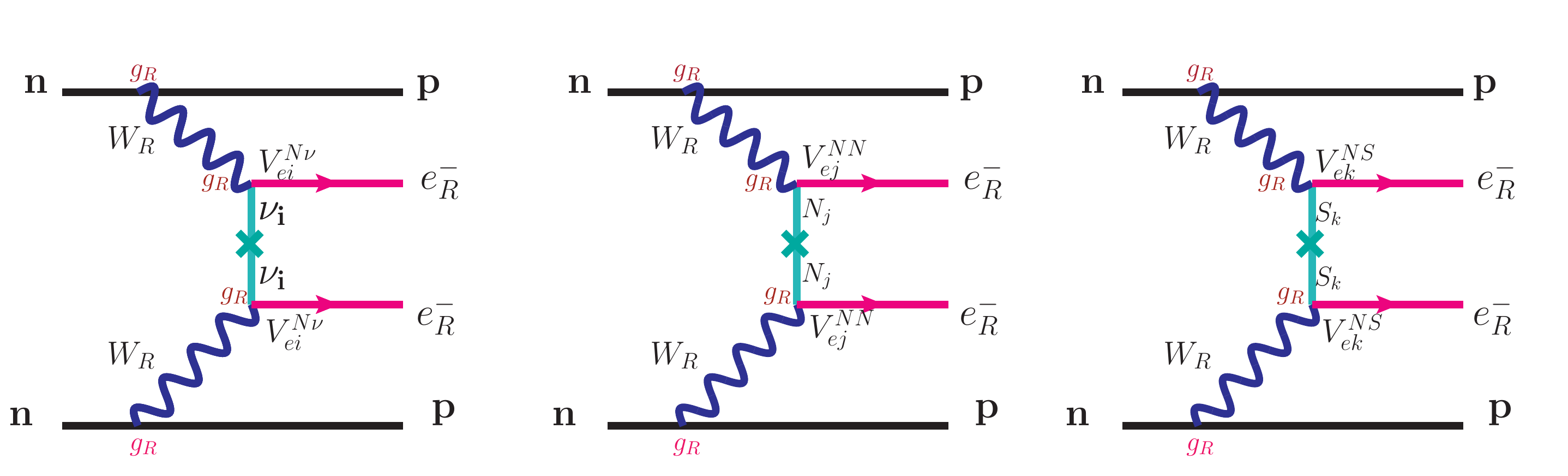}
 \caption{Feynman diagrams for neutrinoless double beta decay ($0\, \nu\, \beta \beta$) via $W^-_R - W^-_R$ mediation 
          with the exchange of virtual Majorana neutrinos $\nu_i$, $N_j$ and $S_k$.}
\label{feyn:lrsm-WRR}
\end{figure}
%
\subsection{Feynman amplitudes for $0\nu\beta\beta$ decay due to purely right-handed currents}
\label{subsec:ampl-onu2beta_RR}
One of our major contribution in this work is that with $W^-_R - W^-_R$ mediation as shown 
in first one of Fig. \ref{feyn:lrsm-WRR} by the exchange of mainly heavy right-handed neutrinos 
within the type-II seesaw dominance can yield significantly large contribution to $0\nu \beta \beta$ 
decay rate than the standard one. The Feynman amplitudes for these diagrams displayed in 
Fig.\ref{feyn:lrsm-WRR} normalized in terms of $G_F$ are given by
\begin{eqnarray}
\label{eq:amp_RR} 
& &\mathcal{A}_{RR}^{\nu} \propto G^2_F \sum_{i=1,2,3} \left(\frac{M_{W_L}}{M_{W_R}} \right)^4 
            \left(\frac{g_R}{g_L} \right)^4 \frac{{\mbox{V}^{N \nu}_{ei}}^2\, 
                      m_{\nu_i}}{p^2} \,, \nonumber \\
& &\mathcal{A}_{RR}^{N} \propto G^2_F \sum_{j=1,2,3} \left(\frac{M_{W_L}}{M_{W_R}} \right)^4 
            \left(\frac{g_R}{g_L} \right)^4 \,   
                      \left(-\frac{{\mbox{V}^{N N}_{ej}}^2}{M_{N_j}} \right)\,, \nonumber \\
& &\mathcal{A}_{RR}^{S} \propto G^2_F \sum_{k=1,2,3} \left(\frac{M_{W_L}}{M_{W_R}} \right)^4 
            \left(\frac{g_R}{g_L} \right)^4 \,   
                      \left(-\frac{{\mbox{V}^{N S}_{ek}}^2}{M_{S_k}} \right)\,.                     
\end{eqnarray}
The resulting dimensionless LNV particle physics parameters due to $W^-_R - W^-_R$ mediated 
diagrams are as follows
\begin{eqnarray}
\label{eq:eta_RR} 
& &\mid \mathcal{\eta}_{R}^{\nu} \mid = \sum_{i=1,2,3} \left(\frac{M_{W_L}}{M_{W_R}} \right)^4 
            \left(\frac{g_R}{g_L} \right)^4 \frac{{\mbox{V}^{N \nu}_{ei}}^2\, m_{\nu_i}}{m_e} \,, \nonumber \\
& &  \mid \mathcal{\eta}_{R}^{N} \mid = \sum_{j=1,2,3} m_p \left(\frac{M_{W_L}}{M_{W_R}} \right)^4 
            \left(\frac{g_R}{g_L} \right)^4\frac{{\mbox{V}^{N N}_{ej}}^2}{M_{N_j}} \,, \nonumber \\                  
& &  \mid \mathcal{\eta}_{R}^{S} \mid = \sum_{k=1,2,3} m_p \left(\frac{M_{W_L}}{M_{W_R}} \right)^4 
            \left(\frac{g_R}{g_L} \right)^4 \frac{{\mbox{V}^{N S}_{ek}}^2}{M_{S_k}} \,.
\end{eqnarray}

\begin{figure}[htb!]
\centering
\includegraphics[scale=0.55]{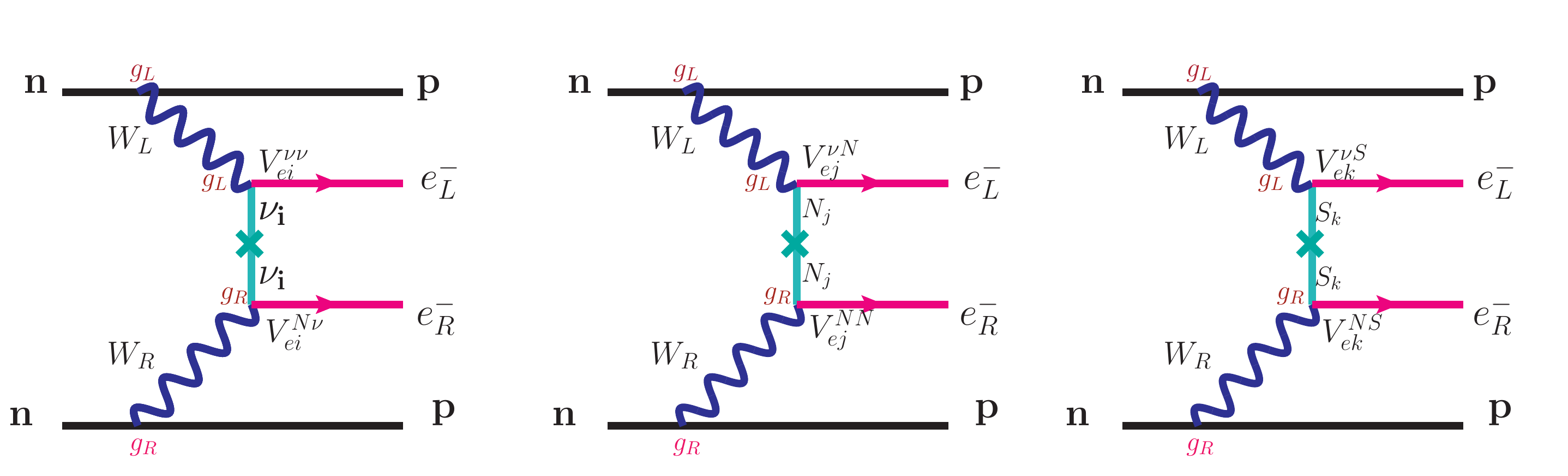}
 \caption{The $\lambda$ diagram for $0\nu\beta\beta$ decay within LRSM via $W_L^--W_R^-$ 
          mediation and by the exchange of virtual Majorana neutrinos $\nu_i$, $N_j$ 
          and $S_k$.}
\label{feyn:lrsm-lambda}
\end{figure}
%
\subsection{Feynman amplitudes for $\lambda$-diagram due to $W^-_L - W^-_R$ mediation}
\label{subsec:ampl-onu2beta_lambda}
%
There are Feynman diagrams for neutrinoless double beta decay due to mixed helicity of emitted 
electrons in the final state via $W^-_L - W^-_R$ mediation and the Feynman amplitudes for these diagrams 
with the exchange of virtual Majorana neutrinos $\nu_i$, $N_j$ and $S_k$ are
\begin{eqnarray}
& & \mathcal{A}^{\nu}_{\lambda} \propto {\bf G_F}^2 \left(\frac{M_{W_L}}{M_{W_R}} \right)^2 
            \left(\frac{g_R}{g_L} \right)^2 \sum_{i=1,2,3} 
\mbox{V}^{\nu \nu}_{e\,i} \mbox{V}^{N \nu}_{e\,i} \frac{1}{|p|} \, , \nonumber \\
& & \mathcal{A}^{N}_{\lambda} \propto {\bf G_F}^2 \sum_{j=1,2,3} \left(\frac{M_{W_L}}{M_{W_R}} \right)^2 
            \left(\frac{g_R}{g_L} \right)^2 \mbox{V}^{\nu N}_{e\,j} \mbox{V}^{N N}_{e\,j} \frac{|p|}{M^2_{N_j}} \, , \nonumber \\
& & \mathcal{A}^{S}_{\lambda} \propto {\bf G_F}^2 \sum_{k=1,2,3} \left(\frac{M_{W_L}}{M_{W_R}} \right)^2 
            \left(\frac{g_R}{g_L} \right)^2 \mbox{V}^{\nu S}_{e\,k} \mbox{V}^{N S}_{e\,k} \frac{|p|}{M^2_{S_k}} \, . 
\end{eqnarray}

\begin{figure}[htb!]
\centering
\includegraphics[scale=0.55]{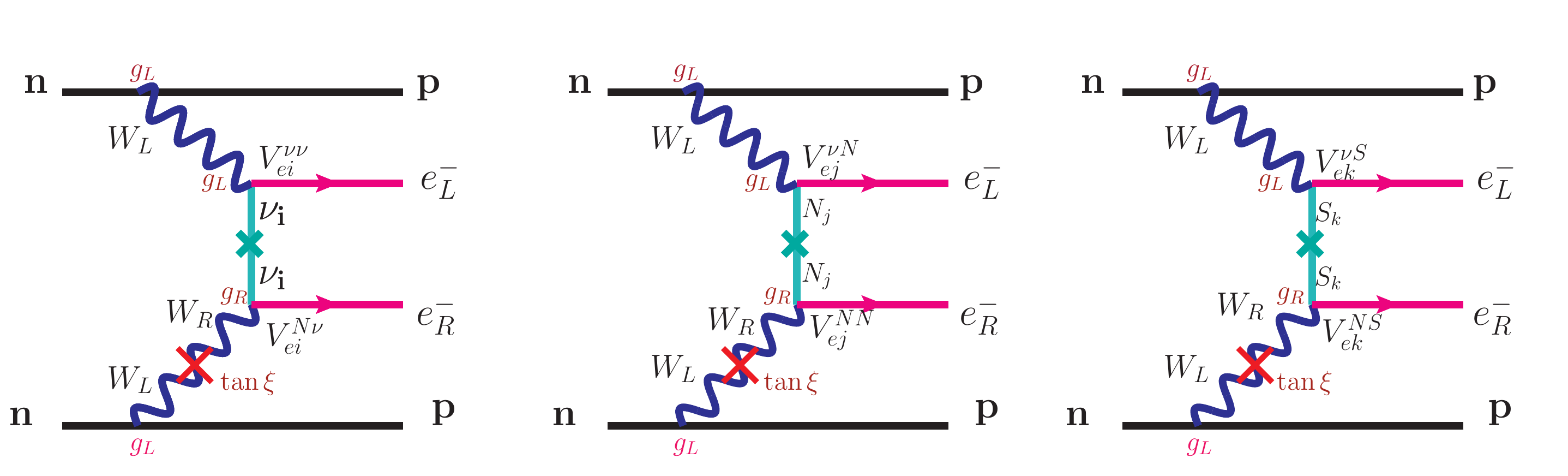}
 \caption{Feynman diagrams for $0\, \nu\, \beta \beta$ decay for standard 
          $\eta$-contributions which involve mixing between $W_L$ and $W_R$, 
          i.e, $\tan \xi$ and $W^-_L - W^-_R$ mediation.}
\label{feyn:lrsm-eta}
\end{figure}
%
\subsection{Feynman amplitudes for $\lambda$-diagram with $W_L-W_R$ mixing}
\label{subsec:ampl-onu2beta_eta}
%
There are Feynman diagrams for neutrinoless double beta decay due to mixed helicity of emitted 
electrons in the final state via $W^-_L - W^-_R$ mediation as well as involves mixing between 
$W_L$ and $W_R$ gauge boson. The Feynman amplitudes for these diagrams with the exchange of 
virtual Majorana neutrinos $\nu_i$, $N_j$ and $S_k$ are given by

\begin{eqnarray}
& & \mathcal{A}^{\nu}_{\eta} \propto {\bf G_F}^2 \sum_{i=1,2,3} \left(\frac{g_R}{g_L} \right) \tan \xi
           \mbox{V}^{\nu \nu}_{e\,i} \mbox{V}^{N \nu}_{e\,i} \frac{1}{|p|} \, , \nonumber \\
& & \mathcal{A}^{N}_{\eta} \propto {\bf G_F}^2 \sum_{j=1,2,3} \left(\frac{g_R}{g_L} \right) \tan \xi
           \mbox{V}^{\nu N}_{e\,j} \mbox{V}^{N N}_{e\,j} \frac{|p|}{M^2_{N_j}} \, , \nonumber \\
& & \mathcal{A}^{S}_{\eta} \propto {\bf G_F}^2 \sum_{k=1,2,3} \left(\frac{g_R}{g_L} \right) \tan \xi
           \mbox{V}^{\nu S}_{e\,k} \mbox{V}^{N S}_{e\,k} \frac{|p|}{M^2_{S_k}} \, . 
\end{eqnarray}

\begin{figure}[htb!]
\centering
\includegraphics[scale=0.55]{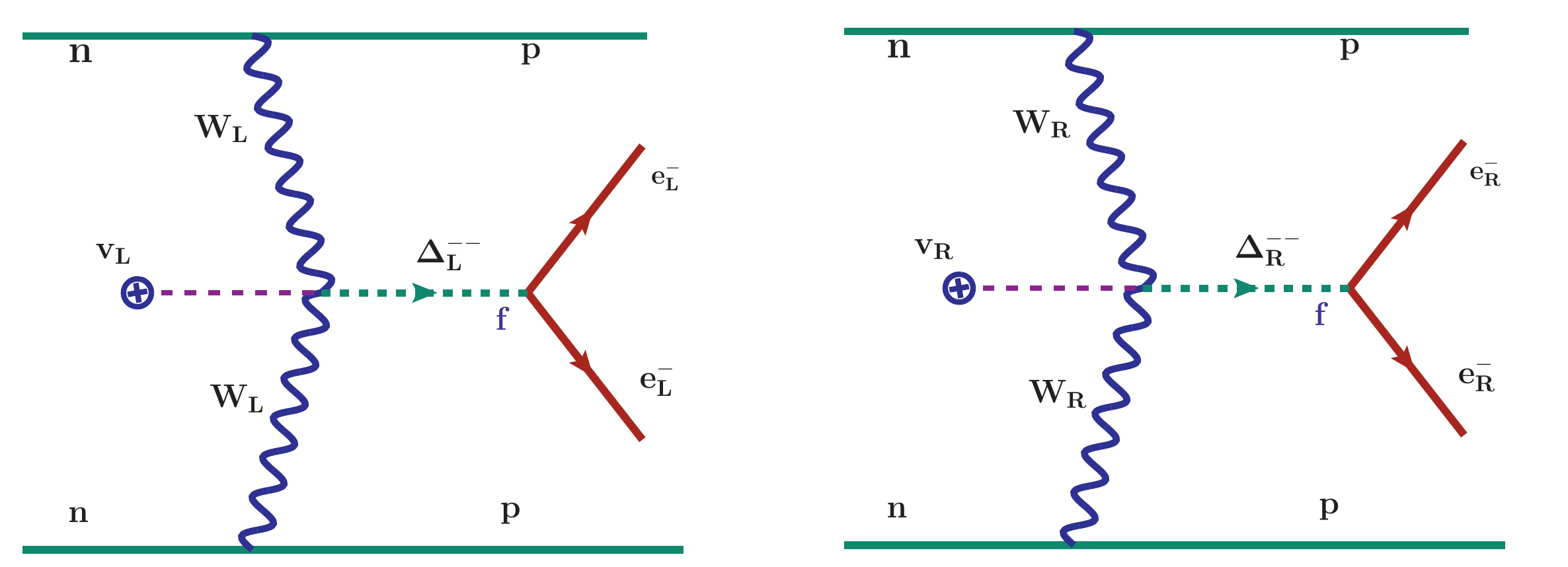}
 \caption{Feynman diagrams for $0\, \nu\, \beta \beta$ decay due to doubly charged 
          scalar triplets.}
\label{feyn:lrsm-eta}
\end{figure}
%
\subsection{Feynman amplitudes for $0\nu\beta\beta$ decay due to doubly charged scalar}
\label{subsec:ampl-onu2beta_triplets}
The Feynman amplitudes due to doubly charged Higgs scalars $\Delta^{--}_L$ ($\Delta^{--}_R$) 
exchanges are given by
\begin{eqnarray}
\label{eq:amp_LL} 
& &\mathcal{A}_{LL}^{\Delta_L} \propto G^2_F 
               \sum_{i=1,2,3} \frac{1}{M_{\Delta_L}^2} {\mbox{V}^{\nu\nu}_{ei}}^2 m_{\nu_i} \,, \nonumber \\
& &\mathcal{A}_{RR}^{\Delta_R} \propto G^2_F \sum_{j=1,2,3} \left(\frac{M_{W_L}}{M_{W_R}} \right)^4 
            \left(\frac{g_R}{g_L} \right)^4 \,   
            \frac{1}{M^2_{\Delta_{R}}} {\mbox{V}^{NN}_{ej}}^2 M_{N_j } \, ,\nonumber                    
\end{eqnarray}
result in lepton number violating dimensionless particle physics parameters.  

\section{Half-life and normalized LNV effective Mass parameters}
\label{sec:halflife-effLNV}

From the earlier discussion, we found that there are various contributions to neutrinoless double beta decay 
arising from purely left-handed currents, from purely right-handed currents, from mixed diagrams with left-handed 
as well as right-handed currents and possible interference effects. In this regard we closely follow the 
refs.\cite{Muto:1989cd,Suhonen:1998ck,Barry:2013xxa} where the QRPA calculations of the matrix elements for the mixed 
diagrams leads to life-time of $0\nu\beta\beta$ transition as 
\begin{align}
 \left[T_{1/2}^{0\nu}\right]^{-1} &= G^{0\nu}_{01}|{\cal M}^{0\nu}_{\rm GT}|^2\left\{\left|X_L\right|^2 + \left|X_R\right|^2 + \tilde{C}_2|\eta_\lambda| |X_L|\cos\psi_1+\tilde{C}_3|\eta_\eta| |X_L|
\cos\psi_2 \right. \notag \\
& \left. + \ \tilde{C}_4|\eta_\lambda|^2 + \tilde{C}_5|\eta_\eta|^2+\tilde{C}_6|\eta_\lambda||\eta_\eta|\cos(\psi_1-\psi_2)+{\rm
Re}\left[\tilde{C}_2X_R\eta_\lambda+\tilde{C}_3X_R\eta_\eta\right]\right\}, \label{eq:half-life_full}
\end{align}
where $G^{0\nu}_{01}$ is the phase space factor, ${\cal M}^{0\nu}_{\rm GT}$ is the matrix element for $0\nu\beta\beta$ transition. 
Here $X_L$ and $X_R$ represent the relevant contributions arising from left-handed and right-handed currents respectively. 
The coefficients $\tilde{C}_i$ stand for combination of matrix elements and integrated kinematically factors and $\psi_i$ 
represents complex phases. The LNV dimensionless particle physics parameters are denoted by $\eta's$. 
In above eq.(\ref{eq:half-life_full}), we omitted the interference terms between left-handed and right-handed currents 
as they are suppressed due to different electron helicities. However the interference terms arising from mixed helicity 
$\lambda$ and $\eta$ diagrams are included.

The neutrino virtual momentum 
$|p^2| \simeq (\mbox{100\, MeV})^2$ plays a crucial role as the formula for $0\nu\beta\beta$ transition could be 
different for $M^2_i \ll p^2$ or $M^2_i \gg p^2$ where $M_i$ denoted as mass of any type 
of neutrinos.
It is observed that the light neutrinos having $m_i \ll p^2$ contributing 
to nuclear matrix element different than the mediating particles having masses $M^2_i \gg p^2$. 
To demonstrate this, we assume that the $0\nu\beta\beta$ transition is only mediated by light neutrinos 
and heavy neutrinos while neglecting the right-handed current effects. In this scenario, the 
analytic formula for inverse half-life for a given isotope from purely left-handed currents 
due to exchange of light $\nu$ and heavy $N$ neutrinos is given by 
\begin{eqnarray}
 \left[T_{1/2}^{0\nu}\right]^{-1} &=& G^{0\nu}_{01}\bigg[ |{\cal M}^{0\nu}_\nu \cdot \eta_\nu|^2 + 
 |{\cal M}^{0\nu}_N \cdot \eta_{N}\big|^2 \bigg] \nonumber \\ 
 &=& G^{0\nu}_{01}\bigg[ |{\cal M}^{0\nu}_\nu|^2 \frac{\langle m^{\nu}_{ee} \rangle^2}{m^2_e} + 
 |{\cal M}^{0\nu}_N|^2 \bigg| \left(\frac{m_p}{\langle M_N \rangle}\right) \bigg|^2 \bigg]
\label{eq:Hlife-a}
\end{eqnarray}
Here 
\begin{eqnarray}
\langle m^{\nu}_{ee} \rangle = \sum_{i} U^2_{ei} m_i\, \quad {\rm and} \quad \frac{1}{\langle M_N \rangle} = - \sum_i \frac{V^2_{ei}}{M_i}\, .
\end{eqnarray}
Actually, we normalized here the inverse half-life for standard mechanism due to exchange of 
light neutrino mechanism as
\begin{eqnarray}
\left[T_{1/2}^{0\nu}\right]^{-1} = G^{0\nu}_{01} \left| \frac{{\cal M}^{0\nu}_\nu}{m_e} \right|^2\, 
|{m}_{\rm ee}^{\nu}|^2 \, .
\end{eqnarray}
Now we take $G^{0\nu}_{01} \left| \frac{{\cal M}^{0\nu}_\nu}{m_e} \right|^2$ as common factor and normalized others 
with respect to this common factor. Then using Eq.(\ref{eq:Hlife-a}), one can express
\begin{eqnarray}
 \left[T_{1/2}^{0\nu}\right]^{-1}  
 &=& G^{0\nu}_{01} \bigg| \frac{{\cal M}^{0\nu}_\nu}{m_e} \bigg|^2 
 \bigg[   \big| \sum_{i} U^2_{ei} m_i \big|^2 
        + \big| \left(- m_p m_e \frac{{\cal M}^{0\nu}_N}{{\cal M}^{0\nu}_\nu}\right) \sum_{i} \frac{V^2_{ei}}{M_i}\,\big|^2 \bigg] \nonumber \\
 &=& G^{0\nu}_{01} \bigg| \frac{{\cal M}^{0\nu}_\nu}{m_e} \bigg|^2 \left|m^{\nu+N}_{ee} \right|^2
\label{eq:Hlife-b}
\end{eqnarray}
where
\begin{eqnarray}
&&\left|m^{\nu+N}_{ee} \right|^2 \equiv \left|m^{\rm eff}_{ee} \right|^2 = \left|m^{\nu}_{ee} \right|^2 + \left|m^{N}_{ee} \right|^2 \nonumber \\
&& m^{N}_{ee} = \left(- m_p m_e \frac{{\cal M}^{0\nu}_N}{{\cal M}^{0\nu}_\nu} \right) \sum_{i} \frac{V^2_{ei}}{M_i}
             \equiv \langle p \rangle^2 \sum_{i} \frac{V^2_{ei}}{M_i} \, .
\end{eqnarray}
It is clear now that the virtual momentum can be expressed in terms of known masses and nuclear matrix elements. 
\begin{eqnarray}
\langle p \rangle^2 = - m_e\,m_p\, \frac{{\cal M}^{0\nu}_N}{{\cal M}^{0\nu}_\nu} \simeq (\mbox{100\, MeV})^2\,.
\end{eqnarray}

We discuss here another situation to get a clear idea about how heavy Majorana neutrinos contribute to neutrinoless double beta decay 
mediated by purely left-handed currents and purely right-handed currents and difference between them. 
Since we have already discussed heavy neutrino contributions to $0\nu\beta\beta$ transition, one can express 
inverse half-life formula arising from right-handed currents due to exchange of heavy neutrinos $N_R$ as
\begin{subequations}
\begin{eqnarray}
\frac{1}{T_{1/2}^{0\nu}}&=&
	G^{0\nu}_{01} \bigg[\left| \mathcal{M}_N^{0\nu} \cdot \eta^{R}_N \right|^2 \bigg] \nonumber \\
	&=&G^{0\nu}_{01} \bigg[ \left|{\cal M}^{0\nu}_N\right|^2 \big| \left(\frac{g_R}{g_L}\right)^4 \left(\frac{M_{W_L}}{M_{W_R}}\right)^4 
	                       \left(\frac{m_p}{\langle M_N \rangle}\right) \big|^2 \bigg]\, .
\end{eqnarray}
\label{eq:half-life_typeII}
\end{subequations}

Again following Eq.(1.2), Eq.(1.3) and normalized with respect to standard factor 
$G^{0\nu}_{01} \left| \frac{{\cal M}^{0\nu}_\nu}{m_e} \right|^2$, 
one can express 
\begin{eqnarray}
 \left[T_{1/2}^{0\nu}\right]^{-1}  
 &=& G^{0\nu}_{01} \bigg| \frac{{\cal M}^{0\nu}_\nu}{m_e} \bigg|^2 
 \bigg[ \big| \left(- m_p m_e \frac{{\cal M}^{0\nu}_N}{{\cal M}^{0\nu}_\nu}\right) 
            \left(\frac{g_R}{g_L}\right)^4 \left(\frac{M_{W_L}}{M_{W_R}}\right)^4  
            \sum_{i} \frac{V^2_{ei}}{M_i}\,\big|^2 \bigg] \nonumber \\
 &=& G^{0\nu}_{01} \bigg| \frac{{\cal M}^{0\nu}_\nu}{m_e} \bigg|^2 \left|m^{N}_{ee,R} \right|^2
\label{eq:Hlife-b}
\end{eqnarray}
where 
$$m^{N}_{ee,R} = \left(- m_p m_e \frac{{\cal M}^{0\nu}_N}{{\cal M}^{0\nu}_\nu} \right) 
              \left(\frac{g_R}{g_L}\right)^4 \left(\frac{M_{W_L}}{M_{W_R}}\right)^4 
              \sum_{i} \frac{V^2_{ei}}{M_i}
             \equiv \langle p \rangle^2 
             \left(\frac{g_R}{g_L}\right)^4 \left(\frac{M_{W_L}}{M_{W_R}}\right)^4  
             \sum_{i} \frac{V^2_{ei}}{M_i}. $$
It is seen that the proton mass $m_p$ appears whenever neutrinoless double beta decay is mediated by heavy particles 
regardless of left-handed or right-handed currents. However, with right-handed current an additional factor of 
$\left(\frac{g_R}{g_L}\right)^4 \left(\frac{M_{W_L}}{M_{W_R}}\right)^4$ appears. {\bf Similarly, one can express 
half-life for mixed helicity $\lambda$ and $\eta$ diagrams and their interference terms in terms of effective Majorana 
mass parameters.}

In order to arrive at a common normalization factor for all types of contributions, at first we use 
the expression for inverse half-life for $0\nu 2\beta$ decay process due to only light active Majorana neutrinos, 
$\left[T_{1/2}^{0\nu}\right]^{-1} = G^{0\nu}_{01}\left|{\cal M}^{0\nu}_\nu \right|^2|\eta_\nu|^2$. 
Using the numerical values given in Table.\ref{tab:nucl-matrix}, we rewrite the inverse half-life 
in terms of effective mass parameter
$$\left[T_{1/2}^{0\nu}\right]^{-1} = G^{0\nu}_{01} \left| \frac{{\cal M}^{0\nu}_\nu}{m_e} \right|^2\, 
|{\large \bf m}^{\rm ee}_{\nu}|^2 = 1.57 \times 10^{-25}\, \mbox{yrs}^{-1}\, \mbox{eV}^{-2} |{\large 
\bf m}^{\rm ee}_{\nu}|^2 = \mathcal{K}_{0\nu}\, |{\large \bf m}^{\rm ee}_{\nu}|^2 $$
where ${\large \bf m}^{\rm ee}_{\nu} = \sum^{}_{i} \left(\mbox{V}^{\nu \nu}_{e\,i}\right)^2\, m_{\nu_i}$ and 
$\mathcal{K}_{0\nu}\, \simeq 1.57 \times 10^{-25}\, \mbox{yrs}^{-1}\, \mbox{eV}^{-2}$.

We present here the analytic formula for half-life and normalized effective mass parameters 
for neutrinoless double beta decay for a given isotope for all relevant contributions are as follows
\begin{eqnarray}
\left[T_{1/2}^{0\nu}\right]^{-1}&=& \mathcal{K}_{0\nu}\, 
\bigg[ |{\large \bf m}_{\rm ee}^{\nu}|^2 + |{\large \bf m}_{\rm ee,L}^{S,N}|^2 
  + |{\large \bf m}_{\rm ee,R}^{S,N}|^2+ |{\large \bf  m}_{\rm ee}^{\lambda}|^2 
  + |{\large \bf m}_{\rm ee}^{\eta}|^2 \bigg] + \cdots \nonumber \\
&=&\mathcal{K}_{0\nu}\, \bigg[ 
    \bigg\{ |{\large \bf  m}_{\rm ee}^{\nu}|^2 + |{\large \bf m}_{\rm ee,L}^{S}+{\large \bf m}_{\rm ee,L}^{N}|^2 \bigg\}
  + \bigg\{ |{\large \bf  m}_{\rm ee,R}^{S}+{\large \bf  m}_{\rm ee,R}^{N}|^2 \bigg\}  \nonumber \\
  &+&\bigg\{ |{\large \bf  m}_{\rm ee}^{\lambda,\nu}+{\large \bf  m}_{\rm ee}^{\lambda,S}
            +{\large \bf  m}_{\rm ee}^{\lambda,N}|^2 \bigg\} 
   +\bigg\{ |{\large \bf  m}_{\rm ee}^{\eta,\nu}+{\large \bf  m}_{\rm ee}^{\eta,S}
            +{\large \bf  m}_{\rm ee}^{\eta,N}|^2 \bigg\}  \bigg] + \mbox{Interference terms}   \nonumber \\
\label{eqn:halflife-eff}
\end{eqnarray}
In the above expression for inverse half-life, $G^{0\nu}_{01}$ is the the phase space factor and 
the other nuclear matrix elements defined for different chiralities of the weak currents 
such as $\left(\mathcal{M}^{0\nu}_{\nu} \right)$, $\left(\mathcal{M}^{0\nu}_{N} \right)$, 
$\left(\mathcal{M}^{0\nu}_{\lambda} \right)$ and $\left(\mathcal{M}^{0\nu}_{\eta} \right)$ are 
presented in Table.\ref{tab:nucl-matrix}. The effective Majorana mass parameters due to purely left handed currents 
are presented in Table.\ref{tab:mee_LL} while Table.\ref{tab:mee_RR} represents the effective Majorana mass parameters 
due to purely right handed currents and Table.\ref{tab:mee_LR} shows the contributions due to involvement of both left 
handed as well as right handed currents. However we do not take into account the interference terms in this work. 

\begin{table}[t]
 \centering
\vspace{10pt}
 \begin{tabular}{c|c}
 \hline \hline
Effective Mass Parameters & Analytic formula  \\
\hline
${\large \bf  m}_{\rm ee,L}^{\nu}$ & $\sum_{i=1}^3 {\mbox{V}^{\nu \nu}_{e\,i}}^2\, m_{\nu_i} $  \\
${\large \bf  m}_{\rm ee,L}^{N}$   & $\sum_{i=1}^3 {\mbox{V}^{\nu N}_{e\,i}}^2\, \frac{|p|^2}{M_{N_i}}$  \\
${\large \bf  m}_{\rm ee,L}^{S}$   & $\sum_{i=1}^3 {\mbox{V}^{\nu S}_{e\,i}}^2\, \frac{|p|^2}{M_{S_i}}$   \\
\hline \hline
 \end{tabular}
 \caption{Effective Majorana mass parameters due to purely left-handed currents}
 \label{tab:mee_LL}
\end{table}

\begin{table}[htb!]
 \centering
\vspace{10pt}
 \begin{tabular}{c|c}
 \hline \hline
Effective Mass Parameters & Analytic formula  \\
\hline
${\large \bf  m}_{\rm ee,R}^{\nu}$ & $\left(\frac{M_{W_L}}{M_{W_R}} \right)^4 
            \left(\frac{g_R}{g_L} \right)^4\, \sum_{i=1}^3  {\mbox{V}^{N \nu}_{e\,i}}^2\, m_{\nu_i}$   \\
${\large \bf  m}_{\rm ee,R}^{N}$   & $\left(\frac{M_{W_L}}{M_{W_R}} \right)^4 
            \left(\frac{g_R}{g_L} \right)^4\, \sum_{i=1}^3 {\mbox{V}^{N N}_{e\,i}}^2\, \frac{|p|^2}{M_{N_i}} $  \\
${\large \bf  m}_{\rm ee,R}^{S}$ & $\left(\frac{M_{W_L}}{M_{W_R}} \right)^4 
            \left(\frac{g_R}{g_L} \right)^4\, \sum_{i=1}^3  {\mbox{V}^{N S}_{e\,i}}^2\, \frac{|p|^2}{M_{S_i}}$  \\
\hline \hline
 \end{tabular}
 \caption{Effective Majorana mass parameters due to purely right-handed currents}
 \label{tab:mee_RR}
\end{table}

\begin{table}[h!]
 \centering
\vspace{10pt}
 \begin{tabular}{c|c}
 \hline \hline
Effective Mass Parameters  &  Analytic formula  \\
\hline
${\large \bf  m}_{\rm ee,\lambda}^{\nu}$   & $10^{-2}\, \left(\frac{M_{W_L}}{M_{W_R}} \right)^2 
            \left(\frac{g_R}{g_L} \right)^2\, \sum_{i=1}^3  \mbox{V}^{\nu \nu}_{e\,i} \mbox{V}^{N \nu}_{e\,i}\, |p|$  \\
${\large \bf  m}_{\rm ee,\lambda}^{N}$   & $10^{-2}\, \left(\frac{M_{W_L}}{M_{W_R}} \right)^2 
            \left(\frac{g_R}{g_L} \right)^2\, \sum_{j=1}^3  \mbox{V}^{\nu N}_{e\,j} \mbox{V}^{N N}_{e\,j}\, \frac{|p|^3}{M^2_{N_j}}$  \\
${\large \bf  m}_{\rm ee\lambda}^{S}$   & $10^{-2}\, \left(\frac{M_{W_L}}{M_{W_R}} \right)^2 
            \left(\frac{g_R}{g_L} \right)^2\, \sum_{k=1}^3  \mbox{V}^{\nu S}_{e\,k} \mbox{V}^{N S}_{e\,k}\, \frac{|p|^3}{M^2_{S_k}}$  \\
\hline
${\large \bf  m}_{\rm ee,\eta}^{\nu}$   & $ \left(\frac{g_R}{g_L}\right)\, \sum_{i=1}^3
 \mbox{V}^{\nu \nu}_{e\,i} \mbox{V}^{N \nu}_{e\,i}\, \tan \zeta_{LR}\, |p|$  \\
${\large \bf  m}_{\rm ee,\eta}^{N}$   & $\left(\frac{g_R}{g_L}\right)\, \sum_{j=1}^3 
 \mbox{V}^{\nu N}_{e\,j} \mbox{V}^{N N}_{e\,j}\, \tan \zeta_{LR}\, \frac{|p|^3}{M^2_{N_j}}$  \\
${\large \bf  m}_{\rm ee,\eta}^{S}$   & $\left(\frac{g_R}{g_L}\right)\, \sum_{k=1}^3 
 \mbox{V}^{\nu S}_{e\,k} \mbox{V}^{N S}_{e\,k}\, \tan \zeta_{LR}\, \frac{|p|^3}{M^2_{S_k}}$  \\
\hline \hline
 \end{tabular}
 \caption{Effective Majorana mass parameters due to so called $\lambda$ and $\eta$ type diagrams. 
          It is to be noted that the suppression factor $10^{-2}$ arises in the $\lambda-$diagram 
          because of normalization w.r.t to the standard mechanism.}
 \label{tab:mee_LR}
\end{table}

\newpage
\section{Numerical results within natural type-II seesaw dominance}
\label{sec:numerics}
%
\subsection{Input Model Parameters}
Before moving towards the numerical estimation of various contributions to neutrinoless double beta decay, 
it is desirable to know the model parameters and thus we list them below. 
%

The method of diagonalization is given in Sec.III and the resulting physical masses for 
all neutral fermions in terms of $U_{PMNS}$ matrix and mass of light neutrinos are give by
\begin{eqnarray}
& &m_\nu = U_{\rm PMNS} m^{\rm diag}_\nu U^T_{\rm PMNS} \, , \nonumber \\[2mm]
& &M_N \equiv M_R = \frac{v_R}{v_L} U_{\rm PMNS} m^{\rm diag}_\nu U^T_{\rm PMNS} \,, \nonumber  \\[1mm]
& &M_S = -m^2_S\, \frac{v_L}{v_R} U^*_{\rm PMNS} {m^{\rm diag}_\nu}^{-1} U^\dagger_{\rm PMNS} \,.
\end{eqnarray}
The flavor basis of active LH neutrinos $\nu_L$, active RH neutrinos $N_R$ and heavy sterile neutrinos 
$S_L$ in terms of mass basis and mixing are given as follows
\begin{eqnarray}
\begin{pmatrix}
\nu_{L} \\ S_{L} \\ N^c_{R}
\end{pmatrix}_\alpha 
&=&
\begin{pmatrix}
{\mbox V}^{\nu\nu} & {\mbox V}^{\nu{S}} & {\mbox V}^{\nu {N}} \\
{\mbox V}^{S\nu} & {\mbox V}^{SS} & {\mbox V}^{SN} \\
{\mbox V}^{N\nu} & {\mbox V}^{NS} & {\mbox V}^{NN} 
\end{pmatrix}_{\alpha i} 
\begin{pmatrix}
\nu \\ S \\ N
\end{pmatrix}_i
\end{eqnarray}
in order to express in terms of know neutrino oscillation parameters and light neutrino masses. Where 
\begin{eqnarray}
&&{\mbox V}^{\nu\nu} = U_{\rm PMNS} \,, \quad  {\mbox V}^{\nu{S}} = \frac{1}{m_S} M_D U^*_{\rm PMNS}\, , \quad 
{\mbox V}^{\nu {N}} = \frac{v_L}{v_R} M_D U^{-1}_{\rm PMNS} {m^{\rm diag.}_{\nu}}^{-1}\,, \nonumber  \\
&&{\mbox V}^{S\nu} = \frac{1}{m_S} M^\dagger_D U_{\rm PMNS}\,, \quad {\mbox V}^{SS} = U^*_{\rm PMNS} \, ,\quad 
{\mbox V}^{SN} = \frac{v_L}{v_R} m_S U^{-1}_{\rm PMNS} {m^{\rm diag.}_{\nu}}^{-1} \,, \nonumber \\
&&{\mbox V}^{N\nu} = \mathbb{O}\,, \quad   {\mbox V}^{NS}=\frac{v_L}{v_R} m_S U^{-1}_{\rm PMNS} {m^{\rm diag.}_{\nu}}^{-1}  \,, \quad  
{\mbox V}^{NN} =U_{\rm PMNS}\,.
\end{eqnarray}
Here we consider Dirac neutrino mass matrix motivated from $SO(10)$ GUT and assumed heavy $N-S$ 
mixing matrix $M$ to be diagonal and degenerate i.e, $M=m_S\, \mbox{diag}\{1,1,1\}$. We fix $m_S$ 
at $500~$GeV for all our numerical estimations. If we assume that the present TeV scale left-right 
symmetric model is originated from Pati-Salam symmetry~\cite{Pati:1974yy} or $SO(10)$ GUT~\cite{Fritzsch:1974nn}, 
then the Dirac neutrino mass matrix $M_D$ can be approximated as up-type quark mass matrix
\footnote{RG effects modifies the value of Dirac neutrino mass matrix at left-right breaking scale 
as discussed in refs.~\cite{Dev:2009aw,LalAwasthi:2011aa,Awasthi:2013ff}}. This can be reconstructed 
using masses of up, charm \& top quarks and the corresponding 
CKM mixing matrix in the quark sector~\cite{Agashe:2014kda, Awasthi:2013we} as
\begin{eqnarray}
M_D&=& V_{CKM}\,M_u\,V^{T}_{CKM} \nonumber \\
&=&\left(
\begin{array}{ccc}
 0.067-0.004\,i & 0.302-0.022\,i  & 0.550-0.530\,i \\
 0.302-0.022\,i & 1.480           & 6.534-0.001\,i \\
 0.550-0.530\,i & 6.534-0.0009\,i & 159.72
\end{array}
\right)\text{GeV}\, . \nonumber
\end{eqnarray}
In the above matrix we use the PDG~~\cite{Agashe:2014kda} value of up-type quark mass matrix and the corresponding 
CKM mixing matrix as
\begin{eqnarray}
&&M_u= \mbox{diag}\{2.3~\mbox{MeV},1.275~\mbox{GeV},173.210~\mbox{GeV}\}\, , \nonumber \\
&&V_{CKM}=
\bmt 
0.97427 & 0.22534 & 0.00351-i0.0033 \\
-0.2252+i0.0001 & 0.97344 & 0.0412 \\
0.00876-i0.0032 & -0.0404-i0.0007 & 0.99912 
\emt\, .
\end{eqnarray}

The bound derived from quark flavor changing neutral current processes is $v_R > 6$~TeV
~\cite{Bertolini:2014sua,Maiezza:2010ic,Zhang:2007da,Beall:1981ze} nonetheless we fix 
it at greater than $8~$TeV. The electroweak $\rho$ parameter gives bounds on left-handed scalar triplet VEV as 
$v_L < 2$ GeV~\cite{Agashe:2014kda} whereas we consider $v_L$ to be $0.1$~eV. The light neutrino masses are 
diagonalised by the Pontecorvo-Maki-Nakagawa-Sakata (PMNS) mixing matrix $U_{\rm PMNS}$ as 
$$m^{\rm diag}_\nu=U^\dagger_{\rm PMNS} m_\nu U^*_{\rm PMNS}= \mbox{diag.}(m_1, m_2, m_3)$$ 
where

\bea
U_{\rm {PMNS}}&=& \begin{pmatrix} c_{13}c_{12}&c_{13}s_{12}&s_{13}e^{-i\delta}\\
-c_{23}s_{12}-c_{12}s_{13}s_{23}e^{i\delta}&c_{12}c_{23}-s_{12}s_{13}s_{23}e^{i\delta}&s_{23}c_{13}\\
s_{12}s_{23}-c_{12}c_{23}s_{13}e^{i\delta}&-c_{12}s_{23}-s_{12}s_{13}c_{23}e^{i\delta}&c_{13}c_{23}
\end{pmatrix}\cdot \mbox{P}\,.
\label{PMNS} 
\eea
Here, we have denoted $s_{ij}=\sin \theta_{ij}$, $c_{ij}=\cos \theta_{ij}$ 
and diagonal phase matrix $\mbox{P}=\mbox{diag}\left(1, e^{i\alpha}, e^{i \beta} \right)$, 
where $\delta$ is the Dirac CP phase and $\alpha$, $\beta$ are Majorana phases varied 
from $0 \to 2 \pi$. From now onwards, we adopt the notations like $(c_\alpha, s_\alpha)\equiv 
(\cos\theta_\alpha, \sin\theta_\alpha)$ where the atmospheric mixing angle is defined as 
$\theta_a\equiv \theta_{23}$, the solar mixing angle is defined as $\theta_s\equiv \theta_{12}$, 
and the reactor mixing angle is defined as $\theta_r\equiv \theta_{13}$. 
The atmospheric, solar and reactor based neutrino oscillation experiments 
provide the values of mixing angles $\theta_{23}$, $\theta_{12}$ and ($\theta_{13}$) 
and mass squared differences like ($\Delta\,m_{\rm atm}^2$) and ($\Delta\,m_{\rm sol}^2$). 
Since the precise measurement of the sign of $\Delta\,m_{\rm atm}^2$ is not confirmed, 
one can have different possibilities in the arrangement of light neutrino masses like
      
\textbf{Normal hierarchy (NH)}: $\Delta\,m_{\rm atm}^2 \equiv \Delta m_{31}^2 > 0$, which gives $ m_1 < m_2 < m_3$ with
      \begin{equation*}
         m_2 = \sqrt{m_1^2 +\Delta\,m_{\rm sol}^2}\;,\qquad 
         m_3 = \sqrt{m_1^2 +\Delta\,m_{\rm atm}^2}\;, \label{eq1:NH_m2_m3}
      \end{equation*}
  
\textbf{Inverted hierarchy (IH)}: $\Delta\,m_{\rm atm}^2 \equiv \Delta m_{31}^2 < 0$, implying $m_3 < m_1 < m_2$ with
      \begin{equation*}
         m_1 = \sqrt{m_3^2 +\Delta\,m_{\rm atm}^2}\;,\qquad 
         m_2 = \sqrt{m_3^2 +\Delta\,m_{\rm atm}^2+\Delta\,m_{\rm sol}^2}\;. \label{eq1:IH_m1_m2}
      \end{equation*}
      
\begin{table}[htb!]
\begin{tabular}{cccc}
        \hline 
Oscillation Parameters & Within 3$\sigma$ range         & within 3$\sigma$ range & within 3$\sigma$ range  \\
           &   ({\it Schwetz et al.}\cite{GonzalezGarcia:2012sz})   &   ({\it Fogli et al.}\cite{Fogli:2012ua}) & Gonzalez-Garcia et al 
           (\cite{Gonzalez-Garcia:2014bfa}) \\
        \hline \hline
$\Delta m^2_{\rm {21}} [10^{-5} \mbox{eV}^2]$              & 7.00-8.09   & 6.99-8.18  & 7.02 - 8.09   \\
$|\Delta m^2_{\rm {31}}(\mbox{NH})| [10^{-3} \mbox{eV}^2]$ & 2.27-2.69   & 2.19-2.62  & 2.317 - 2.607   \\
$|\Delta m^2_{\rm {31}}(\mbox{IH})| [10^{-3} \mbox{eV}^2]$ & 2.24-2.65   & 2.17-2.61  & 2.307 - 2.590   \\
\hline
$\sin^2\theta_{s}$                                        & 0.27-0.34   & 0.259-0.359 & 0.270 - 0.344  \\
$\sin^2\theta_{a}$                                        & 0.34-0.67   & 0.331-0.637 & 0.382 - 0.643  \\
$\sin^2\theta_{r}$                                        & 0.016-0.030 & 0.017-0.031 & 0.0186 - 0.0250  \\
        \hline
\end{tabular}
\caption{The oscillation parameters like mass squared differences and mixing angles 
         within $3\sigma$ range. However we adopt the values given in ref.\cite{Gonzalez-Garcia:2014bfa}.}
\label{table-osc}
\end{table}
%

\subsection{$0\nu\beta\beta$ contributions from purely left-handed currents:-}
The analytic expression for inverse of half-life for neutrinoless double beta decay due to purely left-handed current via
$W_L-W_L$ mediation with the exchange of Majorana neutrinos $\nu_L, S_L$ \& $N_R$ is given by
\bea
 \left[T_{1/2}^{0\nu}\right]^{-1} &=& G^{0\nu}_{01}\bigg[ |{\cal M}^{0\nu}_\nu|^2|\eta^\nu_{L}|^2 
 + |{\cal M}^{0\nu}_N|^2 \big| \eta^{N}_{L} +  \eta^{S}_{L} \big|^2 \bigg]  \nonumber \\
 &=& \mathcal{K}_{0\nu}\, \bigg[ |{\large \bf  m}_{\rm ee}^{\nu}|^2 
 + |{\large \bf  m}_{\rm ee,L}^{N}|^2+|{\large \bf  m}_{\rm ee,L}^{S}|^2 \bigg]\, \nonumber \\
&=&\mathcal{K}_{0\nu}\, \bigg[ \bigg|\sum_{i=1}^3 {\mbox{V}^{\nu \nu}_{e\,i}}^2\, m_i \bigg|^2 
                                 +\bigg|\sum_{j=1}^3 {\mbox{V}^{\nu N}_{e\,j}}^2\, \frac{|p|^2}{M_{N_j}}\bigg|^2
                                 + \bigg|\sum_{k=1}^3 {\mbox{V}^{\nu S}_{e\,k}}^2\, \frac{|p|^2}{M_{S_k}}   \bigg|^2\bigg]
\label{NH-largeMWR-rate}
\eea
\subsubsection{For Standard Mechanism ${\large \bf  m}_{\rm ee}^{\nu}$ and $T_{1/2}^{0\nu} \big|_{\nu}$}
The LNV effective Majorana mass parameter $m^\nu_{ee}$ and corresponding half-life 
$T_{1/2}^{0\nu}\big|_{\nu}$ due to standard mechanism by the exchange of light neutrinos 
is given by 
\begin{subequations}
\begin{eqnarray}
\left| m^\nu_{ee} \right|&=&
  \bigg| |U^2_{e1}|\, m_1 + |U^2_{e2}|\, m_2 e^{i \alpha} + |U^2_{e3}|\, m_3 e^{i \beta} \bigg|  \\ &=&
  \left| c^2_s c^2_r m_1 + s^2_s c^2_r m_2 e^{i \alpha} + s^2_r m_3 e^{i \beta} \right| \,,
\label{eq:mee-std} \\ 
T_{1/2}^{0\nu}\big|_{\nu} &=& \bigg[ G^{0\nu}_{01} \left| \frac{{\cal M}^{0\nu}_\nu}{m_e} \right|^2 
\left| c^2_s c^2_r m_1 + s^2_s c^2_r m_2 e^{i \alpha} + s^2_r m_3 e^{i \beta} \right|^2 \bigg]^{-1}
\label{eq:hlife-std}
\end{eqnarray}
\end{subequations}
\begin{figure}[h!]
\centering
\includegraphics[width=0.49\textwidth]{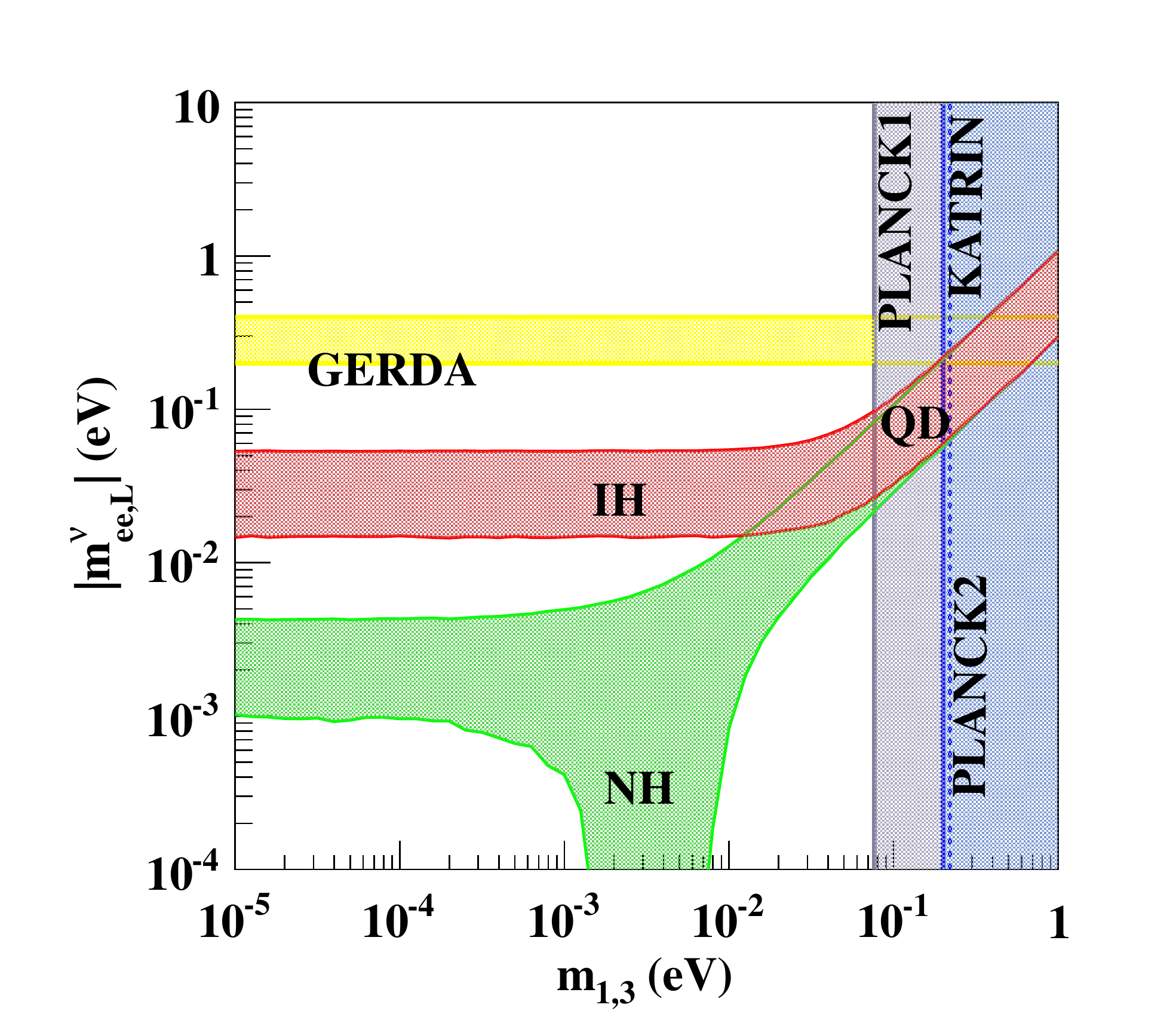}
\includegraphics[width=0.49\textwidth]{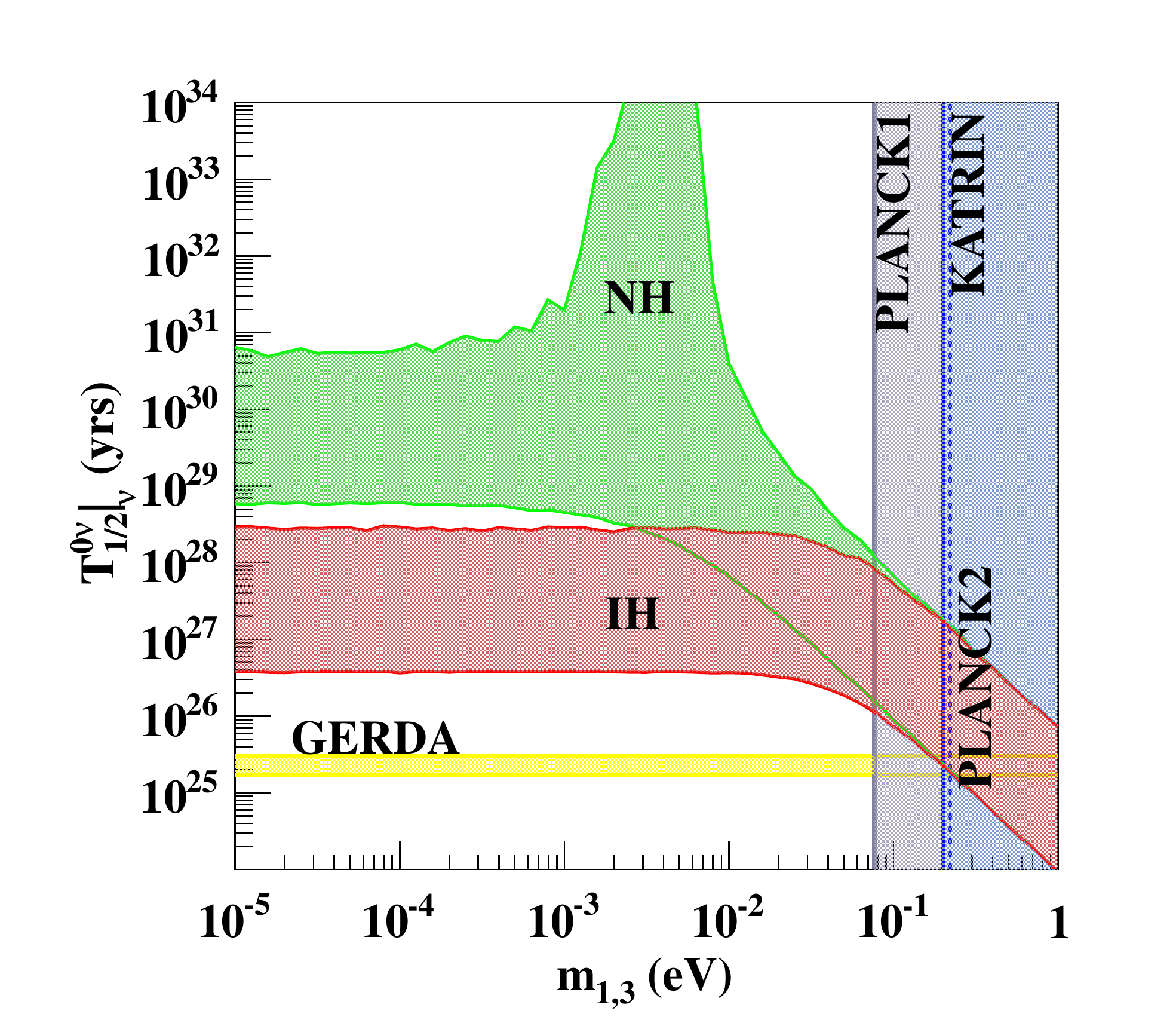}
\caption{Left Panel: The SM contribution to the plot of effective neutrino mass as a function of the 
lightest neutrino mass, m$_1$ (m$_3$) for NH (IH). Right Panel: The SM contribution to the half life 
of $0\nu\beta\beta$ vs lightest neutrino mass, m$_1$ (m$_3$) for NH (IH). The NH contributions are 
displayed by green band while the IH contributions are given by red band. 
The vertical lines and the corresponding 
shaded areas are for constraints on the sum of light neutrino masses from recent cosmological data (PLANCK1 and PLANCK2) 
and KATRIN detector. The yellow band horizontal lines and the respective shaded areas are for the 
limits in effective Majorana mass parameter and half life by GERDA and EXO+KamLAND-Zen experiments}
\label{nuLL}
\end{figure}

Using $m_e=0.51~$MeV, $G^{0\nu}_{01}$ and ${\cal M}^{0\nu}_\nu$ from Table.\ref{tab:nucl-matrix} and 
$3\,\sigma$ ranges of oscillation parameters like mixing angles and mass squared differences from 
Table.\ref{table-osc}, we examine the variation of effective mass and half-life vs. lightest neutrino 
mass $m_{\rm lightest}= m_1 \mbox{(NH)}, m_3 \mbox{(IH)}$. We plot effective Majorana mass parameter $m^\nu_{ee}$ 
in left-panel of Fig.\ref{nuLL} and half-life $T_{1/2}^{0\nu} \big|_{\nu}$ in the right-panel of Fig.\ref{nuLL} 
as a function of the mass of the lightest neutrino. It is observed that quasi degenerate (QD) pattern of light neutrinos 
i.e., $m_1\simeq m_2 \simeq m_3$ and $m_{\rm lightest} = \sum_i m_i/3$ is disfavoured by current bound on 
the sum of light neutrino mass $m_\Sigma < 0.23$ derived from Planck+WP+highL+BAO data (PLANCK1) at 95 \% 
C.L. while $m_\Sigma < 1.08$ derived from Planck+WP+highL (PLANCK2) at 95 \% C.L.~\cite{Ade:2013zuv}. 
Moreover, IH as well as NH pattern will be difficult to probe within the standard mechanism even for next 
generation experiments. This motivates us to consider all possible new physics contributions to neutrinoless 
double beta decay in the present framework, that might give crucial information about lower limit on absolute 
scale of lightest neutrino mass as well as mass hierarchy. 

\begin{table}[h!]
\begin{center}
	\begin{tabular}{c|c|c|c}
	Isotope & $T_{1/2}^{0 \nu}~[10^{25}\text{ yrs}]$ & $m_\text{ee}^{\nu}~[\text{eV}]$ & Collaboration \\
	\hline
	$^{76}$Ge	& $> 2.1$	& $< (0.2 - 0.4)$ 	& GERDA~\cite{Agostini:2013mzu} 			\\
	$^{136}$Xe	& $> 1.6$	& $< (0.14 - 0.38)$ 	& EXO~\cite{Auger:2012ar} 				\\
	$^{136}$Xe	& $> 1.9$	& n/a 				& KamLAND-Zen~\cite{Gando:2012zm} 		\\
	$^{136}$Xe	& $> 3.6$	& $< (0.12 - 0.25)$ 	& EXO + KamLAND-Zen combined~\cite{Gando:2012zm} 	
	\end{tabular}
\caption{The table shows the lower limits on the half life $T_{1/2}^{0 \nu}$ and upper limits on the effective mass parameter $m_\text{eff}^{0 \nu}$ 
              for $0\nu\beta\beta$ transition for the isotopes $^{76}$Ge and $^{136}$Xe from different collaborations. The range for the effective mass 
              parameter comes from the uncertainties in the nuclear matrix elements.}
\label{tab:0nbb_limits}
\end{center}
\end{table}
      

\subsubsection{Non-standard Mechanism ${\large \bf  m}_{\rm ee}^{N,S}$}
\begin{figure}[h]
\centering
\includegraphics[width=0.49\textwidth]{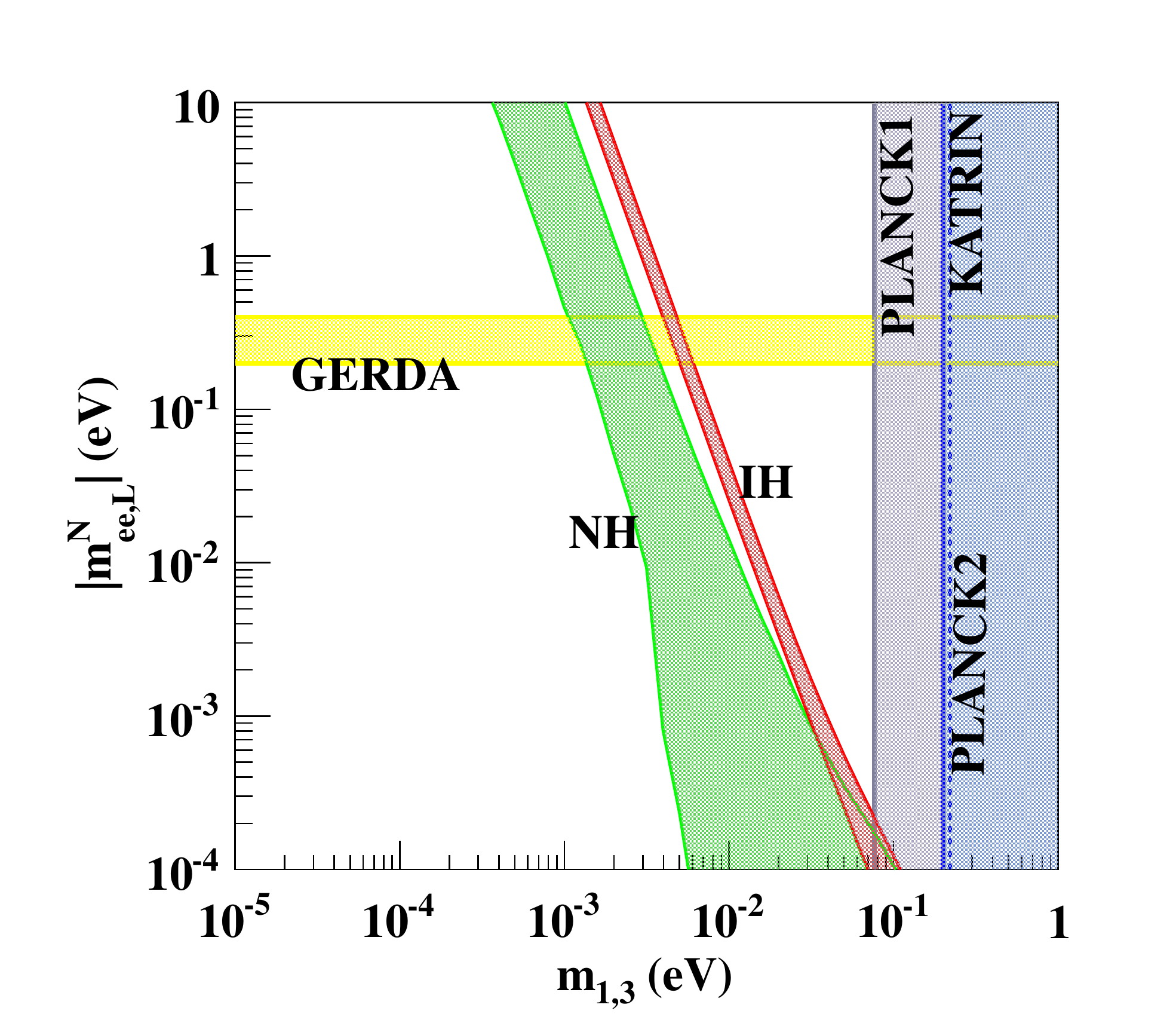}
\includegraphics[width=0.49\textwidth]{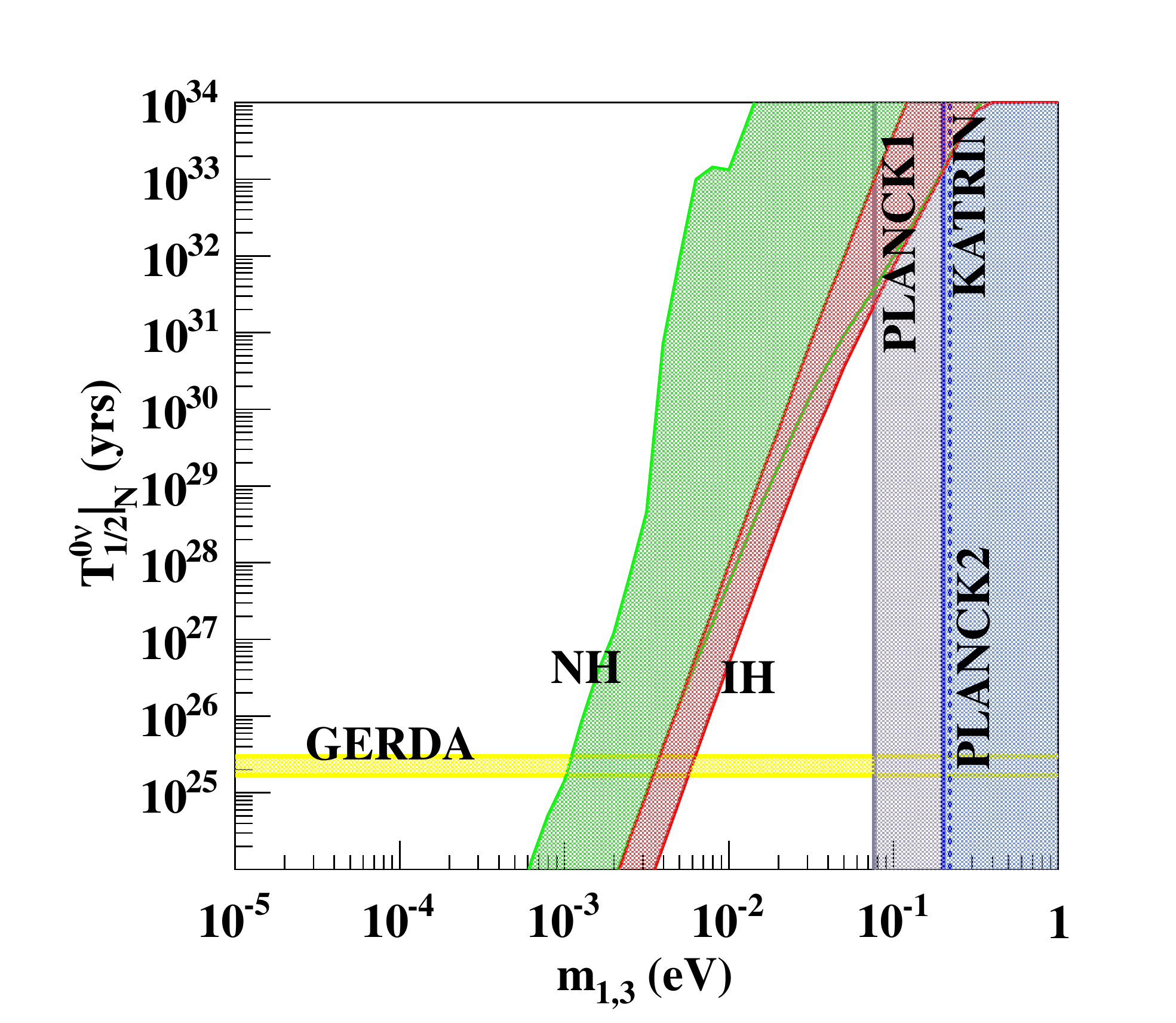}
\caption{Left Panel: The LRSM type-II seesaw dominance contribution to the plot of effective neutrino 
mass as a function of the lightest neutrino mass, m$_1$ (m$_3$) for NH (IH) via $W_L-W_L$ mediation with 
the exchange of virtual RH neutrino (N). Right Panel: The corresponding half life 
of $0\nu\beta\beta$ vs lightest neutrino mass, m$_1$ (m$_3$) for NH (IH).}
\label{NLL}
\end{figure}

\begin{figure}[h]
\centering
\includegraphics[width=0.49\textwidth]{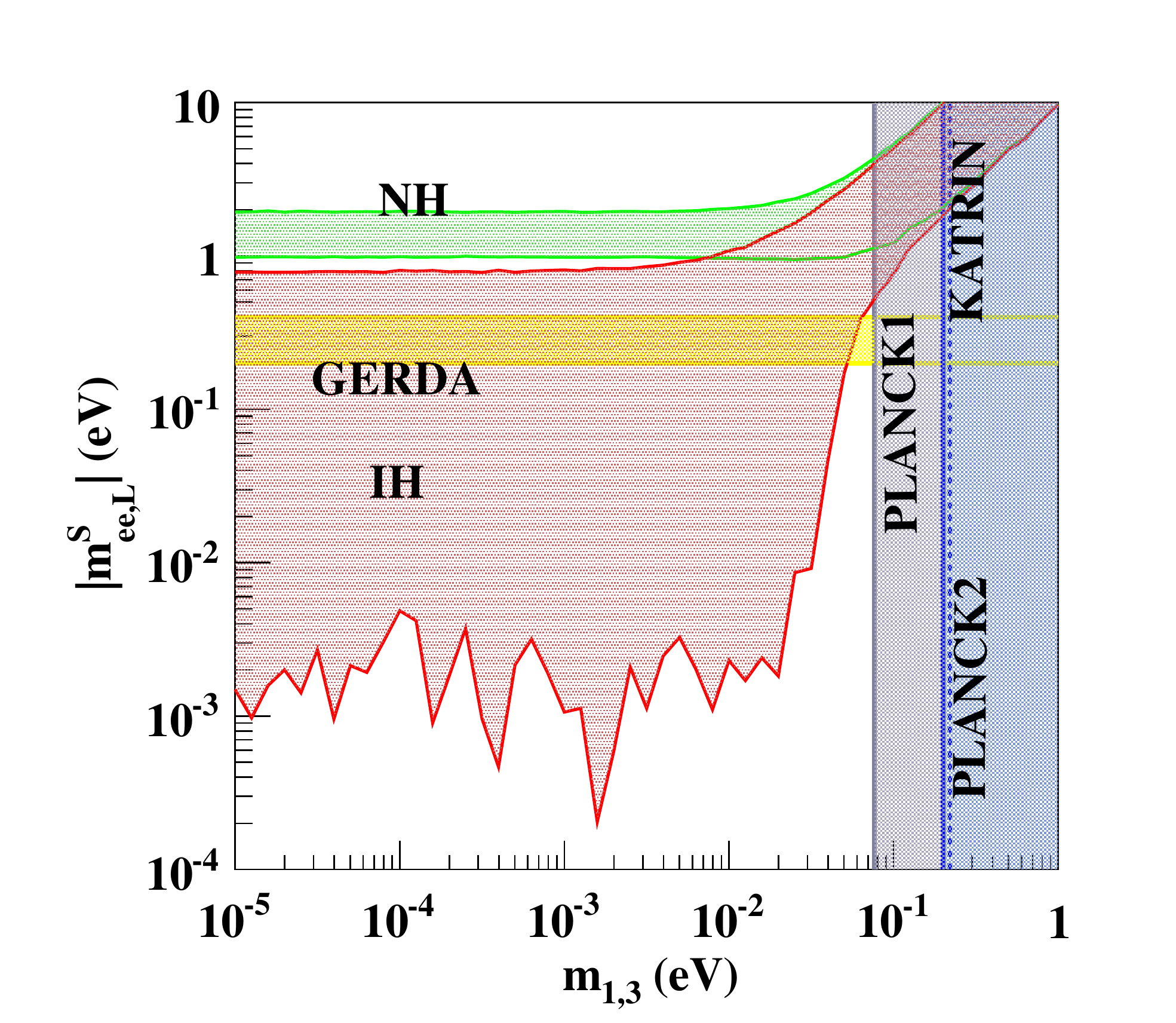}
\includegraphics[width=0.49\textwidth]{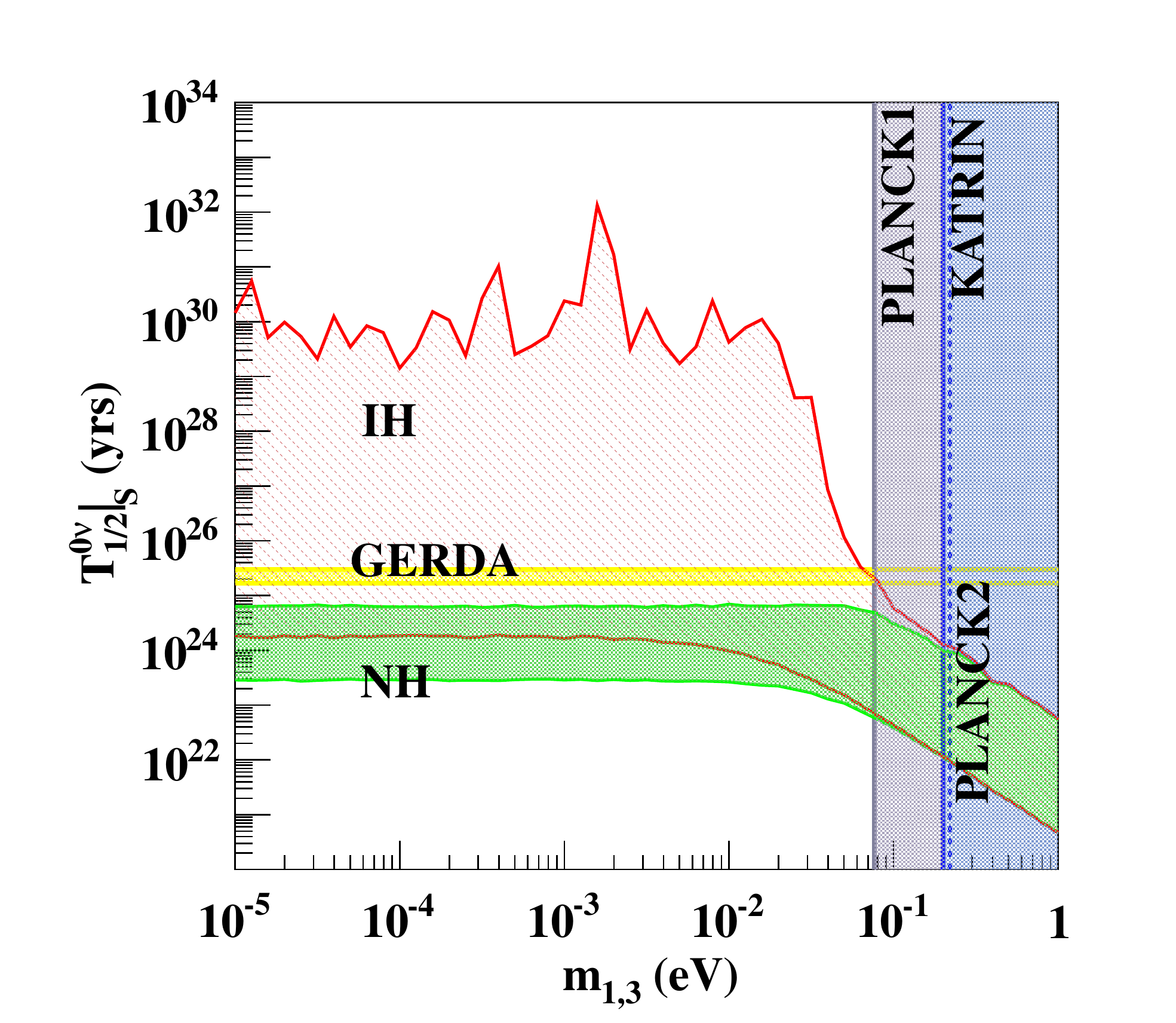}
\caption{Left Panel: The LRSM type-II seesaw dominance contribution to the plot of effective neutrino 
mass as a function of the lightest neutrino mass, m$_1$ (m$_3$) for NH (IH) via $W_L-W_L$ mediation with 
the exchange of virtual sterile neutrino (S). Right Panel: The corresponding half life 
of $0\nu\beta\beta$ vs lightest neutrino mass, m$_1$ (m$_3$) for NH (IH).}
\label{SLL}
\end{figure}

\begin{figure}[!h]
\centering
\includegraphics[width=0.49\textwidth]{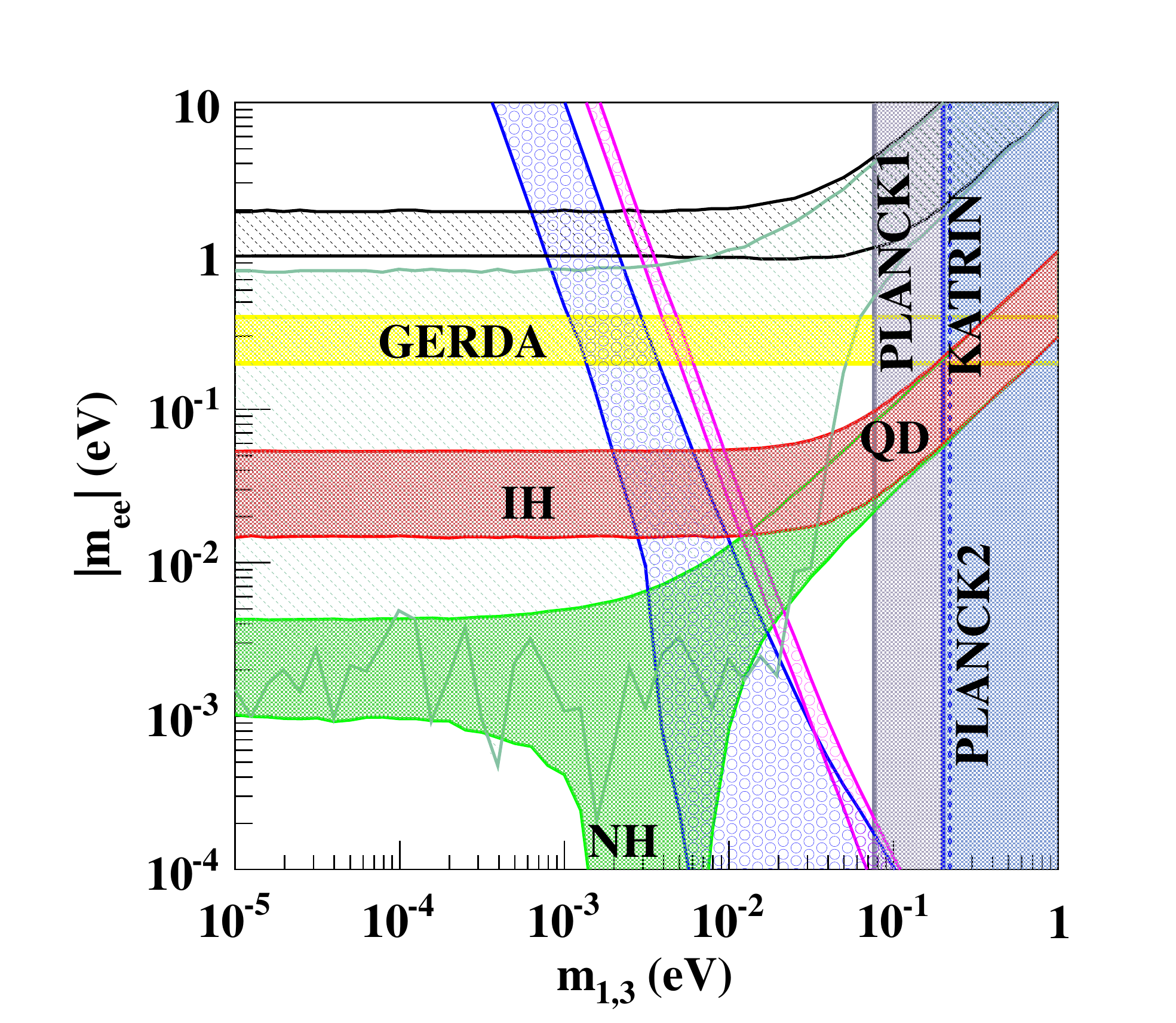}
\includegraphics[width=0.49\textwidth]{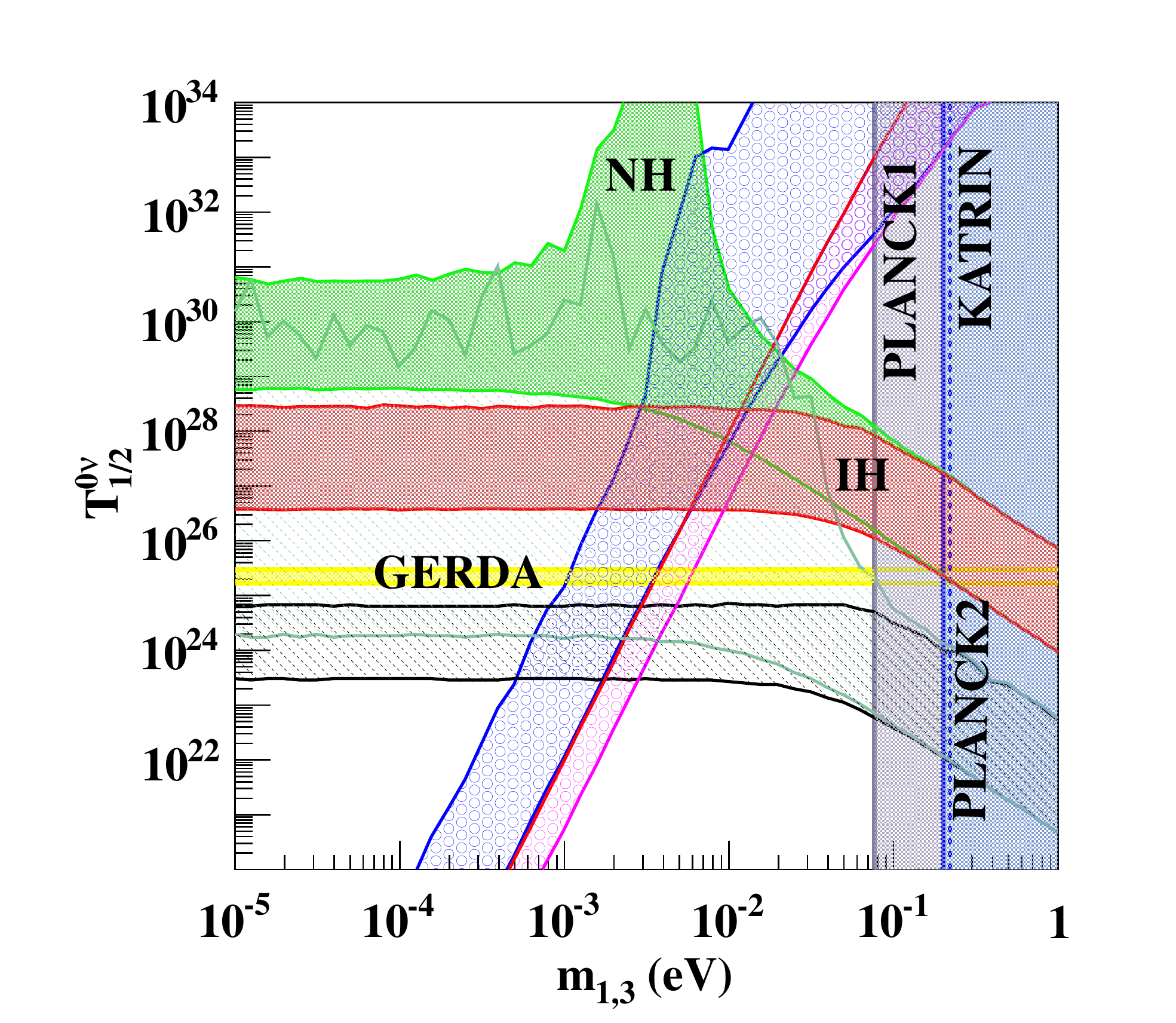}
\caption{Effective Majorana mass (left-panel) and half life (right-panel) as a function of the lightest neutrino mass, 
          $m_1$ ($m_3$) for NH (IH) for combined effect of purely left-handed currents mediated by $\nu$, $N$ and $S$.}
\label{nuNSll}
\end{figure}

The expressions for the effective Majorana mass parameter due to exchange of heavy right-handed 
Majorana neutrinos $N_R$ and extra sterile neutrinos $S_L$ are given by
   \begin{eqnarray}
   && \bigg|{\large \bf  m}_{\rm ee,L}^{N}\bigg| 
        = \bigg|\sum_{k=1}^3 {\mbox{V}^{\nu N}_{e\,k}}^2\, \frac{\langle p \rangle^2}{M_{N_k}} \bigg|
        = \bigg|{\mbox{V}^{\nu N}_{e\,1}}^2\, \frac{\langle p \rangle^2}{M_{N_1}} + {\mbox{V}^{\nu N}_{e\,2}}^2\, \frac{\langle p \rangle^2}{M_{N_2}} 
                + {\mbox{V}^{\nu N}_{e\,3}}^2\, \frac{\langle p \rangle^2}{M_{N_3}} \bigg|\, , \nonumber \\
   && \bigg|{\large \bf  m}_{\rm ee,L}^{S}\bigg| 
        = \bigg|\sum_{k=1}^3 {\mbox{V}^{\nu S}_{e\,k}}^2\, \frac{\langle p \rangle^2}{M_{S_k}} \bigg|
        = \bigg|{\mbox{V}^{\nu S}_{e\,1}}^2\, \frac{\langle p \rangle^2}{M_{S_1}} + {\mbox{V}^{\nu S}_{e\,2}}^2\, \frac{\langle p \rangle^2}{M_{S_2}} 
                + {\mbox{V}^{\nu S}_{e\,3}}^2\, \frac{\langle p \rangle^2}{M_{S_3}} \bigg|\, .
   \end{eqnarray}
The variation of LNV effective mass parameters and corresponding half life with the lightest neutrino mass 
are displayed in left-panel and right-panel of Fig.\ref{NLL} due to exchange of $N_R$. Similarly, we have shown 
${\large \bf  m}_{\rm ee,L}^{S}$ and $T_{1/2}^{0\nu} \big|_{S}$ vs. lightest neutrino mass in Fig.\ref{SLL} and 
the sum of these two new physics contributions is presented in Fig.\ref{nuNSll}.

\begin{table}[h]
 \centering
\vspace{10pt}
 \begin{tabular}{lcccccc}
 \hline \hline
 \\ \multirow{2}{*}{Isotope} & $G^{0\nu}_{01}$ ${\rm yrs}^{-1}]$  
 & \multirow{2}{*}{${\cal M}^{0\nu}_\nu$} & \multirow{2}{*}{${\cal M}^{0\nu}_N$} 
 & \multirow{2}{*}{${\cal M}^{0\nu}_\lambda$} & \multirow{2}{*}{${\cal M}^{0\nu}_\eta$} \\[3mm] 
\hline \\
$^{76}$Ge  & $5.77 \times 10^{-15}$  & 2.58--6.64 & 233--412 & 1.75--3.76 & 235--637 \\ 
$^{136}$Xe  & $3.56 \times 10^{-14}$ & 1.57--3.85 & 164--172 & 1.96--2.49 & 370--419 \\ 
\hline \hline
 \end{tabular}
 \caption{Phase space factor $G^{0\nu}_{01}$ and Nuclear Matrix Elements taken from ref.~\cite{Meroni:2012qf}}
 \label{tab:nucl-matrix}
\end{table}   
\subsection{$0\nu\beta\beta$ from purely right-handed currents}
In the present framework the right-handed gauge boson ${W_R}$ and right-handed Majorana neutrinos
$N_R$ lie around few TeV thereby leading to new physics contributions to neutrinoless double beta decay 
due to purely right-handed currents via $W_R-W_R$ mediation and exchange of heavy neutrinos $N_R$. 
In addition to this, the type-II seesaw dominance connects light and heavy neutrinos with each other for 
which one can express new physics contributions in terms of oscillation parameters. 
\begin{equation}
  m_\nu = M_L \propto M_R \,.
\end{equation}
As a result of this, both light and heavy neutrino masses are diagonalised simultaneously by the 
$U_{\rm PMNS}$ and the mass eigenvalues are related as follows:

\noindent
\textbf{Normal hierarchy (NH)}: 
   \begin{eqnarray}
    && m_2 = \sqrt{m_1^2 +\Delta\,m_{\rm sol}^2}\;,\qquad 
       m_3 = \sqrt{m_1^2 +\Delta\,m_{\rm atm}^2}\;,\nonumber \\
    && M_{N_1} = \frac{m_{1}}{m_{3}}\, M_{N_3}\,,\quad 
       M_{N_2} = \frac{m_{2}}{m_{3}}\, M_{N_3}\,.
   \end{eqnarray}
where we fixed the heaviest RH Majorana neutrino mass $M_{N_3}$ for NH. 

\noindent  
      \textbf{Inverted hierarchy (IH)}: 
   \begin{eqnarray}
   && m_1 = \sqrt{m_3^2 +\Delta\,m_{\rm atm}^2}\;,\qquad 
     m_2 = \sqrt{m_3^2 +\Delta\,m_{\rm atm}^2+\Delta\,m_{\rm sol}^2}\;\nonumber \\
   && M_{N_1} = \frac{m_{1}}{m_{2}}\, M_{N_2}\,,\quad 
      M_{N_3} = \frac{m_{3}}{m_{2}}\, M_{N_2}\,. 
   \end{eqnarray}
where we fixed the heaviest RH Majorana neutrino mass $M_{N_2}$ for IH. 

\begin{figure}[t]
\centering
\includegraphics[width=0.49\textwidth]{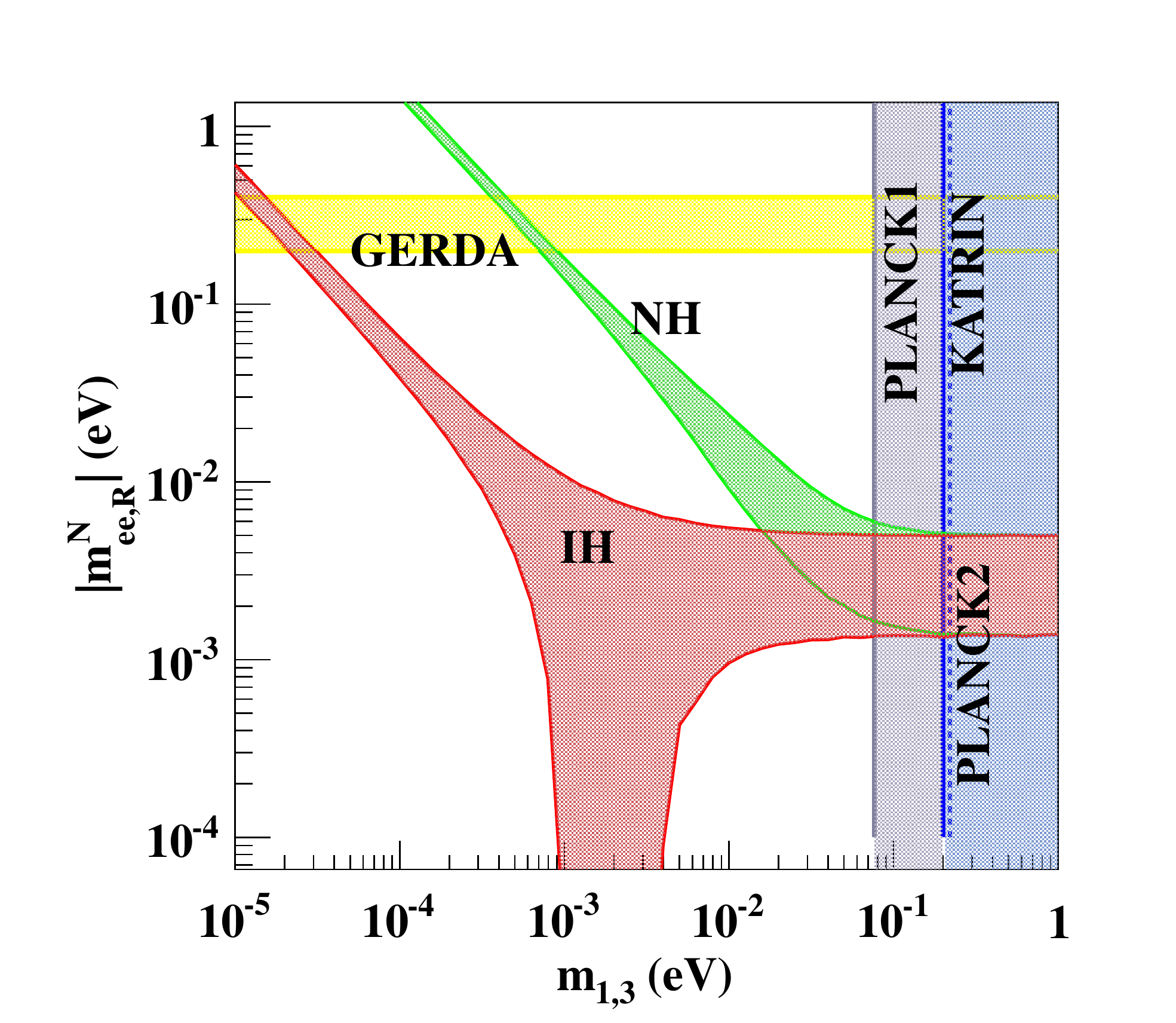}
\includegraphics[width=0.49\textwidth]{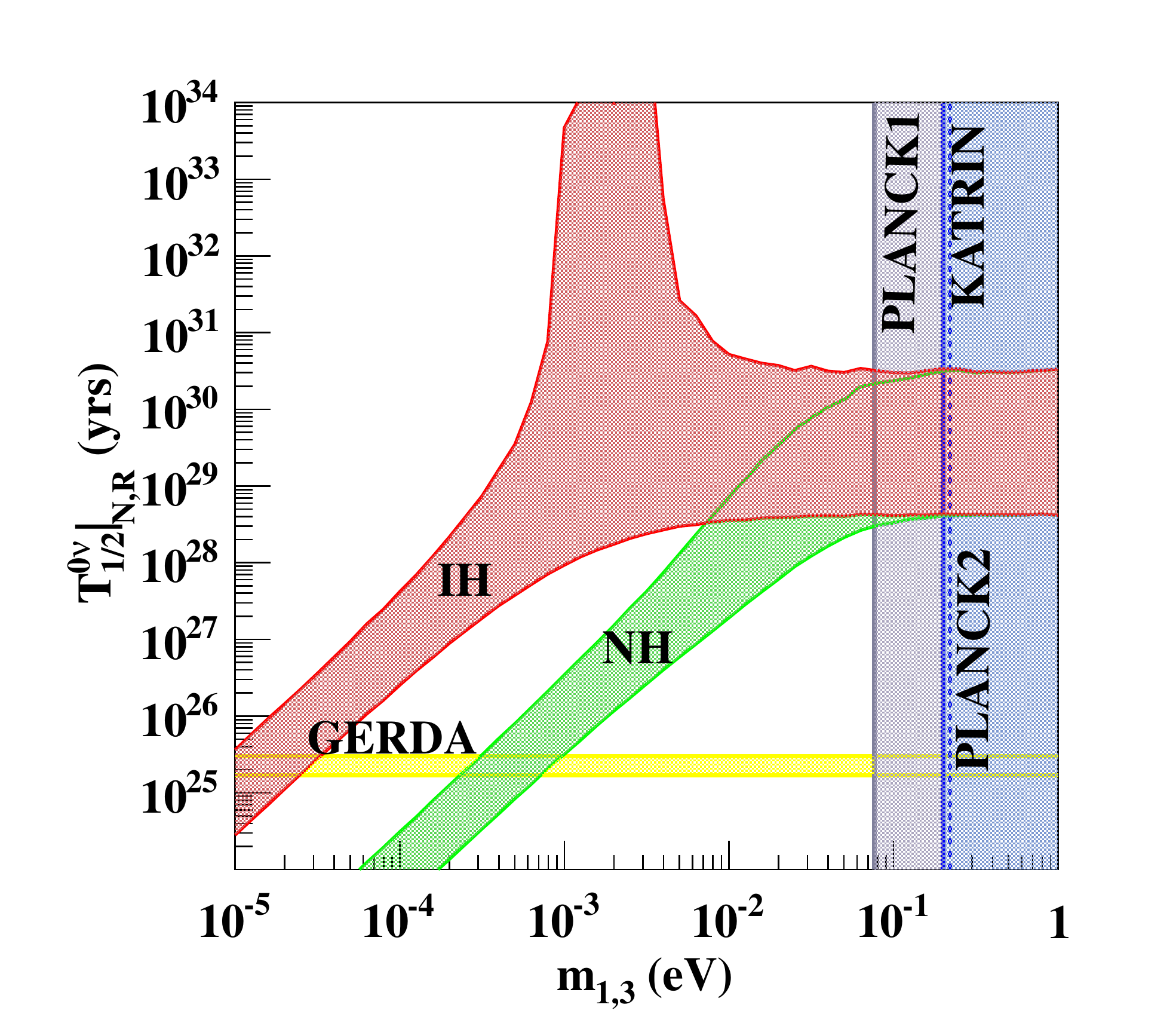}
\caption{Left Panel: The LRSM type-II seesaw dominance contribution to the plot of effective neutrino 
mass as a function of the lightest neutrino mass, m$_1$ (m$_3$) for NH (IH) via $W_R-W_R$ mediation with 
the exchange of virtual RH neutrino (N). Right Panel: The corresponding half life 
of $0\nu\beta\beta$ vs lightest neutrino mass, m$_1$ (m$_3$) for NH (IH).}
\label{NRR}
\end{figure}

The expression for inverse half-life of $0\nu\beta\beta$ transition for a given isotope due to purely 
right-handed currents along with standard mechanism is given by
\begin{subequations}
\begin{eqnarray}
\frac{1}{{T_{1/2}^{0\nu}}\big|_{N,R}}&=&
	G^{0\nu}_{01} \bigg|\frac{\mathcal{M}_\nu^{0\nu}}{m_e} \bigg|^2 \cdot |m^N_{\rm ee,R}|^2 \\
\frac{1}{{T_{1/2}^{0\nu}}\big|_{LR}}&=&
G^{0\nu}_{01} \bigg|\frac{\mathcal{M}_\nu^{0\nu}}{m_e} \bigg|^2 \bigg[|m^\nu_{\rm ee}|^2+ |m^N_{\rm ee,R}|^2 \bigg]
	\\
	&=&
 	G^{0\nu}_{01} \big|\frac{\mathcal{M}_\nu^{0\nu}}{m_e} |m^{\left(\nu+N\right)}_{\rm ee}|^2 ,
\end{eqnarray}
\label{eq:half-life_typeII}
\end{subequations}
where $|m^{\left(\nu+N\right)}_{\rm ee}|^2 =|m^\nu_{\rm ee}|^2+ |m^N_{\rm ee}|^2$. Under this type-II seesaw dominance, 
the expressions for $|m^\nu_{\rm ee}|$ and $|m^N_{\rm ee,R}|$ are given by
\begin{subequations}
\begin{eqnarray}
\hspace{-5mm}
&&m^\nu_{\rm ee}
=\left| c^2_s c^2_r m_1 + s^2_s c^2_r m_2 e^{i \alpha} + s^2_r m_3 e^{i \beta} \right| \,,
\label{eq:mee-std} \\
\hspace{-5mm}
&&m^N_{\rm ee}\bigg|_{NH} 
=\frac{C_N}{M_{3}}  \bigg|c^2_s c^2_r \frac{m_3}{m_1} 
       +  s^2_s c^2_r \frac{m_3}{m_2} \,e^{i \alpha} + s^2_r\,e^{i \beta} \bigg| \,,
\label{eq:NH_mee-typeII} 
 \\
\hspace{-5mm}
&&m^N_{\rm ee}\bigg|_{IH} = \frac{C_N}{M_2} \left[c^2_s c^2_r \frac{m_2}{m_1} 
                               +   s^2_s c^2_r e^{i \alpha} +\frac{m_2}{m_3}  s^2_r e^{i \beta} \right]  \,, 
\quad
\label{eq:IH_mee-typeII} 
\end{eqnarray}
\end{subequations}
where $C_N=\langle p^2 \rangle \left(g_R/g_L\right)^4 \left(M_{W_L}/M_{W_R}\right)^4$. 
We have neglected the other terms arising from purely right-handed currents. However, the right-handed scalar 
triplet contribution can be significant if we consider the triplet mass around $500$~GeV.

This has been discussed in ref.\cite{Ge:2015yqa} and the used model parameters 
are $M_{W_R}\simeq 2~$TeV, $g_R\simeq 2/3 g_L$ and $M_N \simeq 1~$TeV. In the present work, we have considered 
$g_L=g_R\simeq 0.65$, $M_{W_R}\simeq 3~$TeV and $M_N \simeq 5~$TeV for numerical estimation of 
${\large \bf  m}_{\rm ee,L}^{S}$ and $T_{1/2}^{0\nu} \big|_{S}$ vs. lightest neutrino mass in Fig.\ref{SLL}.

\begin{figure}[h]
\centering
\includegraphics[width=0.49\textwidth]{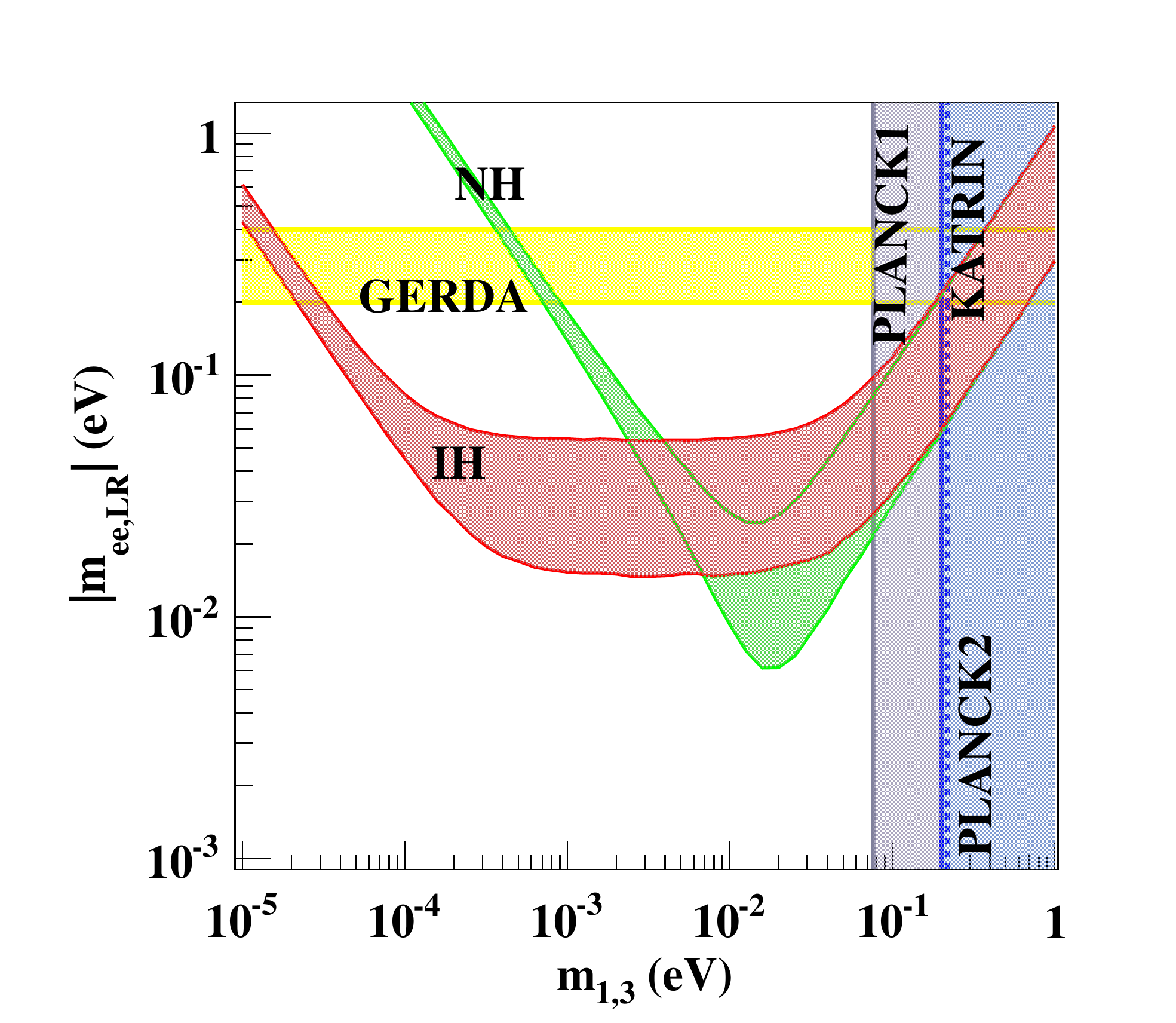}
\includegraphics[width=0.49\textwidth]{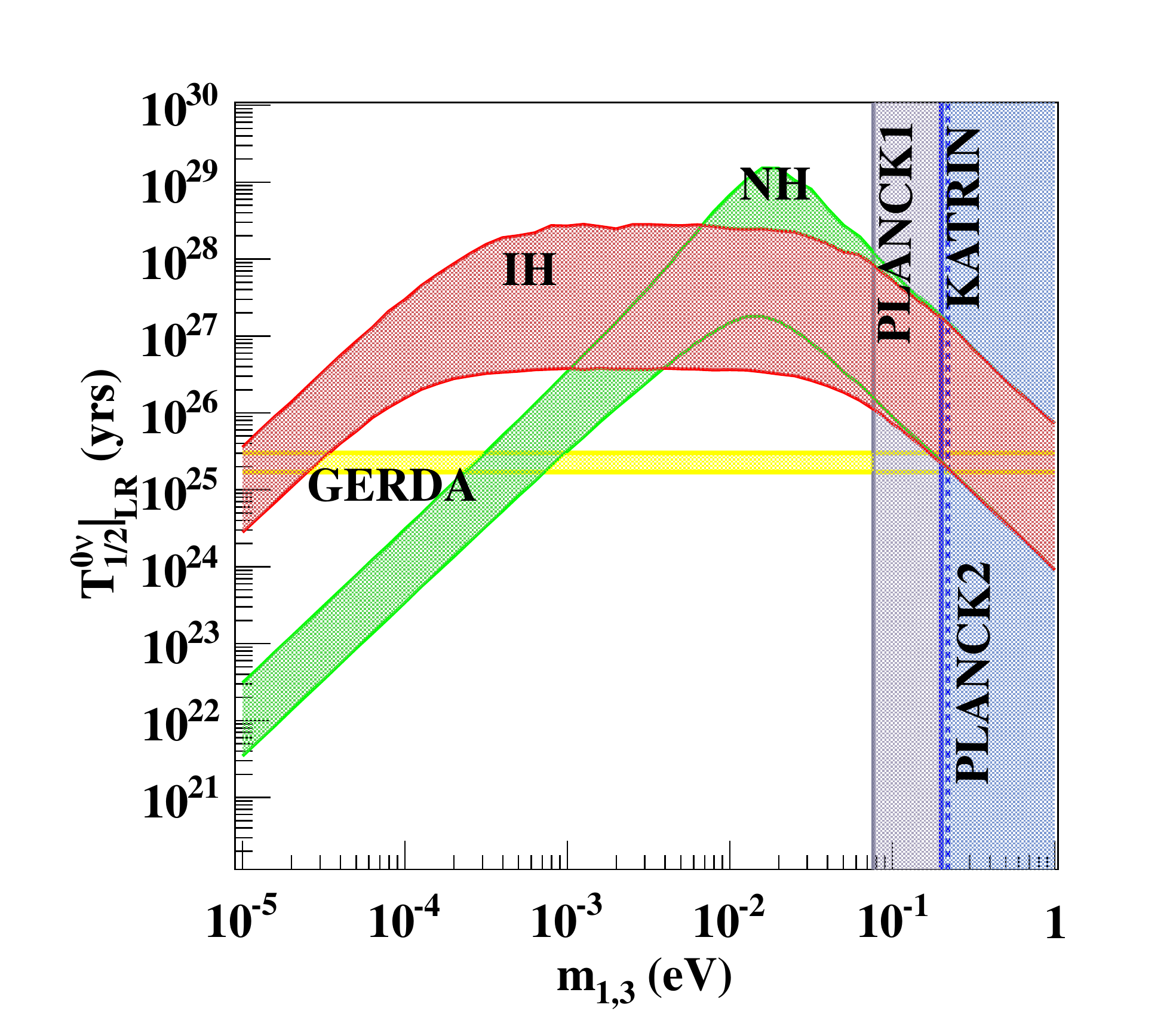}
\caption{Effective Majorana mass (left-panel) and half life (right-panel) as a function of the lightest neutrino mass, 
          $m_1$ ($m_3$) for NH (IH) due to combined effect of standard mechanism and right-handed currents via exchange of 
          heavy right-handed Majorana neutrinos $N$.}
\label{NRR}
\end{figure}

\begin{figure}[h]
\centering
\includegraphics[width=0.49\textwidth]{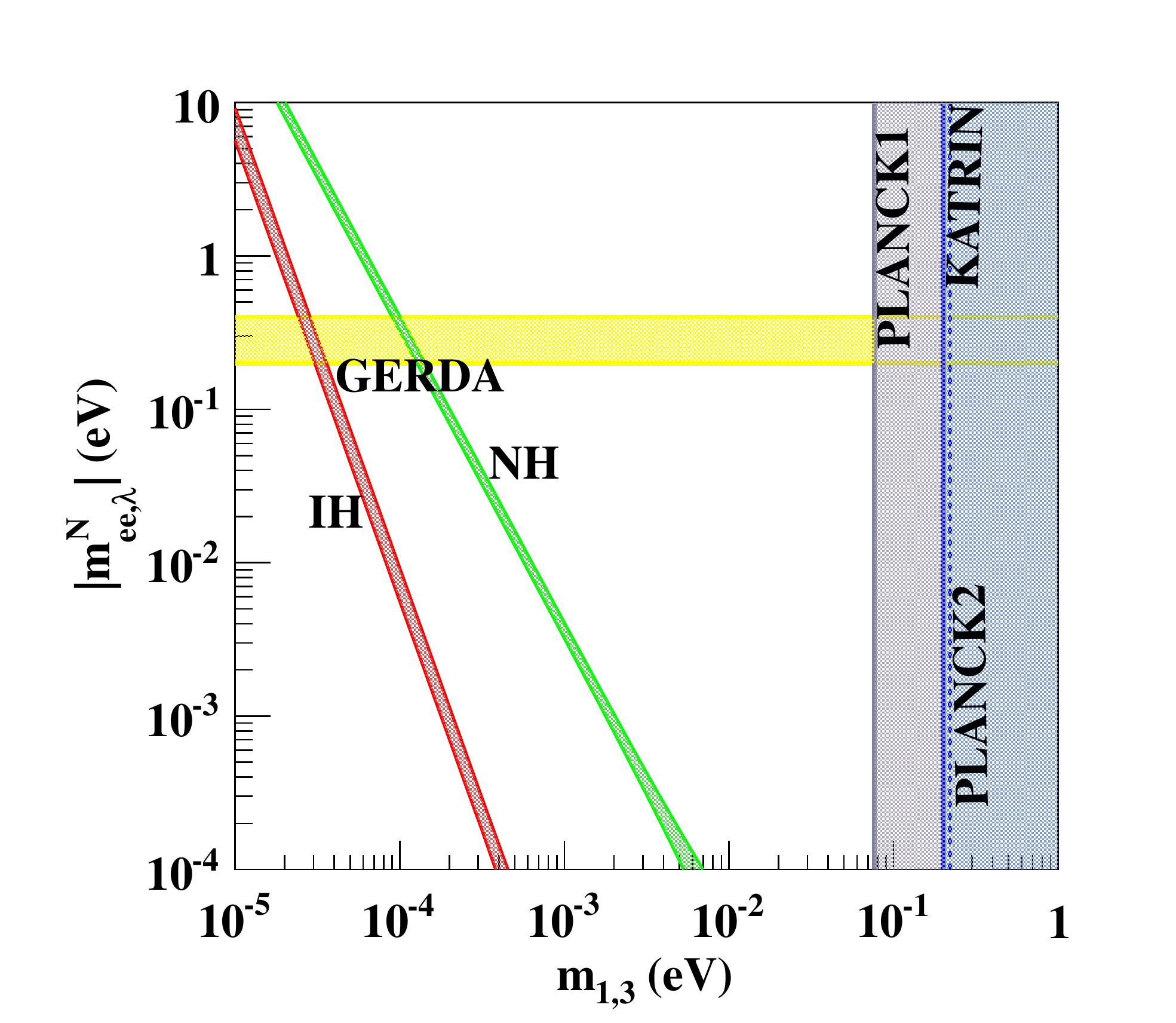}
\includegraphics[width=0.49\textwidth]{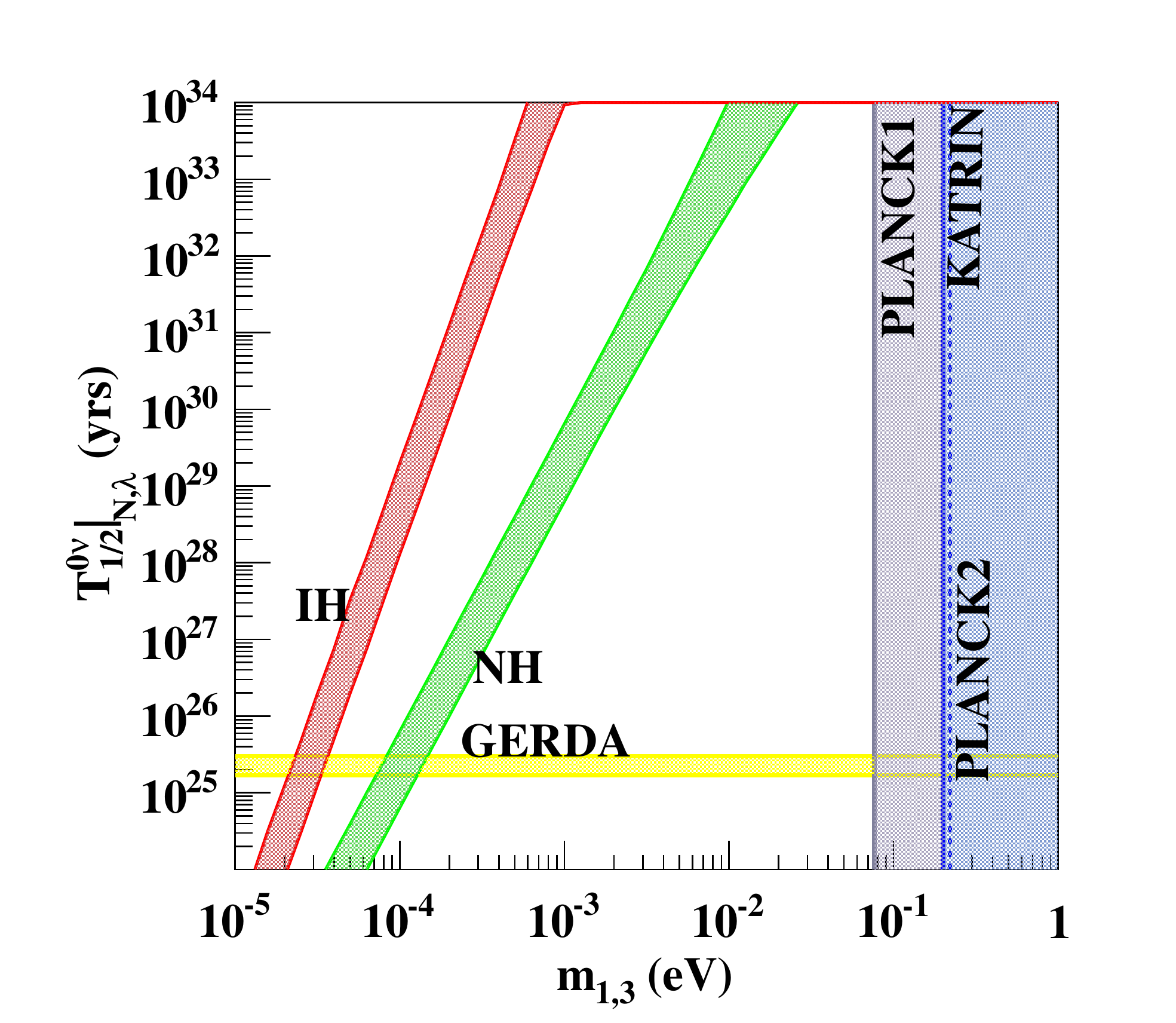}
\caption{Effective Majorana mass (left-panel) and half life (right-panel) as a function of the lightest neutrino mass, 
          $m_1$ ($m_3$) for NH (IH) due to $\lambda$ diagram via exchange of heavy neutrinos.}
\label{NLambda}
\end{figure}
\subsection{$0\nu\beta\beta$ from $\lambda$ and $\eta-$ diagrams} 
In this framework with $g_L = g_R$ and $\mbox{V}^{N\nu}=0$ from the seesaw diagonalization the relevant 
effective Majorana mass parameters due to so called $\lambda$ and $\eta$ diagrams are expressed as follows.

  \begin{eqnarray} 
 &&
 {\large \bf  m}_{\rm ee,\lambda}^{N} =10^{-2}\, \left(\frac{M_{W_L}}{M_{W_R}} \right)^2 
              \sum_{j=1}^3  \mbox{V}^{\nu N}_{e\,j} \mbox{V}^{N N}_{e\,j}\, \frac{|p|^3}{M^2_{N_j}}\\
 &&
 {\large \bf  m}_{\rm ee\lambda}^{S} =10^{-2}\, \left(\frac{M_{W_L}}{M_{W_R}} \right)^2 
              \sum_{k=1}^3  \mbox{V}^{\nu S}_{e\,k} \mbox{V}^{N S}_{e\,k}\, \frac{|p|^3}{M^2_{S_k}}\\
 &&
 {\large \bf  m}_{\rm ee,\eta}^{N} = \sum_{j=1}^3 
  \mbox{V}^{\nu N}_{e\,j} \mbox{V}^{N N}_{e\,j}\, \tan \zeta_{LR}\, \frac{|p|^3}{M^2_{N_j}}\\
 &&
 {\large \bf  m}_{\rm ee,\eta}^{S} = \sum_{k=1}^3 
  \mbox{V}^{\nu S}_{e\,k} \mbox{V}^{N S}_{e\,k}\, \tan \zeta_{LR}\, \frac{|p|^3}{M^2_{S_k}}\\
 \end{eqnarray}
 
\begin{figure}[h]
\centering
\includegraphics[width=0.49\textwidth]{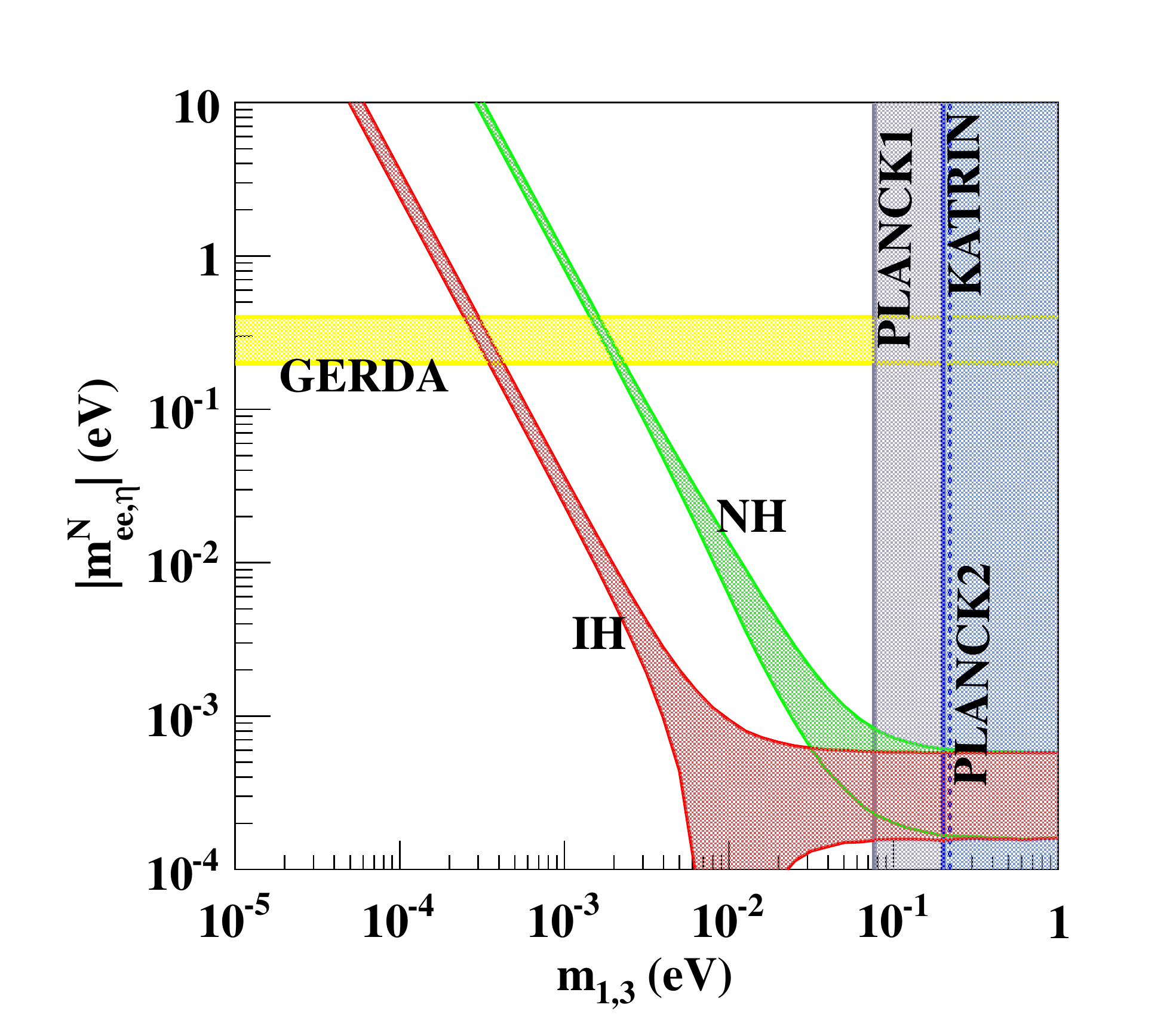}
\includegraphics[width=0.49\textwidth]{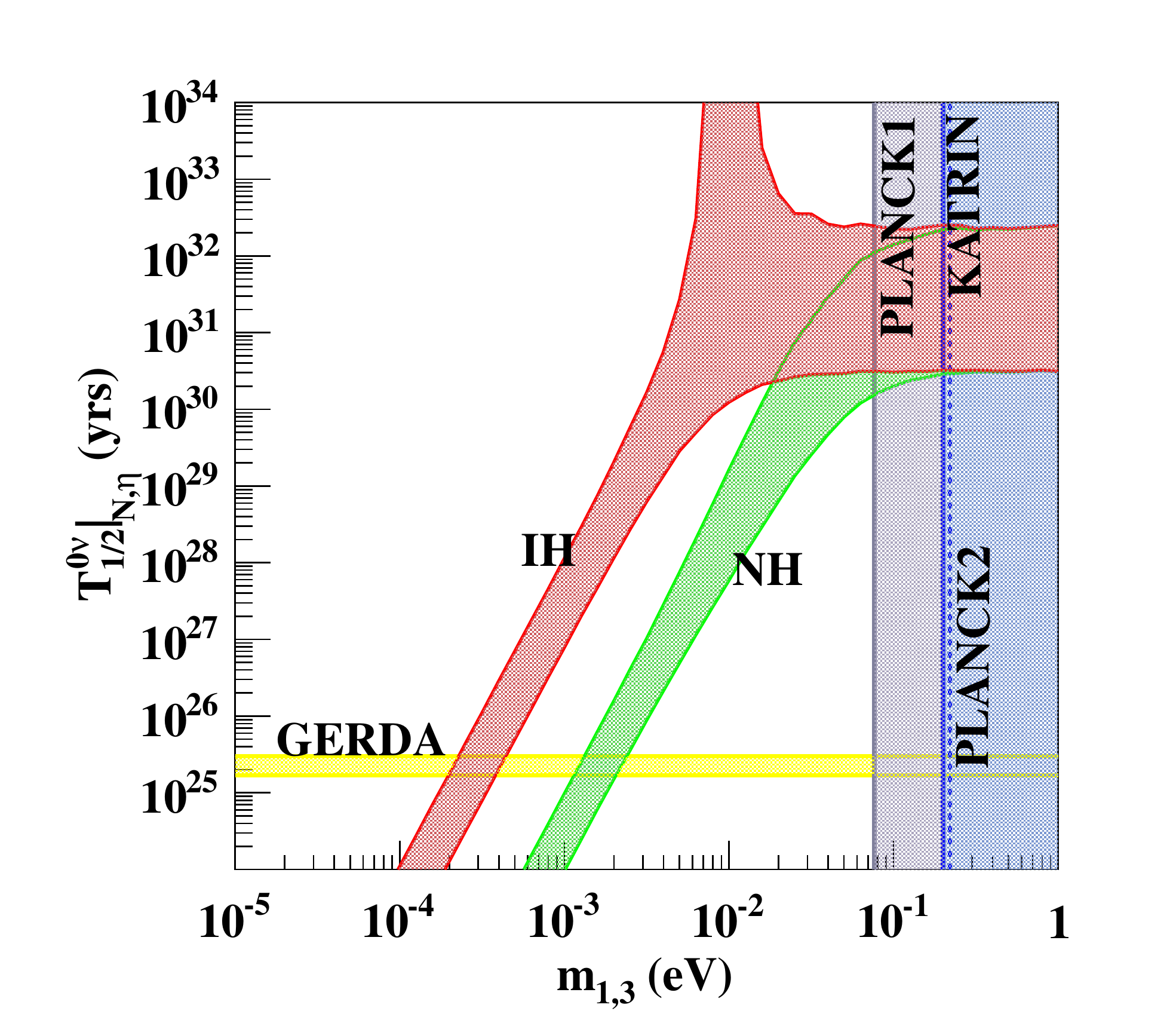}
\caption{Effective Majorana mass (left-panel) and half life (right-panel) as a function of the lightest neutrino mass, 
          $m_1$ ($m_3$) for NH (IH) due to $\eta$ diagram via exchange of heavy right-handed Majorana neutrino $N$.}
\label{NEta}
\end{figure}

\begin{figure}[h]
\centering
\includegraphics[width=0.49\textwidth]{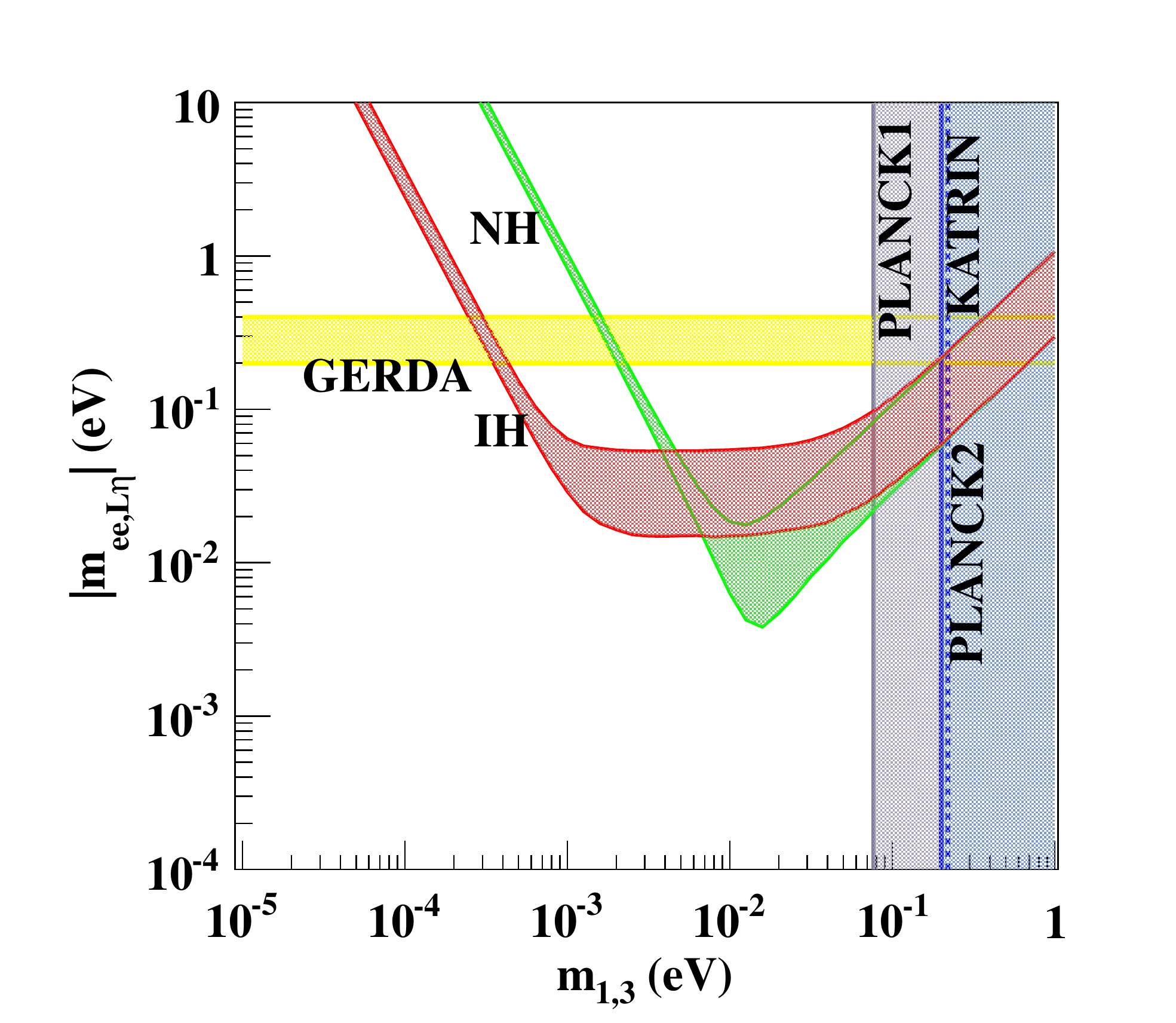}
\includegraphics[width=0.49\textwidth]{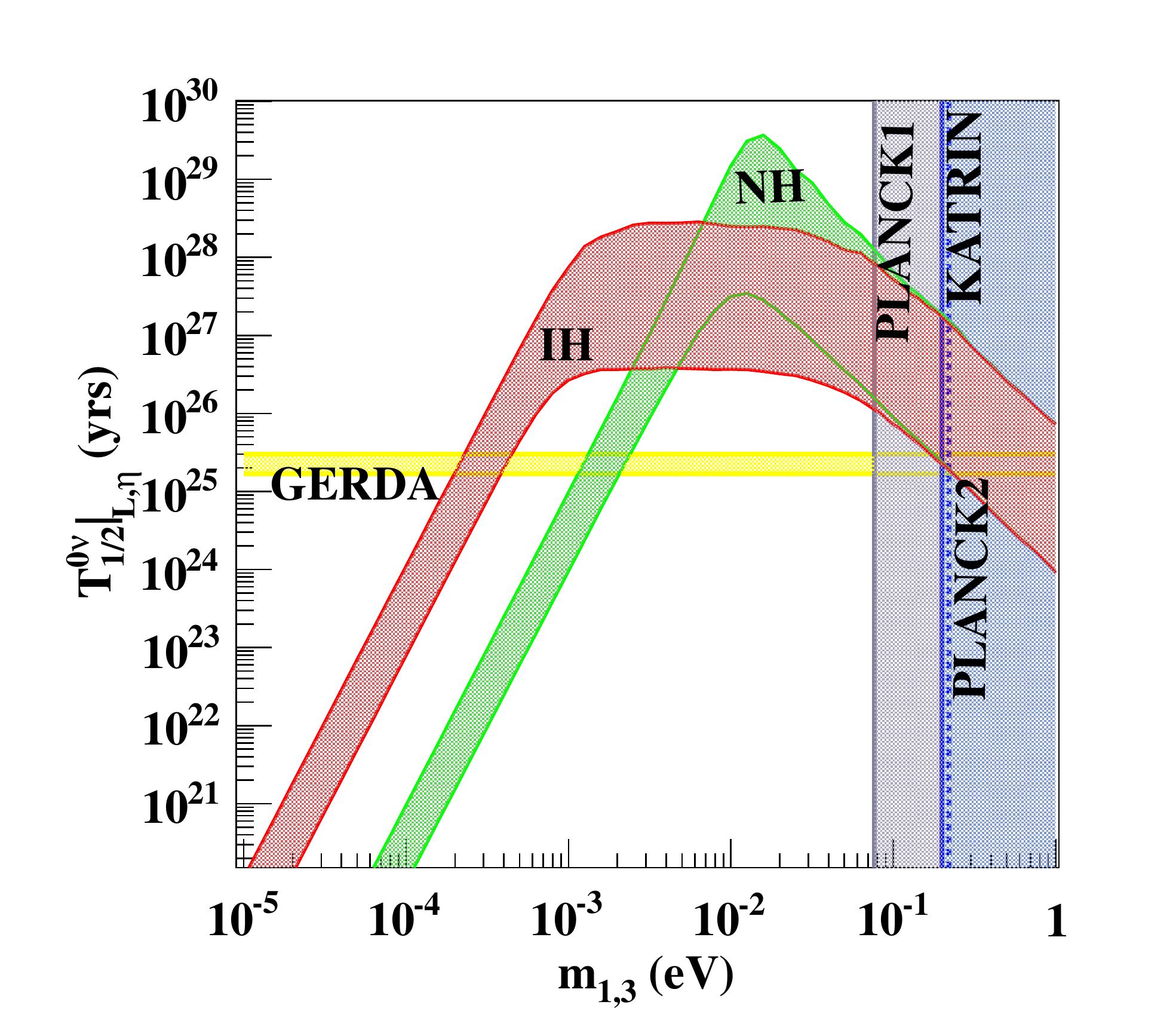}
\caption{Effective Majorana mass (left-panel) and half life (right-panel) as a function of the lightest neutrino mass, 
         $m_1$ ($m_3$) for NH (IH) due to combine effect of standard mechanism and $\eta$ diagram via exchange of 
          $N$ and $S$.}
\label{std_etaNS}
\end{figure}

With $M_D$ similar to up-quark mass matrix and other input model parameters, 
the effective mass and corresponding half-life with the variation of lightest neutrino 
mass $m_1 (NH)$ and $m_3 (IH)$ due to so called $\lambda$ and $\eta$ diagrams to $0\nu\beta\beta$ 
transition are displayed in Fig.\ref{NLambda}, Fig.\ref{NEta} and Fig.\ref{std_etaNS}.

\subsection{Mass hierarchy discrimination within natural type-II seesaw dominance}
\label{sec:mass hierarchy}
\begin{figure}[h]
\centering
\includegraphics[width=0.49\textwidth]{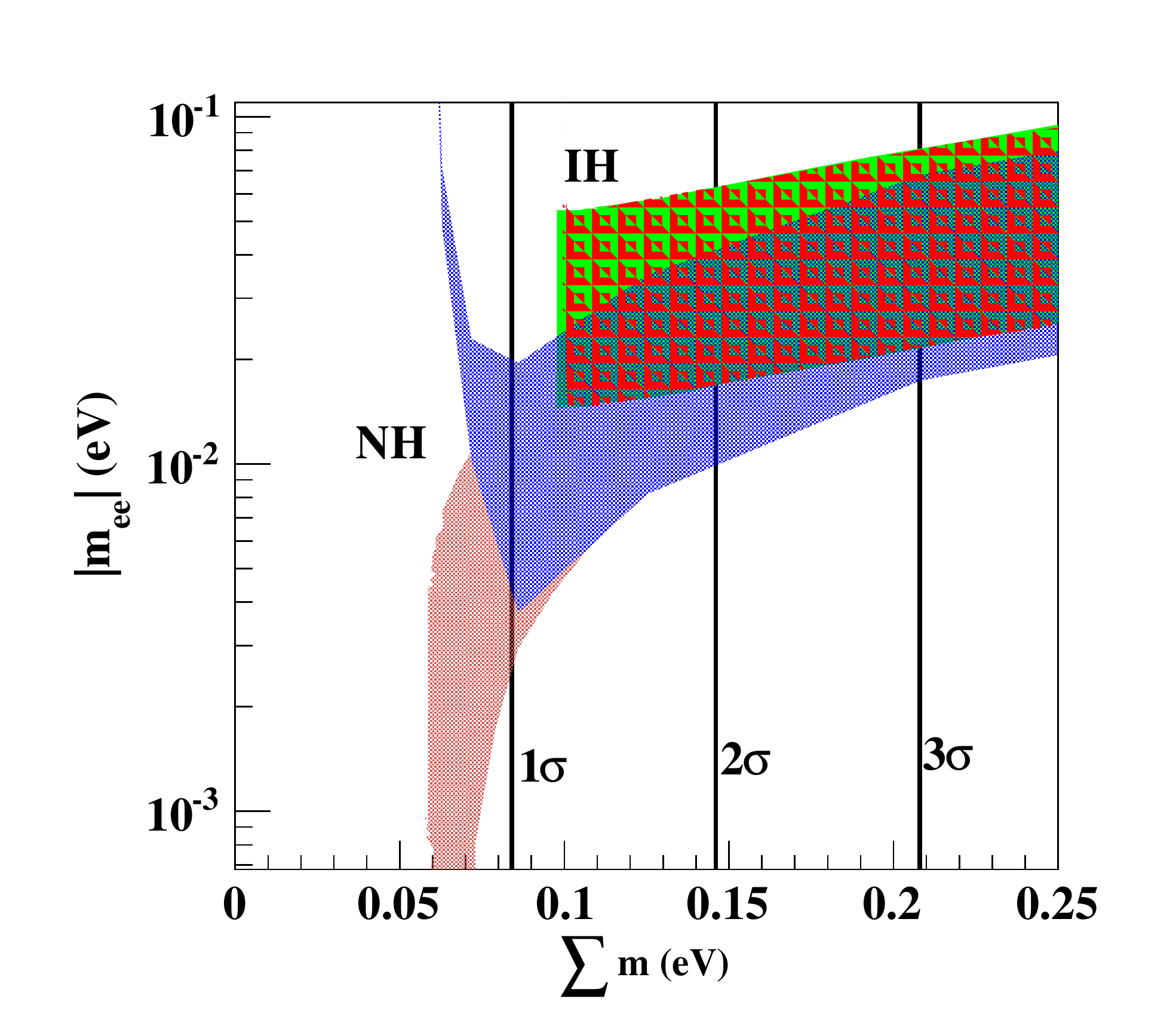}
\includegraphics[width=0.49\textwidth]{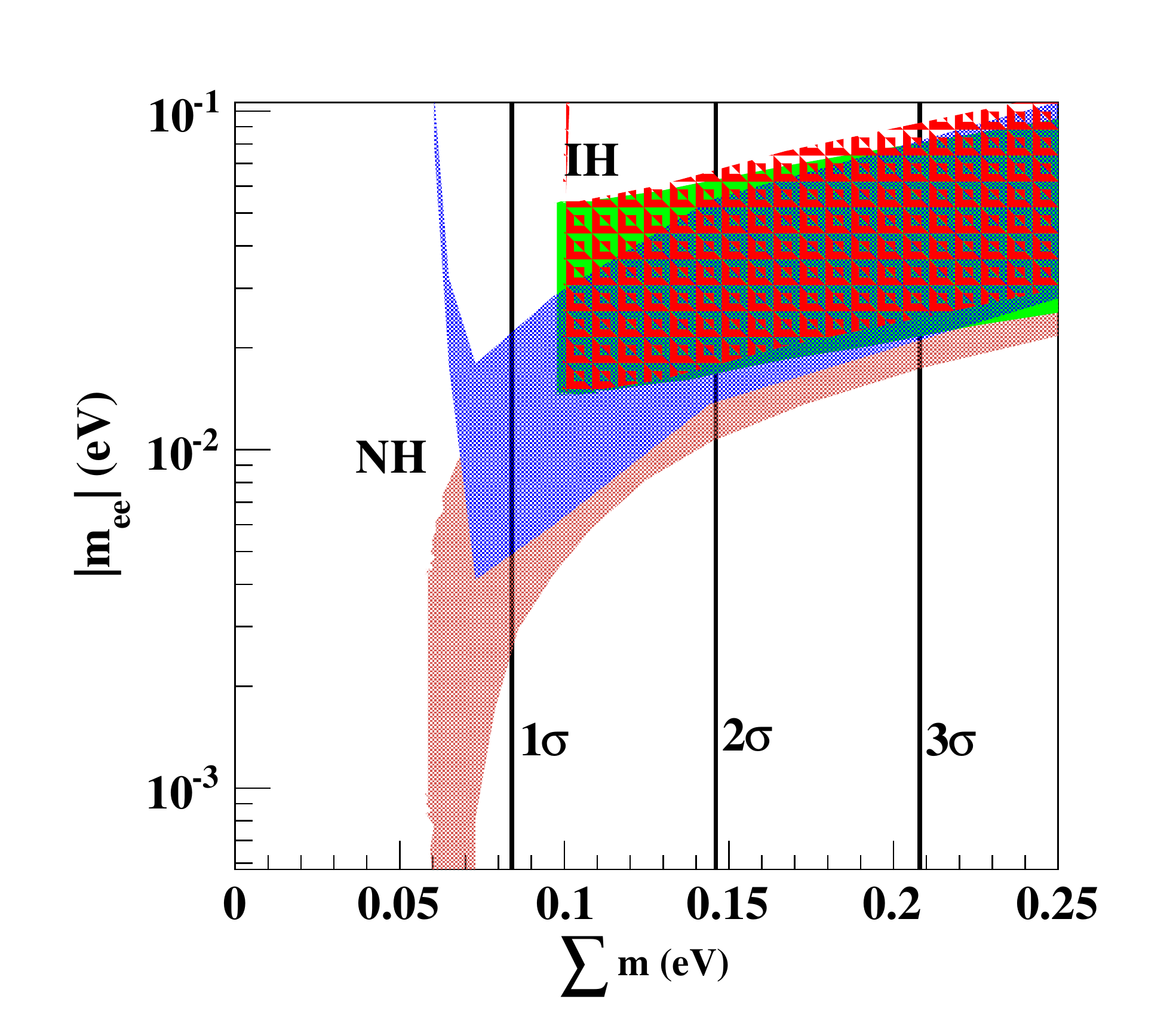}
\caption{Left Panel: Preferred region of effective mass parameter $|m_{ee}|$ for standard mechanism and its 
addition to $W_L-W_L$ mediation with the exchange of heavy $N_R$ and sterile neutrino as a function 
of sum of light neutrino masses ($m_\Sigma$). Right Panel: Allowed region of $|m_{ee}|$ for standard 
mechanism and its addition to $W_R-W_R$ mediation with the exchange of heavy $N_R$ as a function of $m_\Sigma$.}
\label{sumLLR}
\end{figure}

\begin{figure}[h]
\centering
\includegraphics[width=0.49\textwidth]{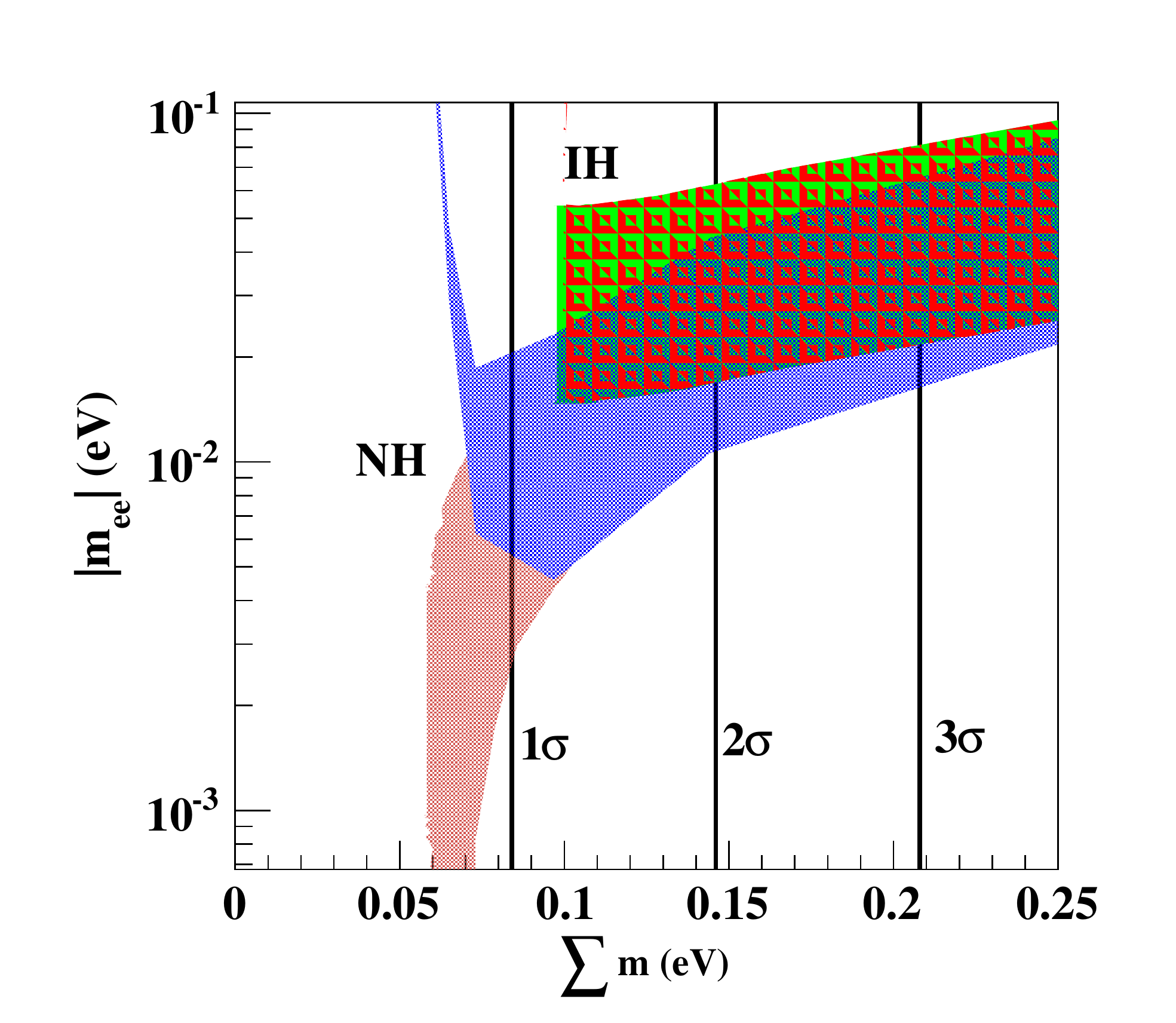}
\caption{Allowed region of $|m_{ee}|$ for standard mechanism and its combined effect to $\lambda$ 
diagram due to exchange of heavy RH neutrino and sterile neutrino as a function of $m_\Sigma$.}
\label{sumLEta}
\end{figure}

Here we discuss the comparison between the standard mechanism and the new physics contributions to $0\nu\beta\beta$ transition 
within the present framework with natural type-II seesaw dominance by plotting effective Majorana mass as a function of sum 
of light neutrino masses ($m_{\Sigma}$) using the cosmological limit on the light neutrino mass sum. The light neutrino 
mass sum is defined as $m_\Sigma=\sum_i m_i=m_1+m_2+m_3$. The lower limits of light neutrino mass sum $m_\Sigma$ 
derived from cosmology as well as measurements from ongoing neutrino less double beta decay experiments at $1\sigma$, $2\sigma$ 
and $3\sigma$ C.L are given by~\cite{Seljak:2004xh,Costanzi:2014tna,Palanque-Delabrouille:2014jca}
\begin{eqnarray}
&&m_{\Sigma} < \mbox{84\, meV} ~\quad \mbox{at $1\sigma$ C.L.}\, , \nonumber \\
&&m_{\Sigma} < \mbox{146\, meV} ~\quad \mbox{at $1\sigma$ C.L.}\, , \nonumber \\
&&m_{\Sigma} < \mbox{208\, meV} ~\quad \mbox{at $1\sigma$ C.L.}\, 
\end{eqnarray}
We plot effective Majorana mass $m_{ee}$ as a function of sum of light neutrino masses ($m_{\Sigma}$)
displayed in Fig.\ref{sumLLR} and Fig.\ref{sumLEta} where standard mechanism is represented by the red band 
for NH and by the green band for IH while the new physics contributions are represented by the
blue band for NH and the red hatched band for IH.

As the contribution of heavy right-handed neutrino and sterile neutrino to effective mass parameter saturate the 
experimental GERDA limit, Fig.\ref{sumLLR} (left-panel) represents their combined effect to 
standard mechanism. The spectrum for IH due to the SM and others are lying within the region 
of cosmological bound and hence disfavoured at $1\sigma$ C.L. Whereas the NH spectrum are lying 
in the privileged region and favored for lower mass of lightest neutrino.
But in case of purely right-handed currents i.e mediation via $W_R-W_R$, the contribution on effective Majorana 
mass parameters due to the exchange of $\nu_L$ and heavy sterile neutrinos are negligible. So the right-panel 
of Fig.\ref{sumLLR} shows the effect of heavy $N_R$ with standard mechanism on sum of light neutrino masses. 
Here also NH is favored over IH both for standard mechanism and new physics. Similarly, Fig.\ref{sumLEta} 
indicates the effect of $\lambda$ diagram due to exchange of heavy $N_R$ and S with standard mechanism on 
effective mass having same characteristic as in Fig.\ref{sumLLR}.  

For all of them, the uncertainty on the effective Majorana mass parameter increases with increase in 
sum of masses in case of normal hierarchy for the contribution of new physics to standard mechanism.

\newpage
\section{\texorpdfstring{Comparison of half-lives for $\obb$ in $^{76}$Ge and $^{136}$Xe}
{Correlation between half-lives for 0nubb in 76Ge and 136Xe}} \label{sect:halflife_corrs}
\begin{figure}[htb!]
\centering
\includegraphics[width=0.49\textwidth]{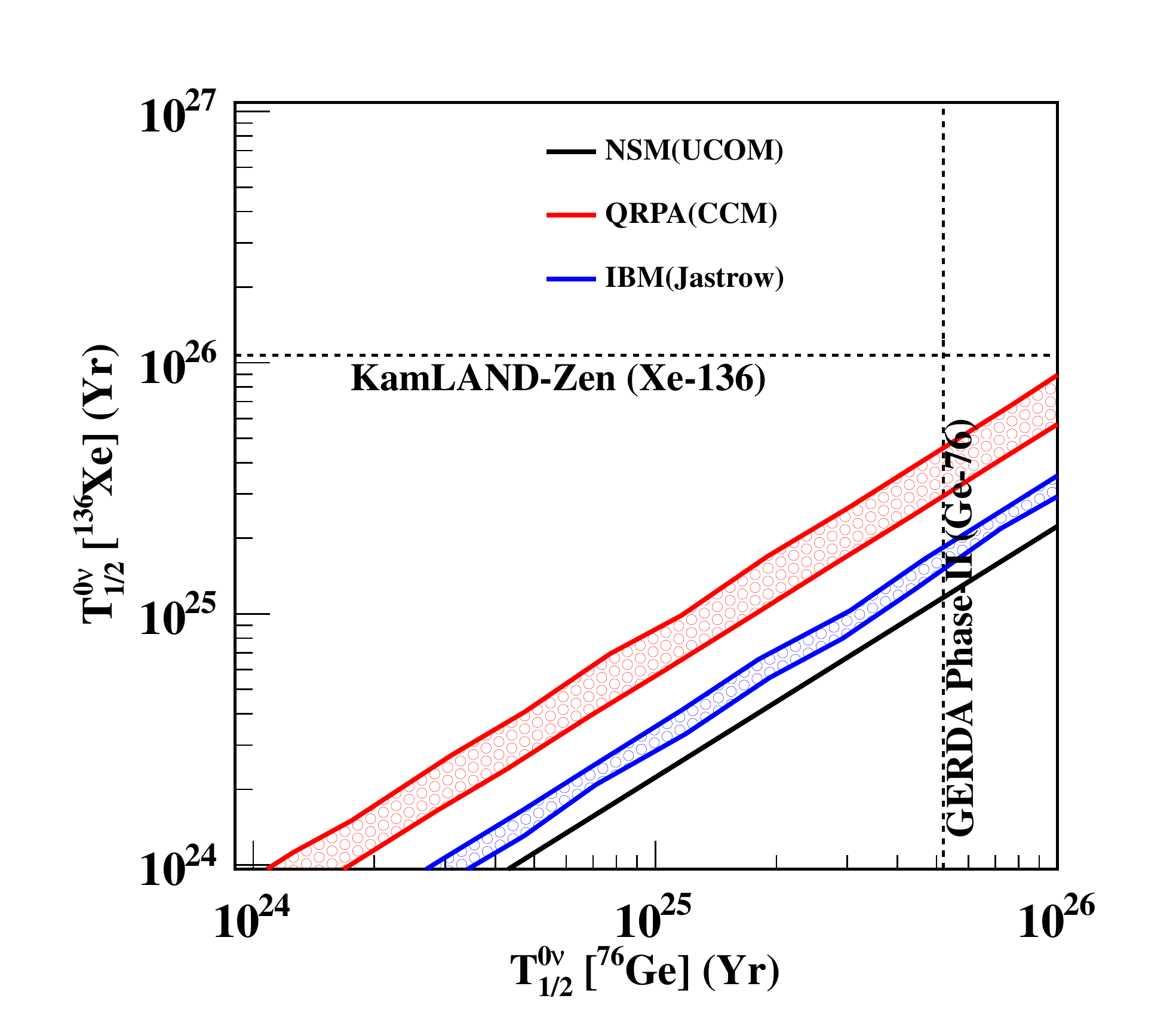}
\includegraphics[width=0.49\textwidth]{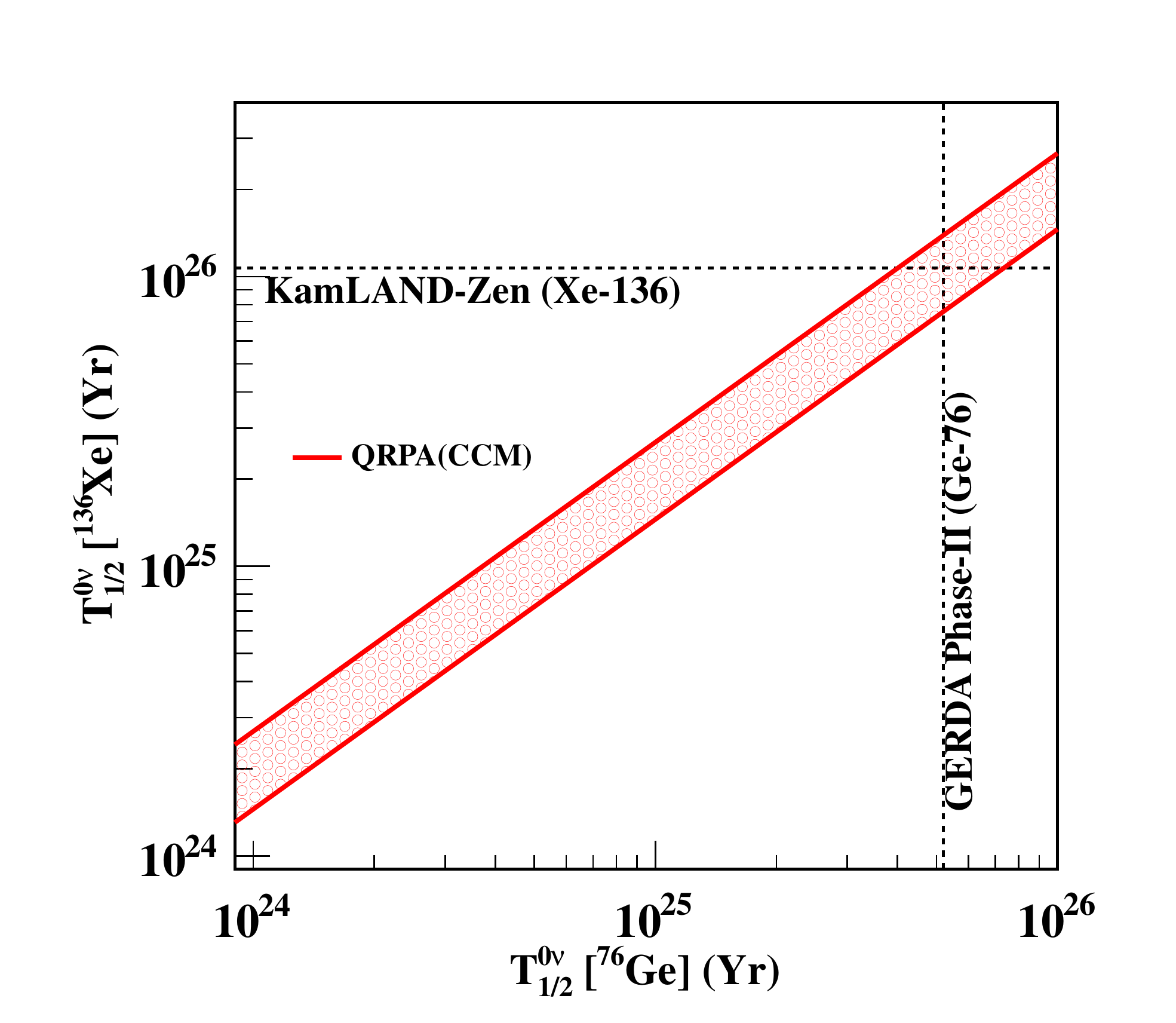}
\caption{The left-panel shows the correlations between the $\obb$ half-lives in $^{136}$Xe and 
         $^{76}$Ge for different matrix element calculations and particle physics contribution due 
         to light neutrino exchange while the right-panel is for heavy neutrino exchange.}
\label{fig:halflife_corrsa}
\end{figure}

\begin{figure}[h!]
\centering
\includegraphics[width=0.49\textwidth]{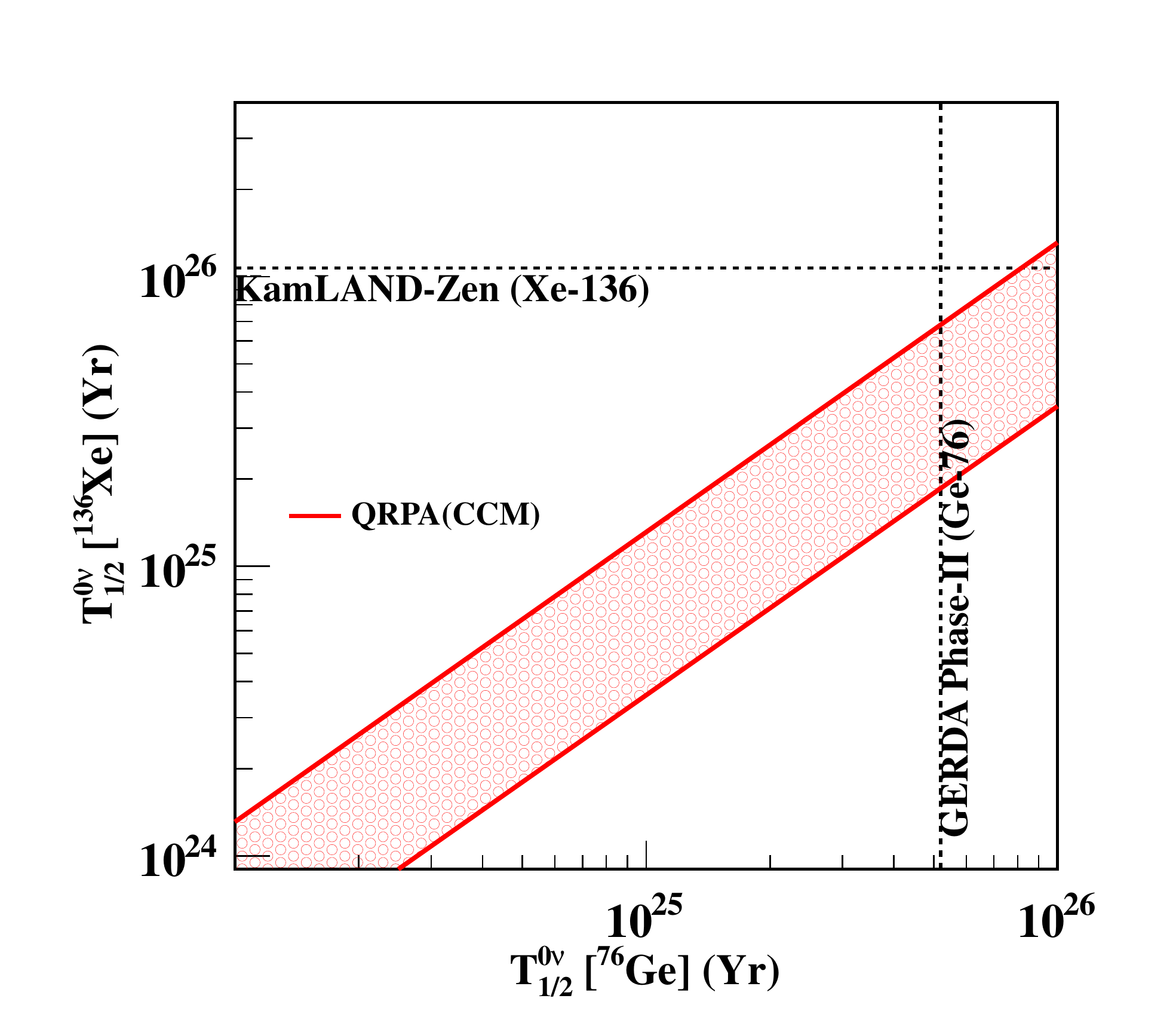}
\includegraphics[width=0.49\textwidth]{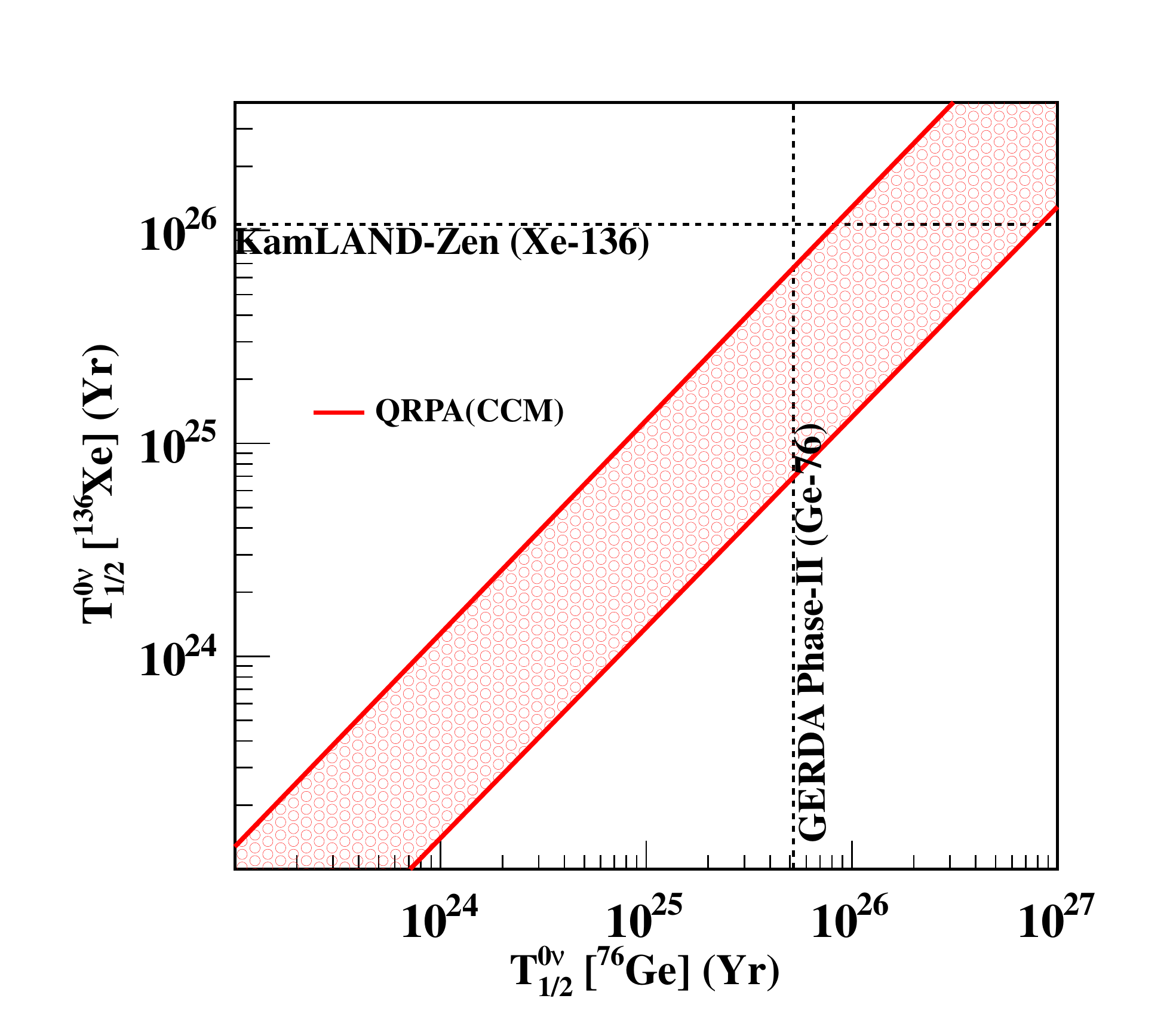}
\caption{The left-panel shows the correlations between the $\obb$ half-lives in $^{136}$Xe and 
         $^{76}$Ge for different matrix element calculations and particle physics contribution due 
         to $\lambda$-diagram while the right-panel is for $\eta-$diagram.}
\label{fig:halflife_corrsb}
\end{figure}

We intend to make here a comparative study of half-lives for neutrinoless double beta decay in
$^{76}$Ge and $^{136}$Xe indicating uncertainties in the nuclear matrix elements (one may refer 
~\cite{Gando:2012zm} for the matrix element calculations). The half-life limits for different experiments 
are $T_{1/2}^{0 \nu} \simeq 1.07 \times 10^{26}\text{yrs}$ (for KamLAND-Zen expt. using $^{136}$Xe) and 
$T_{1/2}^{0 \nu} \simeq 5.52 \times 10^{25}\text{yrs}$ (for GERDA Phase-II using $^{76}$Ge). Using the 
values given in Tables~\ref{table:matrix_elements_2}, \ref{table:matrix_elements_3} and \ref{table:matrix_elements_4} 
for nuclear matrix elements for light and heavy neutrino exchange as well as for $\lambda$- and $\eta$-diagrams 
we have shown the correlation plots between half-lives for $^{76}$Ge and $^{136}$Xe in Fig.~\ref{fig:halflife_corrsa} 
and Fig.~\ref{fig:halflife_corrsb} (these were first introduced in ref.\cite{Barry:2013xxa}). 
The band in each plots shows the measure of uncertainties in different 
nuclear matrix elements while measuring half-life in $^{136}$Xe to one measured in $^{76}$Ge.

\begin{table}[htp]
 \centering
\caption{Values of nuclear matrix elements for light neutrino exchange 
         (${\cal M}^{0\nu}_\nu$) for $^{76}$Ge and $^{136}$Xe.}
\label{table:matrix_elements_2}
\vspace{10pt}
 \begin{tabular}{lccc}
 \hline \hline
 \T Isotope & NSM (UCOM)~\cite{Menendez:2008jp} & QRPA (CCM)~\cite{Simkovic:2009pp} & IBM (Jastrow)~\cite{Barea:2009zza} \\
\hline \T
$^{76}$Ge & 2.58 &4.07--6.64 & 4.25--5.07 \\
$^{136}$Xe & 2.00 & 1.57--3.24 & 3.07 \\
\hline \hline
 \end{tabular}
\end{table}

\begin{table}[htp]
 \centering
 \caption{Values of nuclear matrix elements for heavy neutrino exchange 
         (${\cal M}^{0\nu}_\nu$) for $^{76}$Ge and $^{136}$Xe.}
\label{table:matrix_elements_3}
\vspace{10pt}
 \begin{tabular}{lcc}
 \hline \hline
 \T Isotope & IBM (M-S)~\cite{Barea:2013bz} & QRPA (CCM)~\cite{Faessler:2011rv} \\
\hline \T
$^{76}$Ge & 48.1 & 233--412 \\
$^{136}$Xe & 35.1 & 164--172 \\
\hline \hline
 \end{tabular}
\end{table}

\begin{table}[htp]
 \centering
 \caption{Nuclear matrix elements for the $\lambda$- and $\eta$-diagrams with exchange of light neutrinos. 
         However it should be noted that there are no nuclear matrix elements for lambda and eta diagrams 
         through exchange of heavy neutrinos.}
\label{table:matrix_elements_4}
\vspace{10pt}
 \begin{tabular}{lcc|cc}
 \hline \hline
 \T \multirow{2}{*}{Isotope} & \multicolumn{2}{c|}{${\cal M}^{0\nu}_\lambda$} & \multicolumn{2}{c}{${\cal M}^{0\nu}_\eta$} \\
 & QRPA (CCM)~\cite{Pantis:1996py} & QRPA (HD)~\cite{Deppisch:2012nb} & QRPA (CCM)~\cite{Pantis:1996py} & QRPA (HD)~\cite{Deppisch:2012nb} \\[1mm]
\hline \T
$^{76}$Ge & 1.75--3.76 & 4.47 & 235--637 & 791\\
$^{136}$Xe & 1.96--2.49 & 2.17 & 370--419 & 434\\ 
\hline \hline
 \end{tabular}
\end{table}

\newpage
\section{Conclusion}
\label{sec:conclusion}
We have discussed natural realization of type-II seesaw dominance within a 
class of TeV scale left-right symmetric models where scalar sector comprises of 
scalar doublets $H_{L,R}$, triplets $\Delta_{L,R}$ and a bidoublet $\Phi$, 
the fermion sector consists of usual quarks $q_{L,R}$, leptons $\ell_{L,R}$ 
plus one copy of extra sterile fermion $S_L$ per generation. In order to achieve natural 
type-II seesaw dominance, we have considered negligible VEV for LH scalar doublet i.e, 
$\langle H_L \rangle \to 0$, negligible mass term for extra sterile neutrinos $\mu_S \to 0$ 
and mass hierarchy as $M_R > M > M_D \gg M_L$ where $M_L (M_R)$ is the Majorana mass term for LH (RH) neutrinos, 
$M_D$ is the Dirac mass term connecting light-heavy neutrino and $M$ is the $N-S$ mixing matrix. 
We have also discussed that the type-II seesaw dominance allows any value for $M_D$ and thus, 
new physics contributions to $0\nu\beta\beta$ transition arise from the following channels; 
i) due to purely left-handed currents via exchange of heavy RH Majorana neutrinos $N$ and extra 
sterile neutrinos $S$, and ii) due to so called $\lambda$ and $\eta$ type of diagrams.
We have also demonstrated the effect of right-handed currents to $0\nu\beta\beta$ transition 
via $W_R-W_R$ mediation. 

Most importantly we have expressed all the physical masses and mixing like $\nu_L$, $N_R$ and 
$S_L$ which are completely Majorana in nature mediating LNV processes like neutrinoless 
double beta decay in terms of oscillation parameters and mass of lightest 
neutrino with the assumption that $N-S$ mixing matrix is diagonal and degenerate. We have demonstrated that 
large value of Dirac neutrino mass possibly originating from high scale Pati-Salam symmetry or $SO(10)$ GUT 
plays an important role in resulting dominant contributions to these new non-standard $0\nu\beta\beta$ transition. 
With the model parameters like $M_{W_R}=3~$TeV, $M_N \simeq {\cal O}(\mbox{TeV})$, $M_\Delta \simeq {\cal O}(\mbox{TeV})$, 
$M_D$ as up-type quark mass matrix and using oscillation parameters, we numerically estimated new physics contributions 
to $0\nu\beta\beta$ transition and compared it with that of the standard mechanism. We have derived the lower limit 
on absolute scale of lightest neutrino mass by numerically estimating various new physics contributions to 
$0\nu\beta\beta$ transition by saturating the current experimental limit. We have shown that NH is favored over 
IH pattern of light neutrinos resulting from effective mass as a function of light neutrino mass sum taking into 
account the cosmological data. We have also presented correlation plots between half-lives for $^{76}$Ge and $^{136}$Xe 
isotopes showing the uncertainties in the nuclear matrix elements. 

\section*{Acknowledgements}
\label{sec:ackn}
The authors are thankful to Shao-Feng Ge for collaboration in the early stages of this project. 
PP is supported by the DST INSPIRE Fellowship (No. DST/INSPIRE Fellowship/2014) funded by the Department 
of Science and Technology, India. PP is also thankful to ICHEP Organising Committee for selecting the abstract 
that inspired this work for a poster presentation at the 38th International Conference on High Energy 
Physics to be held on $3-10$ August 2016 at Chicago, USA. 

\noindent \\
\textbf{Note added:}
During the finalization of this work, another work appeared on arXiv~\cite{Parida:2016asc} that discusses 
type-II seesaw dominance in LRSM. However, our work differs widely by expressing all the physical masses and 
mixing of heavy neutrinos in terms of oscillation parameters and lightest neutrino mass. Thus, one can get an 
important information about absolute scale of lightest neutrino mass and mass hierarchy from new physics 
contributions to neutrinoless double beta decay by saturating the current experimental limit.

\bibliographystyle{utcaps_mod}
\bibliography{onubb_LR}

\providecommand{\href}[2]{#2}\begingroup\raggedright\begin{thebibliography}{10}

\bibitem{Majorana:1937vz}
E.~Majorana, ``{\em {Theory of the Symmetry of Electrons and Positrons}},''
\href{http://dx.doi.org/10.1007/BF02961314}{Nuovo Cim. {\normalfont \bfseries
  14} (1937)  171--184}.

\bibitem{Mohapatra:1979ia}
R.~N. Mohapatra and G.~Senjanovi{\'c}, ``{\em {Neutrino Mass and Spontaneous
  Parity Violation}},''
\href{http://dx.doi.org/10.1103/PhysRevLett.44.912}{Phys.Rev.Lett. {\normalfont
  \bfseries 44} (1980)  912}.

\bibitem{Minkowski:1977sc}
P.~Minkowski, ``{\em {$\mu \to e\gamma$ at a Rate of One Out of $10^{9}$ Muon
  Decays?}},''
\href{http://dx.doi.org/10.1016/0370-2693(77)90435-X}{Phys. Lett. {\normalfont
  \bfseries B67} (1977)  421--428}.

\bibitem{Yanagida:1979as}
T.~Yanagida, ``{\em {HORIZONTAL SYMMETRY AND MASSES OF NEUTRINOS}},''
Conf. Proc. {\normalfont \bfseries C7902131} (1979)  95--99.

\bibitem{GellMann:1980vs}
M.~Gell-Mann, P.~Ramond, and R.~Slansky, ``{\em {Complex Spinors and Unified
  Theories}},'' Conf. Proc. {\normalfont \bfseries C790927} (1979)  315--321,
\href{http://arxiv.org/abs/1306.4669}{{\normalfont \ttfamily arXiv:1306.4669}}.

\bibitem{Cheng:1980qt}
T.~P. Cheng and L.-F. Li, ``{\em {Neutrino Masses, Mixings and Oscillations in
  SU(2) x U(1) Models of Electroweak Interactions}},''
\href{http://dx.doi.org/10.1103/PhysRevD.22.2860}{Phys. Rev. {\normalfont
  \bfseries D22} (1980)  2860}.

\bibitem{Lazarides:1980nt}
G.~Lazarides, Q.~Shafi, and C.~Wetterich, ``{\em {Proton Lifetime and Fermion
  Masses in an SO(10) Model}},''
\href{http://dx.doi.org/10.1016/0550-3213(81)90354-0}{Nucl. Phys. {\normalfont
  \bfseries B181} (1981)  287--300}.

\bibitem{Magg:1980ut}
M.~Magg and C.~Wetterich, ``{\em {Neutrino Mass Problem and Gauge
  Hierarchy}},''
\href{http://dx.doi.org/10.1016/0370-2693(80)90825-4}{Phys. Lett. {\normalfont
  \bfseries B94} (1980)  61--64}.

\bibitem{Schechter:1980gr}
J.~Schechter and J.~W.~F. Valle, ``{\em {Neutrino Masses in SU(2) x U(1)
  Theories}},''
\href{http://dx.doi.org/10.1103/PhysRevD.22.2227}{Phys. Rev. {\normalfont
  \bfseries D22} (1980)  2227}.

\bibitem{Wetterich:1981bx}
C.~Wetterich, ``{\em {Neutrino Masses and the Scale of B-L Violation}},''
\href{http://dx.doi.org/10.1016/0550-3213(81)90279-0}{Nucl. Phys. {\normalfont
  \bfseries B187} (1981)  343--375}.

\bibitem{Schechter:1981bd}
J.~Schechter and J.~W.~F. Valle, ``{\em {Neutrinoless Double beta Decay in
  SU(2) x U(1) Theories}},''
\href{http://dx.doi.org/10.1103/PhysRevD.25.2951}{Phys. Rev. {\normalfont
  \bfseries D25} (1982)  2951}.

\bibitem{Agostini:2013mzu}
{\normalfont \bfseries GERDA}, M.~Agostini {\em et al.}, ``{\em {Results on
  Neutrinoless Double-$\beta$ Decay of $^{76}$Ge from Phase I of the GERDA
  Experiment}},'' \href{http://dx.doi.org/10.1103/PhysRevLett.111.122503}{Phys.
  Rev. Lett. {\normalfont \bfseries 111} (2013) no.~12, 122503},
\href{http://arxiv.org/abs/1307.4720}{{\normalfont \ttfamily arXiv:1307.4720}}.

\bibitem{Gando:2012zm}
{\normalfont \bfseries KamLAND-Zen}, A.~Gando {\em et al.}, ``{\em {Limit on
  Neutrinoless $\beta\beta$ Decay of $^{136}$Xe from the First Phase of
  KamLAND-Zen and Comparison with the Positive Claim in $^{76}$Ge}},''
  \href{http://dx.doi.org/10.1103/PhysRevLett.110.062502}{Phys. Rev. Lett.
  {\normalfont \bfseries 110} (2013) no.~6, 062502},
\href{http://arxiv.org/abs/1211.3863}{{\normalfont \ttfamily arXiv:1211.3863}}.

\bibitem{Auger:2012ar}
{\normalfont \bfseries EXO}, M.~Auger {\em et al.}, ``{\em {Search for
  Neutrinoless Double-Beta Decay in $^{136}$Xe with EXO-200}},''
  \href{http://dx.doi.org/10.1103/PhysRevLett.109.032505}{Phys. Rev. Lett.
  {\normalfont \bfseries 109} (2012)  032505},
\href{http://arxiv.org/abs/1205.5608}{{\normalfont \ttfamily arXiv:1205.5608}}.

\bibitem{Mohapatra:1986su}
R.~N. Mohapatra, ``{\em {New Contributions to Neutrinoless Double beta Decay in
  Supersymmetric Theories}},''
\href{http://dx.doi.org/10.1103/PhysRevD.34.3457}{Phys. Rev. {\normalfont
  \bfseries D34} (1986)  3457--3461}.

\bibitem{Babu:1995vh}
K.~S. Babu and R.~N. Mohapatra, ``{\em {New vector - scalar contributions to
  neutrinoless double beta decay and constraints on R-parity violation}},''
  \href{http://dx.doi.org/10.1103/PhysRevLett.75.2276}{Phys. Rev. Lett.
  {\normalfont \bfseries 75} (1995)  2276--2279},
\href{http://arxiv.org/abs/hep-ph/9506354}{{\normalfont \ttfamily
  arXiv:hep-ph/9506354}}.

\bibitem{Hirsch:1995vr}
M.~Hirsch, H.~V. Klapdor-Kleingrothaus, and S.~G. Kovalenko, ``{\em {New
  supersymmetric contributions to neutrinoless double beta decay}},''
  \href{http://dx.doi.org/10.1016/0370-2693(95)00460-3}{Phys. Lett.
  {\normalfont \bfseries B352} (1995)  1--7},
\href{http://arxiv.org/abs/hep-ph/9502315}{{\normalfont \ttfamily
  arXiv:hep-ph/9502315}}.

\bibitem{Hirsch:1995ek}
M.~Hirsch, H.~V. Klapdor-Kleingrothaus, and S.~G. Kovalenko, ``{\em
  {Supersymmetry and neutrinoless double beta decay}},''
  \href{http://dx.doi.org/10.1103/PhysRevD.53.1329}{Phys. Rev. {\normalfont
  \bfseries D53} (1996)  1329--1348},
\href{http://arxiv.org/abs/hep-ph/9502385}{{\normalfont \ttfamily
  arXiv:hep-ph/9502385}}.

\bibitem{Hirsch:1996ye}
M.~Hirsch, H.~V. Klapdor-Kleingrothaus, and S.~G. Kovalenko, ``{\em {New
  leptoquark mechanism of neutrinoless double beta decay}},''
  \href{http://dx.doi.org/10.1103/PhysRevD.54.R4207}{Phys. Rev. {\normalfont
  \bfseries D54} (1996)  4207--4210},
\href{http://arxiv.org/abs/hep-ph/9603213}{{\normalfont \ttfamily
  arXiv:hep-ph/9603213}}.

\bibitem{Deppisch:2012nb}
F.~F. Deppisch, M.~Hirsch, and H.~Pas, ``{\em {Neutrinoless Double Beta Decay
  and Physics Beyond the Standard Model}},''
  \href{http://dx.doi.org/10.1088/0954-3899/39/12/124007}{J. Phys. {\normalfont
  \bfseries G39} (2012)  124007},
\href{http://arxiv.org/abs/1208.0727}{{\normalfont \ttfamily arXiv:1208.0727}}.

\bibitem{Pas:1999fc}
H.~Pas, M.~Hirsch, H.~V. Klapdor-Kleingrothaus, and S.~G. Kovalenko, ``{\em
  {Towards a superformula for neutrinoless double beta decay}},''
\href{http://dx.doi.org/10.1016/S0370-2693(99)00330-5}{Phys. Lett. {\normalfont
  \bfseries B453} (1999)  194--198}.

\bibitem{Humbert:2015yva}
P.~Humbert, M.~Lindner, S.~Patra, and J.~Smirnov, ``{\em {Lepton Number
  Violation within the Conformal Inverse Seesaw}},''
  \href{http://dx.doi.org/10.1007/JHEP09(2015)064}{JHEP {\normalfont \bfseries
  09} (2015)  064},
\href{http://arxiv.org/abs/1505.07453}{{\normalfont \ttfamily
  arXiv:1505.07453}}.

\bibitem{Allanach:2009xx}
B.~C. Allanach, C.~H. Kom, and H.~Pas, ``{\em {LHC and B physics probes of
  neutrinoless double beta decay in supersymmetry without R-parity}},''
  \href{http://dx.doi.org/10.1088/1126-6708/2009/10/026}{JHEP {\normalfont
  \bfseries 10} (2009)  026},
\href{http://arxiv.org/abs/0903.0347}{{\normalfont \ttfamily arXiv:0903.0347}}.

\bibitem{Pas:2000vn}
H.~Pas, M.~Hirsch, H.~V. Klapdor-Kleingrothaus, and S.~G. Kovalenko, ``{\em {A
  Superformula for neutrinoless double beta decay. 2. The Short range part}},''
  \href{http://dx.doi.org/10.1016/S0370-2693(00)01359-9}{Phys. Lett.
  {\normalfont \bfseries B498} (2001)  35--39},
\href{http://arxiv.org/abs/hep-ph/0008182}{{\normalfont \ttfamily
  arXiv:hep-ph/0008182}}.

\bibitem{Deppisch:2006hb}
F.~Deppisch and H.~Pas, ``{\em {Pinning down the mechanism of neutrinoless
  double beta decay with measurements in different nuclei}},''
  \href{http://dx.doi.org/10.1103/PhysRevLett.98.232501}{Phys. Rev. Lett.
  {\normalfont \bfseries 98} (2007)  232501},
\href{http://arxiv.org/abs/hep-ph/0612165}{{\normalfont \ttfamily
  arXiv:hep-ph/0612165}}.

\bibitem{Pas:2015eia}
H.~Päs and W.~Rodejohann, ``{\em {Neutrinoless Double Beta Decay}},''
  \href{http://dx.doi.org/10.1088/1367-2630/17/11/115010}{New J. Phys.
  {\normalfont \bfseries 17} (2015) no.~11, 115010},
\href{http://arxiv.org/abs/1507.00170}{{\normalfont \ttfamily
  arXiv:1507.00170}}.

\bibitem{Helo:2013dla}
J.~C. Helo, M.~Hirsch, S.~G. Kovalenko, and H.~Pas, ``{\em {Neutrinoless double
  beta decay and lepton number violation at the LHC}},''
  \href{http://dx.doi.org/10.1103/PhysRevD.88.011901}{Phys. Rev. {\normalfont
  \bfseries D88} (2013) no.~1, 011901},
\href{http://arxiv.org/abs/1303.0899}{{\normalfont \ttfamily arXiv:1303.0899}}.

\bibitem{Ge:2015bfa}
S.-F. Ge and W.~Rodejohann, ``{\em {JUNO and Neutrinoless Double Beta
  Decay}},''
\href{http://arxiv.org/abs/1507.05514}{{\normalfont \ttfamily
  arXiv:1507.05514}}.

\bibitem{Mohapatra:1974gc}
R.~Mohapatra and J.~C. Pati, ``{\em {A Natural Left-Right Symmetry}},''
\href{http://dx.doi.org/10.1103/PhysRevD.11.2558}{Phys.Rev. {\normalfont
  \bfseries D11} (1975)  2558}.

\bibitem{Pati:1974yy}
J.~C. Pati and A.~Salam, ``{\em {Lepton Number as the Fourth Color}},''
\href{http://dx.doi.org/10.1103/PhysRevD.10.275,
  10.1103/PhysRevD.11.703.2}{Phys.Rev. {\normalfont \bfseries D10} (1974)
  275--289}.

\bibitem{Senjanovic:1975rk}
G.~Senjanovi{\'c} and R.~N. Mohapatra, ``{\em {Exact Left-Right Symmetry and
  Spontaneous Violation of Parity}},''
\href{http://dx.doi.org/10.1103/PhysRevD.12.1502}{Phys.Rev. {\normalfont
  \bfseries D12} (1975)  1502}.

\bibitem{Senjanovic:1978ev}
G.~Senjanovi{\'c}, ``{\em {Spontaneous Breakdown of Parity in a Class of Gauge
  Theories}},''
\href{http://dx.doi.org/10.1016/0550-3213(79)90604-7}{Nucl.Phys. {\normalfont
  \bfseries B153} (1979)  334--364}.

\bibitem{Mohapatra:1980yp}
R.~N. Mohapatra and G.~Senjanovi{\'c}, ``{\em {Neutrino Masses and Mixings in
  Gauge Models with Spontaneous Parity Violation}},''
\href{http://dx.doi.org/10.1103/PhysRevD.23.165}{Phys.Rev. {\normalfont
  \bfseries D23} (1981)  165}.

\bibitem{Tello:2010am}
V.~Tello, M.~Nemevsek, F.~Nesti, G.~Senjanovic, and F.~Vissani, ``{\em
  {Left-Right Symmetry: from LHC to Neutrinoless Double Beta Decay}},''
  \href{http://dx.doi.org/10.1103/PhysRevLett.106.151801}{Phys. Rev. Lett.
  {\normalfont \bfseries 106} (2011)  151801},
\href{http://arxiv.org/abs/1011.3522}{{\normalfont \ttfamily arXiv:1011.3522}}.

\bibitem{Barry:2013xxa}
J.~Barry and W.~Rodejohann, ``{\em {Lepton number and flavour violation in
  TeV-scale left-right symmetric theories with large left-right mixing}},''
  \href{http://dx.doi.org/10.1007/JHEP09(2013)153}{JHEP {\normalfont \bfseries
  1309} (2013)  153},
\href{http://arxiv.org/abs/1303.6324}{{\normalfont \ttfamily arXiv:1303.6324}}.

\bibitem{Deppisch:2014zta}
F.~F. Deppisch, T.~E. Gonzalo, S.~Patra, N.~Sahu, and U.~Sarkar, ``{\em {Double
  beta decay, lepton flavor violation, and collider signatures of left-right
  symmetric models with spontaneous $D$-parity breaking}},''
  \href{http://dx.doi.org/10.1103/PhysRevD.91.015018}{Phys. Rev. {\normalfont
  \bfseries D91} (2015) no.~1, 015018},
\href{http://arxiv.org/abs/1410.6427}{{\normalfont \ttfamily arXiv:1410.6427}}.

\bibitem{Deppisch:2014qpa}
F.~F. Deppisch, T.~E. Gonzalo, S.~Patra, N.~Sahu, and U.~Sarkar, ``{\em {Signal
  of Right-Handed Charged Gauge Bosons at the LHC?}},''
  \href{http://dx.doi.org/10.1103/PhysRevD.90.053014}{Phys. Rev. {\normalfont
  \bfseries D90} (2014) no.~5, 053014},
\href{http://arxiv.org/abs/1407.5384}{{\normalfont \ttfamily arXiv:1407.5384}}.

\bibitem{Awasthi:2013ff}
R.~L. Awasthi, M.~Parida, and S.~Patra, ``{\em {Neutrino masses, dominant
  neutrinoless double beta decay, and observable lepton flavor violation in
  left-right models and SO(10) grand unification with low mass $ W_R, Z_R$
  bosons}},'' \href{http://dx.doi.org/10.1007/JHEP08(2013)122}{JHEP
  {\normalfont \bfseries 1308} (2013)  122},
\href{http://arxiv.org/abs/1302.0672}{{\normalfont \ttfamily arXiv:1302.0672}}.

\bibitem{Patra:2014goa}
S.~Patra and P.~Pritimita, ``{\em {Post-sphaleron baryogenesis and $n$ -
  $\overline{n}$ oscillation in non-SUSY SO(10) GUT with gauge coupling
  unification and proton decay}},''
  \href{http://dx.doi.org/10.1140/epjc/s10052-014-3078-x}{Eur. Phys. J.
  {\normalfont \bfseries C74} (2014) no.~10, 3078},
\href{http://arxiv.org/abs/1405.6836}{{\normalfont \ttfamily arXiv:1405.6836}}.

\bibitem{Borah:2013lva}
D.~Borah, S.~Patra, and P.~Pritimita, ``{\em {Sub-dominant type-II seesaw as an
  origin of non-zero $\theta\_{13}$ in SO(10) model with TeV scale Z' gauge
  boson}},'' \href{http://dx.doi.org/10.1016/j.nuclphysb.2014.02.017}{Nucl.
  Phys. {\normalfont \bfseries B881} (2014)  444--466},
\href{http://arxiv.org/abs/1312.5885}{{\normalfont \ttfamily arXiv:1312.5885}}.

\bibitem{Patra:2012ur}
S.~Patra, ``{\em {Neutrinoless double beta decay process in left-right
  symmetric models without scalar bidoublet}},''
  \href{http://dx.doi.org/10.1103/PhysRevD.87.015002}{Phys.Rev. {\normalfont
  \bfseries D87} (2013) no.~1, 015002},
\href{http://arxiv.org/abs/1212.0612}{{\normalfont \ttfamily arXiv:1212.0612}}.

\bibitem{Chakrabortty:2012mh}
J.~Chakrabortty, H.~Z. Devi, S.~Goswami, and S.~Patra, ``{\em {Neutrinoless
  double-$\beta$ decay in TeV scale Left-Right symmetric models}},''
  \href{http://dx.doi.org/10.1007/JHEP08(2012)008}{JHEP {\normalfont \bfseries
  08} (2012)  008},
\href{http://arxiv.org/abs/1204.2527}{{\normalfont \ttfamily arXiv:1204.2527}}.

\bibitem{Dev:2013vxa}
P.~Bhupal~Dev, S.~Goswami, M.~Mitra, and W.~Rodejohann, ``{\em {Constraining
  Neutrino Mass from Neutrinoless Double Beta Decay}},''
  \href{http://dx.doi.org/10.1103/PhysRevD.88.091301}{Phys.Rev. {\normalfont
  \bfseries D88} (2013)  091301},
\href{http://arxiv.org/abs/1305.0056}{{\normalfont \ttfamily arXiv:1305.0056}}.

\bibitem{Nemevsek:2011hz}
M.~Nemevsek, F.~Nesti, G.~Senjanovic, and Y.~Zhang, ``{\em {First Limits on
  Left-Right Symmetry Scale from LHC Data}},''
  \href{http://dx.doi.org/10.1103/PhysRevD.83.115014}{Phys. Rev. {\normalfont
  \bfseries D83} (2011)  115014},
\href{http://arxiv.org/abs/1103.1627}{{\normalfont \ttfamily arXiv:1103.1627}}.

\bibitem{Dev:2014iva}
P.~S. Bhupal~Dev, C.-H. Lee, and R.~N. Mohapatra, ``{\em {Leptogenesis
  Constraints on the Mass of Right-handed Gauge Bosons}},''
  \href{http://dx.doi.org/10.1103/PhysRevD.90.095012}{Phys. Rev. {\normalfont
  \bfseries D90} (2014) no.~9, 095012},
\href{http://arxiv.org/abs/1408.2820}{{\normalfont \ttfamily arXiv:1408.2820}}.

\bibitem{Keung:1983uu}
W.-Y. Keung and G.~Senjanovi{\'c}, ``{\em {Majorana Neutrinos and the
  Production of the Right-handed Charged Gauge Boson}},''
\href{http://dx.doi.org/10.1103/PhysRevLett.50.1427}{Phys.Rev.Lett.
  {\normalfont \bfseries 50} (1983)  1427}.

\bibitem{Das:2012ii}
S.~Das, F.~Deppisch, O.~Kittel, and J.~Valle, ``{\em {Heavy Neutrinos and
  Lepton Flavour Violation in Left-Right Symmetric Models at the LHC}},''
  \href{http://dx.doi.org/10.1103/PhysRevD.86.055006}{Phys.Rev. {\normalfont
  \bfseries D86} (2012)  055006},
\href{http://arxiv.org/abs/1206.0256}{{\normalfont \ttfamily arXiv:1206.0256}}.

\bibitem{Bertolini:2014sua}
S.~Bertolini, A.~Maiezza, and F.~Nesti, ``{\em {Present and Future K and B
  Meson Mixing Constraints on TeV Scale Left-Right Symmetry}},''
  \href{http://dx.doi.org/10.1103/PhysRevD.89.095028}{Phys.Rev. {\normalfont
  \bfseries D89} (2014)  095028},
\href{http://arxiv.org/abs/1403.7112}{{\normalfont \ttfamily arXiv:1403.7112}}.

\bibitem{Beall:1981ze}
G.~Beall, M.~Bander, and A.~Soni, ``{\em {Constraint on the Mass Scale of a
  Left-Right Symmetric Electroweak Theory from the $K_L$--$K_S$ Mass
  Difference}},''
\href{http://dx.doi.org/10.1103/PhysRevLett.48.848}{Phys.Rev.Lett. {\normalfont
  \bfseries 48} (1982)  848}.

\bibitem{Lindner:2016lpp}
M.~Lindner, F.~S. Queiroz, and W.~Rodejohann, ``{\em {Dilepton bounds on
  left-right symmetry at the LHC run II and neutrinoless double beta decay}},''
\href{http://arxiv.org/abs/1604.07419}{{\normalfont \ttfamily
  arXiv:1604.07419}}.

\bibitem{Deppisch:2016scs}
F.~F. Deppisch, C.~Hati, S.~Patra, P.~Pritimita, and U.~Sarkar, ``{\em
  {Implications of the diphoton excess on left–right models and gauge
  unification}},''
  \href{http://dx.doi.org/10.1016/j.physletb.2016.03.081}{Phys. Lett.
  {\normalfont \bfseries B757} (2016)  223--230},
\href{http://arxiv.org/abs/1601.00952}{{\normalfont \ttfamily
  arXiv:1601.00952}}.

\bibitem{Dev:2014xea}
P.~S. Bhupal~Dev, S.~Goswami, and M.~Mitra, ``{\em {TeV Scale Left-Right
  Symmetry and Large Mixing Effects in Neutrinoless Double Beta Decay}},''
  \href{http://dx.doi.org/10.1103/PhysRevD.91.113004}{Phys. Rev. {\normalfont
  \bfseries D91} (2015) no.~11, 113004},
\href{http://arxiv.org/abs/1405.1399}{{\normalfont \ttfamily arXiv:1405.1399}}.

\bibitem{Bambhaniya:2015ipg}
G.~Bambhaniya, P.~S.~B. Dev, S.~Goswami, and M.~Mitra, ``{\em {The Scalar
  Triplet Contribution to Lepton Flavour Violation and Neutrinoless Double Beta
  Decay in Left-Right Symmetric Model}},''
  \href{http://dx.doi.org/10.1007/JHEP04(2016)046}{JHEP {\normalfont \bfseries
  04} (2016)  046},
\href{http://arxiv.org/abs/1512.00440}{{\normalfont \ttfamily
  arXiv:1512.00440}}.

\bibitem{Ge:2015yqa}
S.-F. Ge, M.~Lindner, and S.~Patra, ``{\em {New physics effects on neutrinoless
  double beta decay from right-handed current}},''
  \href{http://dx.doi.org/10.1007/JHEP10(2015)077}{JHEP {\normalfont \bfseries
  10} (2015)  077},
\href{http://arxiv.org/abs/1508.07286}{{\normalfont \ttfamily
  arXiv:1508.07286}}.

\bibitem{Deppisch:2015cua}
F.~F. Deppisch, L.~Graf, S.~Kulkarni, S.~Patra, W.~Rodejohann, N.~Sahu, and
  U.~Sarkar, ``{\em {Reconciling the 2 TeV excesses at the LHC in a linear
  seesaw left-right model}},''
  \href{http://dx.doi.org/10.1103/PhysRevD.93.013011}{Phys. Rev. {\normalfont
  \bfseries D93} (2016) no.~1, 013011},
\href{http://arxiv.org/abs/1508.05940}{{\normalfont \ttfamily
  arXiv:1508.05940}}.

\bibitem{Hirsch:1996qw}
M.~Hirsch, H.~V. Klapdor-Kleingrothaus, and O.~Panella, ``{\em {Double beta
  decay in left-right symmetric models}},''
  \href{http://dx.doi.org/10.1016/0370-2693(96)00185-2}{Phys. Lett.
  {\normalfont \bfseries B374} (1996)  7--12},
\href{http://arxiv.org/abs/hep-ph/9602306}{{\normalfont \ttfamily
  arXiv:hep-ph/9602306}}.

\bibitem{Dev:2013oxa}
C.-H. Lee, P.~S. Bhupal~Dev, and R.~N. Mohapatra, ``{\em {Natural TeV-scale
  left-right seesaw mechanism for neutrinos and experimental tests}},''
  \href{http://dx.doi.org/10.1103/PhysRevD.88.093010}{Phys. Rev. {\normalfont
  \bfseries D88} (2013) no.~9, 093010},
\href{http://arxiv.org/abs/1309.0774}{{\normalfont \ttfamily arXiv:1309.0774}}.

\bibitem{Dev:2015pga}
P.~S. Bhupal~Dev and R.~N. Mohapatra, ``{\em {Unified explanation of the
  $eejj$, diboson and dijet resonances at the LHC}},''
  \href{http://dx.doi.org/10.1103/PhysRevLett.115.181803}{Phys. Rev. Lett.
  {\normalfont \bfseries 115} (2015) no.~18, 181803},
\href{http://arxiv.org/abs/1508.02277}{{\normalfont \ttfamily
  arXiv:1508.02277}}.

\bibitem{Dhuria:2015cfa}
M.~Dhuria, C.~Hati, R.~Rangarajan, and U.~Sarkar, ``{\em {Falsifying
  leptogenesis for a TeV scale $W^{\pm}_{R}$ at the LHC}},''
  \href{http://dx.doi.org/10.1103/PhysRevD.92.031701}{Phys. Rev. {\normalfont
  \bfseries D92} (2015) no.~3, 031701},
\href{http://arxiv.org/abs/1503.07198}{{\normalfont \ttfamily
  arXiv:1503.07198}}.

\bibitem{Patra:2015bga}
S.~Patra, F.~S. Queiroz, and W.~Rodejohann, ``{\em {Stringent Dilepton Bounds
  on Left-Right Models using LHC data}},''
  \href{http://dx.doi.org/10.1016/j.physletb.2015.11.009}{Phys. Lett.
  {\normalfont \bfseries B752} (2016)  186--190},
\href{http://arxiv.org/abs/1506.03456}{{\normalfont \ttfamily
  arXiv:1506.03456}}.

\bibitem{Heeck:2015qra}
J.~Heeck and S.~Patra, ``{\em {Minimal Left-Right Symmetric Dark Matter}},''
  \href{http://dx.doi.org/10.1103/PhysRevLett.115.121804}{Phys. Rev. Lett.
  {\normalfont \bfseries 115} (2015) no.~12, 121804},
\href{http://arxiv.org/abs/1507.01584}{{\normalfont \ttfamily
  arXiv:1507.01584}}.

\bibitem{Borah:2016ees}
D.~Borah, A.~Dasgupta, and S.~Patra, ``{\em {Common Origin of $3.55$ keV X-ray
  line and Gauge Coupling Unification with Left-Right Dark Matter}},''
\href{http://arxiv.org/abs/1604.01929}{{\normalfont \ttfamily
  arXiv:1604.01929}}.

\bibitem{Patra:2015vmp}
S.~Patra and S.~Rao, ``{\em {Singlet fermion Dark Matter within Left-Right
  Model}},'' \href{http://dx.doi.org/10.1016/j.physletb.2016.05.098}{Phys.
  Lett. {\normalfont \bfseries B759} (2016)  454--458},
\href{http://arxiv.org/abs/1512.04053}{{\normalfont \ttfamily
  arXiv:1512.04053}}.

\bibitem{Patra:2015qny}
S.~Patra, ``{\em {Dark matter, lepton and baryon number, and left-right
  symmetric theories}},''
  \href{http://dx.doi.org/10.1103/PhysRevD.93.093001}{Phys. Rev. {\normalfont
  \bfseries D93} (2016) no.~9, 093001},
\href{http://arxiv.org/abs/1512.04739}{{\normalfont \ttfamily
  arXiv:1512.04739}}.

\bibitem{Garcia-Cely:2015quu}
C.~Garcia-Cely and J.~Heeck, ``{\em {Phenomenology of left-right symmetric dark
  matter}},'' \href{http://arxiv.org/abs/1512.03332}{{\normalfont \ttfamily
  arXiv:1512.03332}}.
[JCAP1603,021(2016)].

\bibitem{Borah:2016iqd}
D.~Borah and A.~Dasgupta, ``{\em {Charged lepton flavour violcxmation and
  neutrinoless double beta decay in left-right symmetric models with type I+II
  seesaw}},'' \href{http://dx.doi.org/10.1007/JHEP07(2016)022}{JHEP
  {\normalfont \bfseries 07} (2016)  022},
\href{http://arxiv.org/abs/1606.00378}{{\normalfont \ttfamily
  arXiv:1606.00378}}.

\bibitem{Borah:2015ufa}
D.~Borah and A.~Dasgupta, ``{\em {Neutrinoless Double Beta Decay in Type I+II
  Seesaw Models}},'' \href{http://dx.doi.org/10.1007/JHEP11(2015)208}{JHEP
  {\normalfont \bfseries 11} (2015)  208},
\href{http://arxiv.org/abs/1509.01800}{{\normalfont \ttfamily
  arXiv:1509.01800}}.

\bibitem{Awasthi:2016kbk}
R.~L. Awasthi, A.~Dasgupta, and M.~Mitra, ``{\em {Limiting the Effective Mass
  and New Physics Parameters from $0\nu\beta\beta$}},''
\href{http://arxiv.org/abs/1607.03835}{{\normalfont \ttfamily
  arXiv:1607.03835}}.

\bibitem{Awasthi:2015ota}
R.~L. Awasthi, P.~S.~B. Dev, and M.~Mitra, ``{\em {Implications of the Diboson
  Excess for Neutrinoless Double Beta Decay and Lepton Flavor Violation in TeV
  Scale Left Right Symmetric Model}},''
  \href{http://dx.doi.org/10.1103/PhysRevD.93.011701}{Phys. Rev. {\normalfont
  \bfseries D93} (2016) no.~1, 011701},
\href{http://arxiv.org/abs/1509.05387}{{\normalfont \ttfamily
  arXiv:1509.05387}}.

\bibitem{Nayak:2013dza}
B.~P. Nayak and M.~K. Parida, ``{\em {New mechanism for Type-II seesaw
  dominance in SO(10) with low-mass $Z^{\prime }$, RH neutrinos, and verifiable
  LFV, LNV and proton decay}},''
  \href{http://dx.doi.org/10.1140/epjc/s10052-015-3385-x}{Eur. Phys. J.
  {\normalfont \bfseries C75} (2015)  183},
\href{http://arxiv.org/abs/1312.3185}{{\normalfont \ttfamily arXiv:1312.3185}}.

\bibitem{Barry:2011wb}
J.~Barry, W.~Rodejohann, and H.~Zhang, ``{\em {Light Sterile Neutrinos: Models
  and Phenomenology}},'' \href{http://dx.doi.org/10.1007/JHEP07(2011)091}{JHEP
  {\normalfont \bfseries 07} (2011)  091},
\href{http://arxiv.org/abs/1105.3911}{{\normalfont \ttfamily arXiv:1105.3911}}.

\bibitem{Zhang:2011vh}
H.~Zhang, ``{\em {Light Sterile Neutrino in the Minimal Extended Seesaw}},''
  \href{http://dx.doi.org/10.1016/j.physletb.2012.06.074}{Phys. Lett.
  {\normalfont \bfseries B714} (2012)  262--266},
\href{http://arxiv.org/abs/1110.6838}{{\normalfont \ttfamily arXiv:1110.6838}}.

\bibitem{Dev:2009aw}
P.~S.~B. Dev and R.~N. Mohapatra, ``{\em {TeV Scale Inverse Seesaw in SO(10)
  and Leptonic Non-Unitarity Effects}},''
  \href{http://dx.doi.org/10.1103/PhysRevD.81.013001}{Phys. Rev. {\normalfont
  \bfseries D81} (2010)  013001},
\href{http://arxiv.org/abs/0910.3924}{{\normalfont \ttfamily arXiv:0910.3924}}.

\bibitem{Muto:1989cd}
K.~Muto, E.~Bender, and H.~V. Klapdor, ``{\em {Nuclear Structure Effects on the
  Neutrinoless Double Beta Decay}},''
Z. Phys. {\normalfont \bfseries A334} (1989)  187--194.

\bibitem{Suhonen:1998ck}
J.~Suhonen and O.~Civitarese, ``{\em {Weak-interaction and nuclear-structure
  aspects of nuclear double beta decay}},''
\href{http://dx.doi.org/10.1016/S0370-1573(97)00087-2}{Phys. Rept. {\normalfont
  \bfseries 300} (1998)  123--214}.

\bibitem{Fritzsch:1974nn}
H.~Fritzsch and P.~Minkowski, ``{\em {Unified Interactions of Leptons and
  Hadrons}},''
\href{http://dx.doi.org/10.1016/0003-4916(75)90211-0}{Annals Phys. {\normalfont
  \bfseries 93} (1975)  193--266}.

\bibitem{LalAwasthi:2011aa}
R.~Lal~Awasthi and M.~K. Parida, ``{\em {Inverse Seesaw Mechanism in
  Nonsupersymmetric SO(10), Proton Lifetime, Nonunitarity Effects, and a
  Low-mass Z' Boson}},''
  \href{http://dx.doi.org/10.1103/PhysRevD.86.093004}{Phys. Rev. {\normalfont
  \bfseries D86} (2012)  093004},
\href{http://arxiv.org/abs/1112.1826}{{\normalfont \ttfamily arXiv:1112.1826}}.

\bibitem{Agashe:2014kda}
{\normalfont \bfseries Particle Data Group}, K.~A. Olive {\em et al.}, ``{\em
  {Review of Particle Physics}},''
\href{http://dx.doi.org/10.1088/1674-1137/38/9/090001}{Chin. Phys. {\normalfont
  \bfseries C38} (2014)  090001}.

\bibitem{Awasthi:2013we}
R.~L. Awasthi, M.~K. Parida, and S.~Patra, ``{\em {Neutrinoless double beta
  decay and pseudo-Dirac neutrino mass predictions through inverse seesaw
  mechanism}},''
\href{http://arxiv.org/abs/1301.4784}{{\normalfont \ttfamily arXiv:1301.4784}}.

\bibitem{Maiezza:2010ic}
A.~Maiezza, M.~Nemevsek, F.~Nesti, and G.~Senjanovic, ``{\em {Left-Right
  Symmetry at LHC}},''
  \href{http://dx.doi.org/10.1103/PhysRevD.82.055022}{Phys. Rev. {\normalfont
  \bfseries D82} (2010)  055022},
\href{http://arxiv.org/abs/1005.5160}{{\normalfont \ttfamily arXiv:1005.5160}}.

\bibitem{Zhang:2007da}
Y.~Zhang, H.~An, X.~Ji, and R.~N. Mohapatra, ``{\em {General CP Violation in
  Minimal Left-Right Symmetric Model and Constraints on the Right-Handed
  Scale}},'' \href{http://dx.doi.org/10.1016/j.nuclphysb.2008.05.019}{Nucl.
  Phys. {\normalfont \bfseries B802} (2008)  247--279},
\href{http://arxiv.org/abs/0712.4218}{{\normalfont \ttfamily arXiv:0712.4218}}.

\bibitem{GonzalezGarcia:2012sz}
M.~C. Gonzalez-Garcia, M.~Maltoni, J.~Salvado, and T.~Schwetz, ``{\em {Global
  fit to three neutrino mixing: critical look at present precision}},''
  \href{http://dx.doi.org/10.1007/JHEP12(2012)123}{JHEP {\normalfont \bfseries
  12} (2012)  123},
\href{http://arxiv.org/abs/1209.3023}{{\normalfont \ttfamily arXiv:1209.3023}}.

\bibitem{Fogli:2012ua}
G.~L. Fogli, E.~Lisi, A.~Marrone, D.~Montanino, A.~Palazzo, and A.~M. Rotunno,
  ``{\em {Global analysis of neutrino masses, mixings and phases: entering the
  era of leptonic CP violation searches}},''
  \href{http://dx.doi.org/10.1103/PhysRevD.86.013012}{Phys. Rev. {\normalfont
  \bfseries D86} (2012)  013012},
\href{http://arxiv.org/abs/1205.5254}{{\normalfont \ttfamily arXiv:1205.5254}}.

\bibitem{Gonzalez-Garcia:2014bfa}
M.~C. Gonzalez-Garcia, M.~Maltoni, and T.~Schwetz, ``{\em {Updated fit to three
  neutrino mixing: status of leptonic CP violation}},''
  \href{http://dx.doi.org/10.1007/JHEP11(2014)052}{JHEP {\normalfont \bfseries
  11} (2014)  052},
\href{http://arxiv.org/abs/1409.5439}{{\normalfont \ttfamily arXiv:1409.5439}}.

\bibitem{Ade:2013zuv}
{\normalfont \bfseries Planck}, P.~A.~R. Ade {\em et al.}, ``{\em {Planck 2013
  results. XVI. Cosmological parameters}},''
  \href{http://dx.doi.org/10.1051/0004-6361/201321591}{Astron. Astrophys.
  {\normalfont \bfseries 571} (2014)  A16},
\href{http://arxiv.org/abs/1303.5076}{{\normalfont \ttfamily arXiv:1303.5076}}.

\bibitem{Meroni:2012qf}
A.~Meroni, S.~T. Petcov, and F.~Simkovic, ``{\em {Multiple CP Non-conserving
  Mechanisms of $\beta \beta$-Decay and Nuclei with Largely Different Nuclear
  Matrix Elements}},'' \href{http://dx.doi.org/10.1007/JHEP02(2013)025}{JHEP
  {\normalfont \bfseries 02} (2013)  025},
\href{http://arxiv.org/abs/1212.1331}{{\normalfont \ttfamily arXiv:1212.1331}}.

\bibitem{Seljak:2004xh}
{\normalfont \bfseries SDSS}, U.~Seljak {\em et al.}, ``{\em {Cosmological
  parameter analysis including SDSS Ly-alpha forest and galaxy bias:
  Constraints on the primordial spectrum of fluctuations, neutrino mass, and
  dark energy}},'' \href{http://dx.doi.org/10.1103/PhysRevD.71.103515}{Phys.
  Rev. {\normalfont \bfseries D71} (2005)  103515},
\href{http://arxiv.org/abs/astro-ph/0407372}{{\normalfont \ttfamily
  arXiv:astro-ph/0407372}}.

\bibitem{Costanzi:2014tna}
M.~Costanzi, B.~Sartoris, M.~Viel, and S.~Borgani, ``{\em {Neutrino
  constraints: what large-scale structure and CMB data are telling us?}},''
  \href{http://dx.doi.org/10.1088/1475-7516/2014/10/081}{JCAP {\normalfont
  \bfseries 1410} (2014) no.~10, 081},
\href{http://arxiv.org/abs/1407.8338}{{\normalfont \ttfamily arXiv:1407.8338}}.

\bibitem{Palanque-Delabrouille:2014jca}
N.~Palanque-Delabrouille {\em et al.}, ``{\em {Constraint on neutrino masses
  from SDSS-III/BOSS Ly$\alpha$ forest and other cosmological probes}},''
  \href{http://dx.doi.org/10.1088/1475-7516/2015/02/045}{JCAP {\normalfont
  \bfseries 1502} (2015) no.~02, 045},
\href{http://arxiv.org/abs/1410.7244}{{\normalfont \ttfamily arXiv:1410.7244}}.

\bibitem{Menendez:2008jp}
J.~Menendez, A.~Poves, E.~Caurier, and F.~Nowacki, ``{\em {Disassembling the
  Nuclear Matrix Elements of the Neutrinoless beta beta Decay}},''
  \href{http://dx.doi.org/10.1016/j.nuclphysa.2008.12.005}{Nucl. Phys.
  {\normalfont \bfseries A818} (2009)  139--151},
\href{http://arxiv.org/abs/0801.3760}{{\normalfont \ttfamily arXiv:0801.3760}}.

\bibitem{Simkovic:2009pp}
F.~Simkovic, A.~Faessler, H.~Muther, V.~Rodin, and M.~Stauf, ``{\em {The 0 nu
  bb-decay nuclear matrix elements with self-consistent short-range
  correlations}},'' \href{http://dx.doi.org/10.1103/PhysRevC.79.055501}{Phys.
  Rev. {\normalfont \bfseries C79} (2009)  055501},
\href{http://arxiv.org/abs/0902.0331}{{\normalfont \ttfamily arXiv:0902.0331}}.

\bibitem{Barea:2009zza}
J.~Barea and F.~Iachello, ``{\em {Neutrinoless double-beta decay in the
  microscopic interacting boson model}},''
\href{http://dx.doi.org/10.1103/PhysRevC.79.044301}{Phys. Rev. {\normalfont
  \bfseries C79} (2009)  044301}.

\bibitem{Barea:2013bz}
J.~Barea, J.~Kotila, and F.~Iachello, ``{\em {Nuclear matrix elements for
  double-$\beta$ decay}},''
  \href{http://dx.doi.org/10.1103/PhysRevC.87.014315}{Phys. Rev. {\normalfont
  \bfseries C87} (2013) no.~1, 014315},
\href{http://arxiv.org/abs/1301.4203}{{\normalfont \ttfamily arXiv:1301.4203}}.

\bibitem{Faessler:2011rv}
A.~Faessler, G.~L. Fogli, E.~Lisi, A.~M. Rotunno, and F.~Simkovic, ``{\em
  {Multi-Isotope Degeneracy of Neutrinoless Double Beta Decay Mechanisms in the
  Quasi-Particle Random Phase Approximation}},''
  \href{http://dx.doi.org/10.1103/PhysRevD.83.113015}{Phys. Rev. {\normalfont
  \bfseries D83} (2011)  113015},
\href{http://arxiv.org/abs/1103.2504}{{\normalfont \ttfamily arXiv:1103.2504}}.

\bibitem{Pantis:1996py}
G.~Pantis, F.~Simkovic, J.~D. Vergados, and A.~Faessler, ``{\em {Neutrinoless
  double beta decay within QRPA with proton - neutron pairing}},''
  \href{http://dx.doi.org/10.1103/PhysRevC.53.695}{Phys. Rev. {\normalfont
  \bfseries C53} (1996)  695--707},
\href{http://arxiv.org/abs/nucl-th/9612036}{{\normalfont \ttfamily
  arXiv:nucl-th/9612036}}.

\bibitem{Parida:2016asc}
M.~K. Parida and B.~P. Nayak, ``{\em {Sterile Neutrinos, Dominant Seesaw
  Mechanisms, Double Beta Decay, and Other Predictions}},''
\href{http://arxiv.org/abs/1607.07236}{{\normalfont \ttfamily
  arXiv:1607.07236}}.

\end{thebibliography}\endgroup
\end{document}